\pdfoutput=1
\PassOptionsToPackage{svgnames}{xcolor}
\PassOptionsToPackage{table}{xcolor}
\documentclass[letterpaper,12pt,reqno]{article}
\usepackage[markup=underlined]{changes}
\RequirePackage{etex}

\usepackage{etex}
\usepackage{graphicx}
\usepackage{amsmath}
\usepackage{amssymb}
\usepackage{amsfonts}
\usepackage{titlesec}
\usepackage{pictex}
\usepackage{comment}
\usepackage{amsthm}
\usepackage{graphics,epsfig,verbatim,bm,latexsym,url,amsbsy}
\usepackage{rotating}
\usepackage[authoryear,round]{natbib}
\usepackage{float}
\usepackage[position=bottom]{subfig}
\usepackage{mathrsfs}
\usepackage{multirow}
\usepackage{array}
\usepackage{bigints}
\usepackage{bbm}
\usepackage[ruled,vlined]{algorithm2e}
\usepackage{soul}
\usepackage{mathtools,nccmath}
\usepackage{enumitem}
\usepackage[table]{xcolor}
\usepackage{hhline}

\usepackage{hyperref}
\hypersetup{
	colorlinks=true,
	linkcolor={red!70!black},
	citecolor={blue!70!black},
	filecolor=magenta,      
	urlcolor={blue!70!black},
}

\usepackage{mathpazo} 

\DeclarePairedDelimiterX{\inp}[2]{\langle}{\rangle}{#1, #2}

\newcommand{\Ex}{\mathbb{E}}
\newcommand{\N}{\mathbb{N}}

\renewcommand{\Pr}{\mathbb{P}}
\DeclareSymbolFont{largesymbolsA}{U}{txexa}{m}{n}
\DeclareMathSymbol{\varprod}{\mathop}{largesymbolsA}{16}

\theoremstyle{definition} \newtheorem{example}{Example}
\theoremstyle{definition} 
\theoremstyle{definition} \newtheorem{corollary}{Corollary}

\theoremstyle{definition} 
\theoremstyle{definition} \newtheorem{definition}{Definition}
\theoremstyle{definition} 
\theoremstyle{plain} \newtheorem{Lemma}{Lemma}
\theoremstyle{plain} \newtheorem{lemma}{Lemma}

\theoremstyle{plain}\newtheorem{theorem}{Theorem}
\theoremstyle{plain}\newtheorem{proposition}{Proposition}

\theoremstyle{definition} \newtheorem{assumption}{Assumption}
\theoremstyle{definition} \newtheorem*{assumptionnonumber}{Assumption}
\theoremstyle{definition} 
\theoremstyle{definition} 

\theoremstyle{definition} 
\theoremstyle{definition} 
\theoremstyle{definition} 
\theoremstyle{definition} 
\theoremstyle{definition} 
\theoremstyle{definition} 

\theoremstyle{definition} 

\usepackage{mathtools}
\usepackage[a4paper, left=2.5cm,right=2.5cm,top=2cm,bottom=2cm, includefoot,heightrounded]{geometry}
\usepackage{setspace}
\titleformat{\section}
		{\scshape \bfseries\center}
         {\thesection}
        {0.5em}
        {}
        []

\titleformat{\subsection}[runin]
        {\normalfont\bfseries}
        {\thesubsection}
        {0.5em}
        {\addperiod}
        []
\newcommand{\addperiod}[1]{#1.}

\titleformat{\paragraph}[runin]
		{\normalfont\bfseries}
         {}
        {0}
        {}
        []

\titlespacing*{\paragraph}{0pt}{1ex}{2ex}

\expandafter\def\expandafter\normalsize\expandafter{
    \normalsize%
    \setlength\abovedisplayskip{3pt}%
    \setlength\belowdisplayskip{3pt}%
    \setlength\abovedisplayshortskip{-10pt}%
    \setlength\belowdisplayshortskip{10pt}%
}

\title{Attention Capture}
\date{This version: \today}
\author{\makebox[.3\linewidth]{{Andrew Koh}\thanks{MIT Department of Economics; email: \protect\texttt{ajkoh@mit.edu}}}\\{MIT} \and \makebox[.3\linewidth]{{Sivakorn Sanguanmoo}\thanks{MIT Department of Economics; email: \protect\texttt{sanguanm@mit.edu} \newline
First posted version: September 2022. This paper partially subsumes ``Feasible Joint Distributions over Actions and Stopping Times'' by the same authors which was previously circulated as a separate paper. 
\newline \newline 
We are especially grateful to Drew Fudenberg, Stephen Morris, and Parag Pathak for their guidance and support, and for suggestions which substantially improved this paper. We also thank Ian Ball, Alessandro Bonatti, Yi-Chun Chen, Roberto Corrao, Laura Doval, Matt Elliott, Giacomo Lanzani, Ellen Muir, John Quah, Ludvig Sinander, Tomasz Strzalecki, Alex Wolitzky, and Weijie Zhong, as well as audiences at the 2022 MIT Summer Theory Lab, MIT Theory and Industrial Organization Lunches, National University of Singapore, and Singapore Management University for valuable comments.}}\\{MIT}}

\begin{document}
\maketitle
\begin{abstract}

\small 
We develop a unified analysis of how information captures attention. A decision maker (DM) faces a dynamic information structure and decides when to stop paying attention. We characterize the convex-order frontier and extreme points of feasible stopping times, as well as dynamic information structures which implement them. This delivers the form of optimal attention capture as a function of the designer and DM’s relative time preferences. Intertemporal commitment is unnecessary: sequentially optimal information structures always exist by inducing stochastic interim beliefs. We further analyze optimal attention capture under noninstrumental value for information. Our results speak directly to the attention economy.
\end{abstract}

\normalsize 

\setstretch{1.15}

\clearpage 

\section{Introduction}
The modern age is brimming with information. Yet, our access to it is mediated by platforms which funnel our attention toward algorithmically curated streams of content---articles to digest, videos to watch, feeds to scroll---which we have little control over. 
These platforms, in turn, generate revenue through advertisements; their business, therefore, is to capture, repackage, and monetize attention.


Motivated by these developments,\footnote{Platforms are typically paid per `impression' (view) or `click'. In the first quarter of 2022, 97\% of Facebook's revenue, 81\% of Google's revenue, and 92\% of Twitter's revenue came from advertising.} we study the extent to which information can be used to capture attention. A designer (they) controls a decision maker's (DM, she) knowledge of an uncertain state  by choosing some dynamic information structure 
which specifies, for each history of past messages, a distribution over future messages. The DM values information instrumentally\footnote{Information can, of course, have noninstrumental value e.g., \cite{kreps1978temporal}. In Section \ref{sec:noninstrumental} we study optimal attention capture when information is valued noninstrumentally in the sense of \cite*{ely2015suspense}.}
and, at each point in time, chooses between observing more information and stopping to act. On some paths of realized messages, the DM might become quickly become confident and stop to act; on other paths, she might observe contradictory signals and so finds it worthwhile to wait for more information. Hence, dynamic information structures induce distributions over the DM's optimal stopping times. 
How should a designer who values attention provide information over time? 

Our contribution is to give a relatively complete characterization of (i) all stopping times achievable through information and properties of dynamic information structures which implement them; (ii) dynamic information structures which optimally capture attention via a characterization of the convex-order frontier and etxreme-points of feasbile stopping times; (iii) the requisite degree of intertemporal commitment (none); (iv) optimal attention capture when information is valued noninstrumentally; and (v) optimal dynamic information when the designer values both attention and persuasion separably.  

\subsection{Outline of contribution} 
\begin{figure}[h!]
    \centering    
       \caption[Caption for LOF]{Connections between aspects of attention capture
       }
{\includegraphics[width=0.999\textwidth]{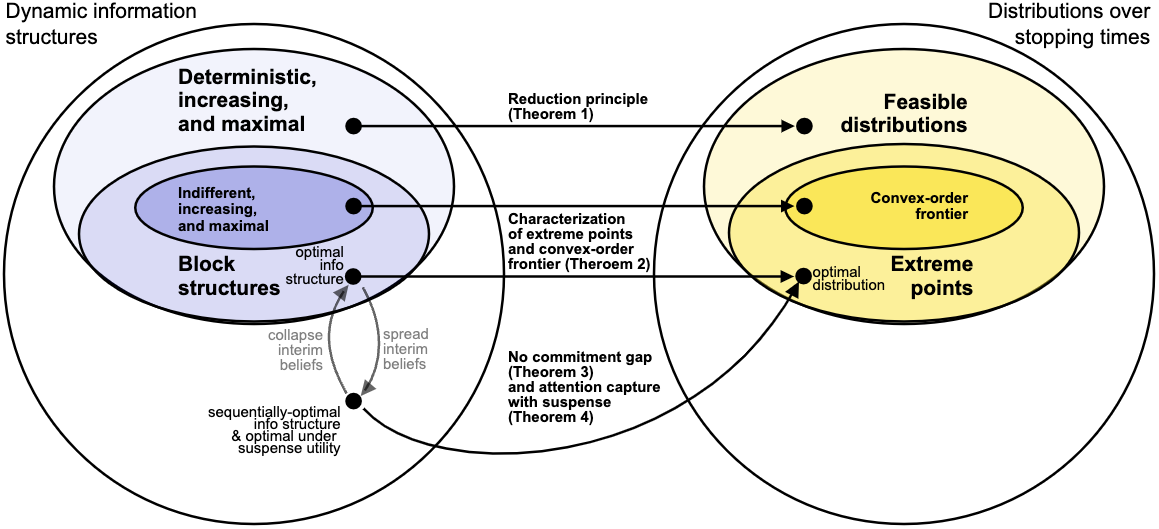}} 
\label{fig:contribution}
\end{figure}

\mbox{} 
\paragraph{Reduction to deterministic, increasing, and maximal belief paths.} Our first result (Theorem \ref{thrm:reduction}) establishes a reduction principle: any feasible distribution over stopping times can be implemented with a special class of dynamic information structures such that upon stopping, DM learns the state perfectly and upon continuing, the DM's continuation beliefs follow a {deterministic path} which is (i) \emph{increasing}: the longer she waits, the more uncertain she becomes; and (ii) \emph{maximal}: at each time step her continuation beliefs move as much as the martingale condition allows. This is represented by the top arrow in Figure \ref{fig:contribution}.

Conceptually, the reduction principle highlights the key role that increasing continuation belief paths play in shaping dynamic continuation incentives. A distinctive feature of attention capture is that the DM's value for information over time depends on her interim belief which is itself endogenous to the dynamic information structure---this is a margin along which the designer exploits. Practically, the reduction principle ensures continuation histories do not branch: conditional on continuing until time $t$, DM's beliefs are uniquely pinned down by the continuation belief path. This simplifies the designer's problem to whether we can find a continuation belief path supporting a target distribution over stopping times.

\noindent \paragraph{Optimal attention capture via convex-orders and extreme points.} We next turn to the structure of optimal attention capture. We completely characterization of the convex-order frontier over the set of feasible stopping times (Theorem \ref{thrm:convex_ext} (i)). Distributions on the convex-order frontier are supported by increasing and maximal belief paths which keep the DM indifferent between continuing and stopping at each point in time. This is depicted by the second arrow in Figure \ref{fig:contribution}. Thus, increasingness and maximality are both \emph{sufficient} to achieve any feasible stopping time and \emph{necessary} to achieve the convex frontier. 

Our results offer a transparent perspective on the form of optimal attention with an arbitrary number of states, arbitrary DM decision problems, and arbitrary (but additively separable) designer and DM time preferences. When the designer's time preferences are more convex than that of the DM's, optimal dynamic information structures must generate full information at a time-inhomogeneous geometric rate which keeps the DM indifferent and induces increasing and maximal continuation belief paths. Conversely, when the designer's preferences are more concave than that of the DM's, optimal dynamic information structures induces no movement in interim beliefs until full information arrives at a fixed time.

We further analyze optimal attention capture with arbitrary designer value and arbitrary DM costs. This allows us to understand environments beyond convex/concave preferences e.g., `S-shaped' response curves commonly studied in the marketing literature. We show that extreme points of the set of feasible stopping times\footnote{When the DM's prior is such that her value for information is maximized.} coincide exactly with those induced by dynamic information with \emph{block structures} such that the support of the stopping times are the times at which the DM is indifferent between stopping and continuing (Theorem \ref{thrm:convex_ext} (ii)).\footnote{With the possible exception of a pair of boundary times.} This is depicted by the third arrow in Figure \ref{fig:contribution}. That is, time is partitioned into blocks: in the interior of each block, the DM never receives information; at the boundaries, the DM has some chance to learn the state but, if she does not, is indifferent between continuing and stopping. Our tight characterization of the extreme points of this set are distinct from those of majorization \citep*{kleiner2021extreme}: feasibility of stopping times is strictly more stringent than either being majorized by, or majorizing another feasible stopping time.\footnote{It is more evidently distinct from the set of extreme points of first-order stochastic dominance \citep{yang2023economics}.} This might be of interest in future work on dynamic mechanism or information design where constraints on the conditional distribution feature prominently.\footnote{For instance, the convex-order frontier of majorization developed in \cite*{kleiner2021extreme} has been productively used in security design \citep*{gershkov2024optimal}.}

Economically, block structures have the natural interpretation as the policy of a platform that controls the duration of advertisements (length of each block). Over the duration of each advertisement (interior of each block), the DM receives no information; at the end of each advertisement, the designer gives the DM full information with some probability, otherwise it shows the DM yet another advertisement at which point the DM is indifferent between continuing and stopping.\footnote{For instance, this is the form of advertisements (ads) on popular streaming platforms such as Youtube, Spotify, and Tiktok which frequently show users `unskippable' ads. Such platforms conduct extensive optimization over the type, duration, and number of ads users see.} Crucially, since the DM has already sunk her attention costs, conditional on being shown yet another advertisement, she finds it weakly optimal to continue paying attention although she might have been better-off not paying attention in the first place. 

\paragraph{Sequential optimality.} In much of the paper we will assume the designer can commit to provision of future information. This promise, in turn, incentivizes the DM to pay attention in the present. We show that intertemporal commitment is unnecessary (Theorem \ref{thrm:time_con}): optimal information structures \emph{always} have \emph{sequentially optimal modifications}. This is represented by the bottom arrow in Figure \ref{fig:contribution}. We give an explicit procedure of such modifications: the designer spreads \emph{interim beliefs} randomly such as to raise the DM's interim outside option. This, in turn, ensures that at future histories the designer finds it optimal to follow through with the promised information. Sequentially optimal modifications exploit the irreversibility of information: once the cat is out of the bag, it cannot be coaxed back in since, as soon as the DM's beliefs have moved, there is no way to systematically undo it---any further movement in the DM's belief is subject to a martingale constraint relative to her new belief. This is a distinctive property of information with no analog in mechanism design and it allows designer to bind their future selves to follow through with the promised information.

\paragraph{Attention capture with non-instrumental value of information.} We next turn to the question of how attention should be optimally captured when the DM has noninstrumental value for information. We suppose that the DM obtains flow utility from suspense \citep*{ely2015suspense} which is the expected variation in beliefs between periods. Just as a sports fan might turn off the television after one team accumulates an insurmountable lead, the DM optimally chooses when to stop paying attention by weighing her future suspense utility against the additional cost of waiting. We show that optimal attention under suspense utility has a tight and surprising connection with sequentially optimal attention capture when information is valued instrumentally (Theorem \ref{thrm:suspenseoptimal}). Exploiting this connection, we explicitly characterize optimal dynamic information structures when the designer has convex or concave value of attention. When the designer's value for attention is concave, optimal structures have {stochastic continuation belief paths} and {deterministically portions out suspense over time}. When the designer's value is convex, optimal structures have {deterministic continuation belief paths} and {stochastically portions out suspense over time}.

\paragraph{Attention capture with persuasion motives.} For much of the paper we focus on a designer who purely values attention. Nonetheless, we show that attention capture remains a prominent force even in environments with additively separable persuasion motives e.g., a platform with multiple revenue steams from selling ads (attention) and commissions (persuasion). We show that the optimal dynamic information structure is \emph{bang-bang} (Theorem \ref{thrm:attention_persuasion_separable}): they either focuses on extracting attention (such that our results on optimal attention capture apply exactly), or persuades in one-shot at a fixed, deterministic time $T$.\footnote{Methodologically, we develop a \emph{switching lemma} which shows how to modify the correlation between stopping beliefs and stopping times while preserving the marginals and dynamic obedience constraints; and a \emph{pasting lemma} which shows how to dynamically perturb continuation beliefs. These techniques are similar in spirit to path perturbation methods in the recent mathematics literature on optimal Skorokhod embeddings \citep*{beiglbock2017optimal}.}

\subsection{Related literature} \mbox{}\\

\vspace{-1em} 
\paragraph{Dynamic information design.} Our paper most closely relates to the recent literature on dynamic information design with forward-looking agents \citep*{ely2020moving,smolin2021dynamic,
zhao2020contracting,ball2023dynamic,orlov2020persuading,
knoepfle2020dynamic,hebert2022engagement}.\footnote{See also \cite*{renault2017optimal} and \cite{ely2017beeps} for early work on dynamic information design with myopic agents. \cite{doval2020sequential} develop a framework for information design in extensive form games.
\cite{de2023rationalizing} develop the notion of deviation rules to study which dynamic choices can be rationalized by information.}
Our contribution is distinct in several regards.

First, the reduction principle we develop is a tool for understanding all stopping times which can arise through information.\footnote{Indeed, the ``all-or-nothing'' information structure of \cite{knoepfle2020dynamic}, and Poisson structures used in \cite*{orlov2020persuading} and \cite{hebert2022engagement} can be viewed as within this class although our models are non-nested.} 
This allows us to fully solve the problem of a designer with arbitrary nonlinear preferences over DM's stopping times, and stands in contrast to much of the recent literature which studies the linear case by examining the time-$0$ obedience constraint \citep{knoepfle2020dynamic,hebert2022engagement}.\footnote{Our work characterizing optimal structures in nonlinear settings complements the recent progress in static persuasion when designer preferences over DM's actions are nonlinear \citep*{dworczak2022persuasion,kolotilin2022persuasion}. 
} 
To do so, we develop a tight characterization of the convex-order frontier and extreme points over the set of feasible stopping times. This set imposes constraints on the {conditional} distribution over stopping times and is distinct from majorization or first-order stochastic dominance \citep*{kleiner2021extreme,yang2023economics}. Second, we solve the problem of a designer with additive preferences over both the DM's action as in \citep{kamenica2011bayesian} as well as stopping time. To our knowledge, this has not been previously studied.\footnote{
Several papers study dynamic persuasion with constraints on the information structure and the designer cares only about DM's action \citep*{che2023dynamicpersuasion,escude2023slow}.} 
Third, we show there is no commitment gap which highlights a novel role of information not simply as a carrot to incentivize attention, but as a stick for the designer to discipline their future self. Finally, we study dynamic information design when information is valued noninstrumentally. This builds on an important paper of \cite*{ely2015suspense} who develop the notion of suspense and surprise. They characterize dynamic information structures which maximize suspense utility for a fixed time horizon $T$. We complement their analysis by characterizing optimal dynamic information to extract attention when the DM stops paying attention optimally such that the time horizon is endogenous. 

Subsequent work has built on our analysis \citep*{koh2024persuasion,saeedi2024getting}; we discuss each paper in turn. \cite*{koh2024persuasion} focuses on implementing joint distributions over actions, states, and stopping times. This is a more general environment but our results are non-nested and complementary.\footnote{We emphasize that in \cite*{koh2024persuasion} we write that the \emph{environment} there ``fully nests'' that of the present paper, but the \emph{results} are non-nested.}  In in this paper we focus on aspects distinct to attention capture which allows us to better understand the (i) structural properties of dynamic information: e.g., how interim beliefs are steered, whether they should be deterministic vs random; and (ii) convex-order frontier and extreme points over all feasible stopping times: we use this to obtain general results on the structure of optimal attention capture which (to our knowledge) cannot be obtained from the duality approaches developed in both \cite*{koh2024persuasion} and \cite*{saeedi2024getting}.

More related is \cite*{saeedi2024getting} who also study attention capture, albeit with exponential discounting. Our papers differ in at least two substantial regards. {First}, our results deliver sharp insights into the \emph{structural properties} of dynamic information structures such as belief paths which cannot be obtained via duality methods. Belief paths are fundamental since (endogenous) interim beliefs shape interim continuation incentives; we develop a relatively complete analysis of these objects, and show how they matter for capturing attention.\footnote{We identify key properties of belief paths which are sufficient to implement any feasible stopping time as well as necessary to achieve the convex-order frontier. We also show that whether interim beliefs are degenerate vs random matters crucially for sequentially optimality as well as attention capture under suspense utility.} {Second}, we study the set of \emph{all feasible stopping times} and characterize its convex-order frontier as well as extreme points. 
 We focus on a DM with additively separable time costs which allows us to obtain general and transparent insights into optimal dynamic information for many states and arbitrary decision problems without reliance on functional forms. The results in \cite*{saeedi2024getting} on optimal dynamic information hold for binary states and exponential discounting, but are qualitatively similar to Corollaries \ref{cor:convexconcave}-\ref{cor:binarybinary} of this paper. Beyond the common prior case, we pursue different behavioral extensions: \cite*{saeedi2024getting} study the important case of non-common priors while we study noninstrumental value for information.

\paragraph{Sequential learning.} We also contribute to the literature on sequential learning, experimentation, and optimal stopping starting from \cite{wald1947foundations} and \cite*{arrow1949bayes}. One set of papers explore settings in which the DM's attention is optimally allocated across several exogenous and stationary information sources \citep*{austen2018optimal, che2019optimal, gossner2021attention, liang2021dynamically}.\footnote{See also \cite{zhong2022optimal} who studies a setting in which the DM can flexibly acquire any dynamic information structure at some cost which is increasing in how quickly information reduces uncertainty.} In our setting, the DM only faces a single source of information and our motivation is to understand how distributions of stopping times vary with the information structure. \cite*{fudenberg2018speed} study the tradeoff between a DM's speed and accuracy within an uncertain-difference drift-diffusion model driven by Brownian signals. The reduction principle as well as our characterization of the extreme points over distributions of feasible stopping times pave the way to more general analyses of speed vs accuracy tradeoff over all possible information structures.


\paragraph{Outline.}
The rest of the paper is organized as follows. Section \ref{sec:model} develops the model; Section \ref{sec:general_results} develops the reduction principle; Section \ref{sec:stopping_time} analyzes optimal attention capture and, en-route, characterizes the convex-order frontier and extreme points of feasbile stopping times; Section \ref{sec:time_con} establishes that there is no commitment gap; Section \ref{sec:noninstrumental} studies attention capture under noninstrumental value of information; Section \ref{sec:bangbang} characterizes optimal structures to jointly extract attention and persuade; Section \ref{sec:concluding} concludes.

\section{Model}\label{sec:model}

Time is discrete and infinite, indexed by $\mathcal{T} = 0,1,2\ldots$. The state space $\Theta$ and action space $A$ are finite. State $\theta \in \Theta$ is drawn from a common full-support prior $\mu_0 \in \Delta(\Theta)$.

\paragraph{Decision maker's preferences.} 
A decision maker (DM)'s payoff from taking action $a \in A$ under state $\theta \in \Theta$ at time $\tau \in \mathcal{T}$ is 
\[
v(a,\theta, \tau) := u(a,\theta) - c\cdot \tau 
\]
where $u:A \times \Theta \to \mathbb{R}$ is an arbitrary decision problem and $c > 0$ is a constant per-unit cost of waiting. Constant waiting costs are standard in the literature on optimal stopping (see, e.g., \cite*{fudenberg2018speed,che2019optimal,che2023dynamicpersuasion,auster2024prolonged} among many others). More importantly, we think additively separabie costs are {particularly apt} for modelling attention capture: the cost of attention is not merely delay (which might shrink the value of the decision problem) but an {opportunity cost} since attending to one thing entails neglecting another. 

We also emphasize that our use of constant cost per-unit time is purely a normalization; our results extend to nonlinear but additively separable costs via a simple time-change. Hence, statements about the shape of $f$ can be equivalently be interpreted as statements about \emph{relative} shapes of the designer's value function $f(t)$ vis-a-vis the DM's nonlinear cost $c(t)$; we formalize this in Online Appendix \ref{appendix:nonlinearcost}.\footnote{For instance, statements holding for convex $f(t)$ can equivalently be interpreted as statements about $f \circ c^{-1}$ convex for some strictly increasing cost function $c: \mathcal{T} \to \mathbb{R}$.}

\paragraph{Dynamic information structures.} A dynamic information structure specifies, for each history of past beliefs, a distribution over the next period's beliefs subject to a martingale constraint. 
Let $I \in \Delta\big(\prod_{t \geq 1}\Delta(\Theta) \big)$
be a distribution over belief paths where $(\mu_t)_{t}$ is a typical realization. Call $H_t := (\mu_s)_{s \leq t}$ a time-$t$ history. Let $I_{t+1}(\cdot |H_t)$ denote the conditional distribution over the next period's beliefs. Say the history $H_t$ realizes with positive probability if $
\{(\mu_s)_{s}: (\mu_s)_{s\leq t} = H_t\} \subseteq \text{supp } (I)$ i.e., if paths where the first $t$ periods agrees with $H_t$ are contained in the support of $I$. $I$ is a dynamic information structure if it is the law of a martingale (with respect to the natural filtration generated by its histories).\footnote{Note future information could depend on past decisions \citep{makris2023information}. In our setting, at time $t$ there is a unique sequence of decisions (wait until $t$) so our formulation is without loss. Further, although we have associated dynamic information structures directly with belief martingales, it is well-known that dynamic information can convey information about both the state (via beliefs) as well as the continuation information structure. All our results hold for this slightly wider class; see Appendix. Nonetheless, optimal dynamic information will not require this distinction so we work directly with belief martingales (which remain optimal with respect to this wider class).} Let $\mathcal{I}$ be the set of all dynamic information structures. 

\paragraph{Measures under different dynamic information structures.} We will often vary the dynamic information structure to understand how variables of interest (e.g., probabilities of histories, incentive compatibility constraints etc.) change. To this end, we will use $\Ex^I[\cdot]$ and $\Pr^I(\cdot)$ to denote expectations and probabilities under dynamic information structure $I \in \mathcal{I}$. Throughout this paper we will typically use superscripts to track dynamic information structures, and subscripts to track time periods.

\paragraph{Decision maker's problem.} 
Facing $I \in \mathcal{I}$, DM solves
\[
    \sup_{\tau , a_{\tau}} \mathbb{E}^I[ v(a_\tau,\theta,\tau) ],
\]
where $\tau$ is a stopping time and $a_{\tau}$ is a (stochastic) action under the natural filtration. Throughout we will assume (i) the DM breaks indifferences in favour of not stopping to ensure that the set of feasible distributions over stopping times is closed; and (ii) $\tau$ is almost-surely finite which is guaranteed by weak regularity assumptions on $v$. 

Thus, each information structure $I \in \mathcal{I}$ induces a joint distribution over outcomes which arises as a solution to the DM's problem.
We will sometimes use $\tau(I)$ to emphasize the dependence of the agent's stopping time on $I$.




\begin{definition} [Feasible distributions] 
Define $d(I) \in \Delta(\mathcal{T} )$ as the distribution over stopping times induced by $I$. For a set $\mathcal{I}' \subseteq \mathcal{I}$, define $\mathcal{D}(\mathcal{I}') := \{d(I): I \in \mathcal{I}'\}$
as set of distributions induced by structures in $\mathcal{I}'$. Say that $d \in \Delta(\mathcal{T})$ is feasible if $d \in \mathcal{D}(\mathcal{I})$.
\end{definition}

\paragraph{Designer's problem.} There is a designer with preferences over the DM's stopping time  $f: \mathcal{T}  \to \mathbb{R}$ which is strictly increasing. The designer's problem is 
\[
\sup_{I \in \mathcal{I}}\Ex^I\big[f\big(\tau(I)
\big)\big], 
\]
noting that the supremum is taken over the whole space of dynamic information structures. The set of feasible stopping times is closed and we will assume enough regularity assumptions on $f$ to ensure the supremum is obtained.\footnote{See Online Appendix \ref{onlineappendix:topology} for a formal discussion.}

    
\paragraph{Discussion of model.} We briefly discuss aspects our model.
\begin{itemize}
    \item \emph{Intertemporal commitment.} Our formulation of the designer's problem is such that the designer first chooses $I$, then the DM stops optimally given information generated by $I$. In Section \ref{sec:time_con} we show that this can be relaxed completely. We do so by showing constructively that every optimal structure has a {sequentially optimal modification} which induces random interim beliefs.  
    \item \emph{Instrumental value of information.} In our baseline model, the DM values information instrumentally to make a better decision. In Section \ref{sec:noninstrumental} we solve for designer-optimal structures when the DM values suspense. 
    \item \emph{Pure attention capture.} In our baseline model, we analyze the case in which the designer aims to extract attention. We view this as a first-order concern in many environments e.g., platforms where attention is the primary source of revenue. In Section \ref{sec:bangbang} we show that the form of optimal attention capture remains similar when the designer values both attention and persuasion separably.
\end{itemize}

\section{The structure of dynamic information}\label{sec:general_results}

We introduce a special class of information structures which will play a crucial role for analyzing feasible distributions over stopping times.

\begin{definition}[Full revelation with deterministic continuation beliefs]
    $I$ is \emph{full-revelation with deterministic continuation beliefs} if there exists a deterministic belief path $(\mu^C_t)_t \in \Delta(\Theta)^{\mathcal{T}}$ such that for any $H_t$ which realizes with positive probability, 
\begin{enumerate}
    \item[(i)] Continuation beliefs follow the belief path $(\mu^C_t)_t$ and stopping beliefs are full-revelation: 
    \[\text{supp }\,I_{t+1}\big(\cdot|H_t\big) \subseteq \{\mu^C_{t+1}\} \cup \{\delta_{\theta}: \theta \in \Theta\};\] 
\item[(ii)] Recommendations are obedient: for each $t$, DM prefers to continue at history $H_t = (\mu^C_s)_{s \leq t}$ and stop at $H_t = (\mu_0,\mu^C_1, \ldots \mu^C_{t-1},\delta_{\theta})$. 
\end{enumerate}
\end{definition}

\begin{figure}[h!]  
\centering
\captionsetup{width=1\linewidth}
    \caption{Illustration of full revelation and deterministic structures}
    \includegraphics[width=0.9\textwidth]{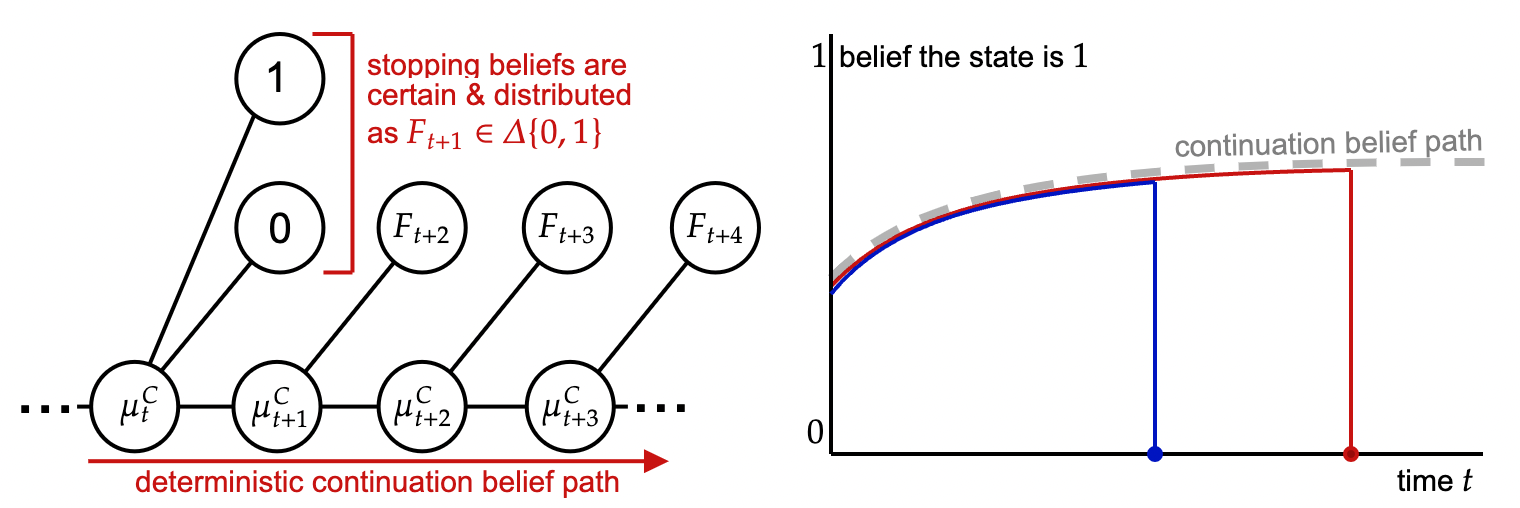}
     \label{fig:deterministic_cont_illust}
\end{figure}

Dynamic information structures which are full-revelation with deterministic continuation beliefs are depicted in Figure \ref{fig:deterministic_cont_illust} which shows histories on the left, and sample belief paths on the right for $\Theta = \{0,1\}$ so that $\mu_t$ is the belief that the state is $1$. 

\begin{definition}[Increasing belief paths and basin of uncertainty] 
The belief path $(\mu^C_t)_t \in \Delta(\Theta)^{\mathcal{T}}$ is \emph{increasing} if $(\phi(\mu^C_t))_t$ is increasing in $t$ where 
\[\phi(\mu) := \underbrace{\Ex_{\mu}[\max_{a \in A}u(a,\theta)]}_{\substack{\text{Expected utility} \\ \text{under full info}}} - \underbrace{\max_{a \in A} \Ex_{\mu}[u(a,\theta)]}_{\substack{\text{Expected utility} \\\text{acting under $\mu$}}}
\] 
is the additional value of full information at belief $\mu$.\footnote{$\phi$ is continuous
and concave since $\mu \mapsto \Ex_{\mu}[\max_{a \in A}u(a,\theta)]$ is linear and $\mu \mapsto \max_{a \in A} \Ex_{\mu}[u(a,\theta)]$ is convex.} Define the maximum value of $\phi$ as $\phi^* := \max_{\mu \in \Delta(\Theta)} \phi(\mu)$. Further define $\Phi^* := \text{argmax}_{\mu \in \Delta(\Theta)} \phi(\mu)$ as the convex set of beliefs for which the DM's utility from obtaining full information relative to stopping immediately is maximized; we will often refer to $\Phi^*$ as the basin of uncertainty. 
\end{definition}

\begin{definition}[Maximal belief paths]
    The belief path $(\mu^C_t)_t \in \Delta(\Theta)^{\mathcal{T}}$ is \emph{maximal} for the stopping time $\tau$ if for each $t \in \mathcal{T}$, 
    \[
    \mu^C_{t+1} \notin \Phi^* \implies \cfrac{\mathbb{P}(\tau > t+1)}{\mathbb{P}(\tau > t)} = \max_{\theta \in \Theta} \cfrac{\mu^C_t(\theta)}{\mu^C_{t+1}(\theta)}.
    \]
\end{definition}

Maximality is a by-product of the martingale condition on beliefs. For instance, consider a full revelation and deterministic dynamic information structure with continuation belief paths $(\mu^C_t)_t$. If the induced stopping time is $\mathbb{P}^I(\tau = t+1|\tau \geq t) = 0$, then continuation beliefs cannot move between times $t$ and $t+1$: since the DM never learns anything at time $t+1$, the absence of full information is entirely uninformative. This is reflected by the condition of maximality since $\mathbb{P}(\tau > t) = \mathbb{P}(\tau > t+1)$ so $\mu^C_t = \mu^C_{t+1}$. 

\begin{figure}[h]
\centering
\caption{Illustration of increasing and maximal paths}
    \subfloat[Increasing paths]{\includegraphics[width=0.49\textwidth]{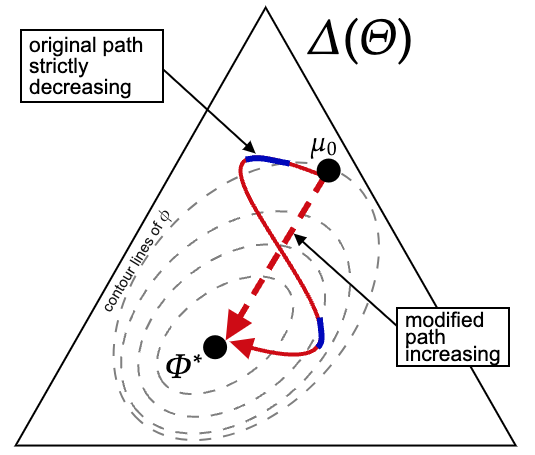}}
    \subfloat[Maximal paths]{\includegraphics[width=0.49\textwidth]{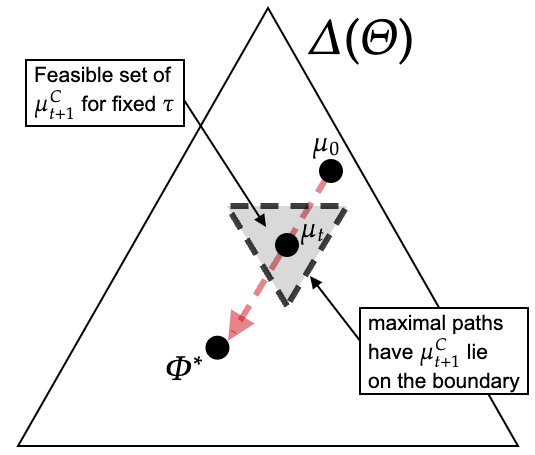}}
    \label{fig:increasing_maximal}
\end{figure}

Increasingness imposes structure on the \emph{direction} that continuation beliefs move: they must steer the DM toward regions of increasing uncertainty; maximality imposes structure on the \emph{magnitude} that continuation beliefs move: they must move as much as possible given the stopping time. Both properties are depicted in Figure \ref{fig:increasing_maximal}. The dotted line in Panel (a) depicts an increasing continuation belief path while the solid line depicts a nonincreasing path where the blue portions correspond to strictly decreasing segments. Panel (b) depicts maximality for $|\Theta| = 3$: the time-$t$ continuation belief $\mu_t^C$ together with the stopping time pin down the degree to which time-$t+1$ continuation beliefs can move; this is depicted by the grey triangle. Maximality requires that $\mu^C_{t+1}$ lie on this boundary.

Let $\mathcal{I}^* \subseteq \mathcal{I}$ be the set of full-revelation structures with deterministic, increasing, and maximal continuation beliefs. 

\begin{theorem}[Reduction principle] \label{thrm:reduction} Every feasible distribution over stopping times can be implemented with full-revelation and deterministic structures with increasing and maximal belief paths i.e., $\mathcal{D}(\mathcal{I}) = \mathcal{D}(\mathcal{I}^{*})$.
\end{theorem}

An attractive property of structures with deterministic continuation beliefs is continuation histories do not branch---conditional on paying attention up to $t$, there is a unique continuation history. Hence, Theorem \ref{thrm:reduction}  drastically prunes the space of dynamic information structures and turns the designer's problem into looking for a target distribution over stopping times and a deterministic and increasing continuation belief path which can support it.\footnote{Since continuation histories do not branch, the number of dynamic incentive constraints grow linearly (rather than exponentially) in time.} We will subsequently use this to understand optimal attention capture.

Conceptually, Theorem \ref{thrm:reduction} highlights the importance of increasing and maximal continuation belief paths for optimal stopping when the space of dynamic information structures is unrestricted. Although continuation beliefs are (on-path) unobserved by an analyst observing the DM's stopping time, it nonetheless plays a crucial role in the background by shaping her interim value for information and hence her continuation incentives. 

\paragraph{Proof sketch of Theorem  \ref{thrm:reduction}.} Suppose that $\Theta = \{0,1\}$ so we associate beliefs with probability that the state is $1$. The DM's decision problem is symmetric, stationary, and her prior is uniform. Fix an iid symmetric information structure $I \in \mathcal{I}$ and suppose the DM's optimal stopping decision is characterized by a stopping boundary which is constant in time.
\begin{figure}[h!]  
\centering
\captionsetup{width=1\linewidth}
    \caption{Illustration of contracting belief paths}
    \includegraphics[width=0.9\textwidth]{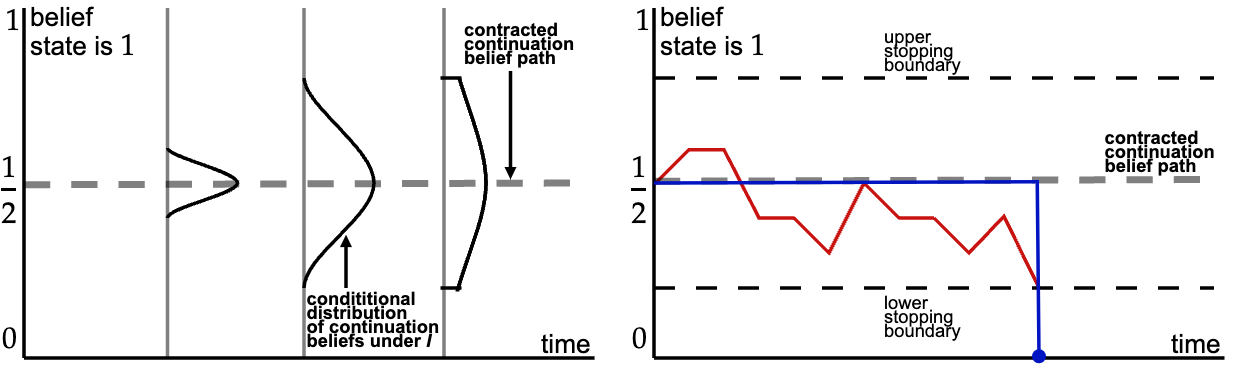}
     \label{fig:matching_paths}
\end{figure} 

Now consider the following modification of $I$: for the histories at which the DM stops (i.e., beliefs under $I$ cross the stopping boundary), modify the information structure to instead deliver full information. Next, for each time $s$ under the original information structure, there will be some some paths which have yet to hit the boundary. On such histories, DM finds it optimal to continue. The distribution of beliefs associated with these histories are depicted on the left of Figure \ref{fig:matching_paths}. Collapse these continuation paths into a single history which, by the symmetry of our environment, is associated with belief $1/2$. Call this new information structure $I'$. 

Observe that for each time $t$, the DM has weakly higher continuation value since, at stopping histories under $I$, she now learns the state perfectly under $I'$. Moreover, the DM found it optimal to pay attention at each continuation history under the original information structure. Hence, since payoffs are convex in beliefs, contracting continuation beliefs must depress the DM's stopping utility so that she also finds it optimal to pay attention at the new collapsed belief. Moreover, $1/2$ remains a mean-preserving contraction of stopping beliefs so $I'$ is a valid dynamic information structure. Finally, observe that $I'$ and $I$ induce the same distribution over stopping times since, by construction, for every belief path on the original structure $I$ (depicted in red on right of Figure \ref{fig:matching_paths}), we instead deliver full information with identical probability under $I'$ (one such path is depicted in blue). 

The argument for increasingness and maximality is more involved. The underlying intuition is that by making the DM progressively more uncertain as much as the martingale condition allows, this endogenously increases the value of all future information which, in turn, slackens future incentives to pay attention. This is depicted in Figure \ref{fig:increasing_maximal} (i) where the grey dashed lines are the level sets of $\phi$. Starting from an initial belief path $(\mu_t^C)_t$ depicted by the solid red line which is strictly decreasing in $\phi$ over the blue portions, we can ``iterative iron'' such paths to construct an alternate path $(\tilde \mu^{C}_t)_t$ which are increasing in $\phi$ while preserving both the stopping time and dynamic continuation incentives.  

The proof of Theorem \ref{thrm:reduction} in Appendix \ref{appendix:reduction_timecon} makes these arguments precise. Our procedure of collapsing continuation histories is conceptually related to revelation-principle type arguments:
collapsing continuation histories pushes down the value of the DM's best deviation; fully revealing the state at stopping histories pushes up the value of continuation.\footnote{See, e.g., \cite{myerson1986multistage, forges1986approach}. More recently, \cite{ely2017beeps} develops an `obfuscation principle' which, roughly, states that the designer can simply tell the DM what her beliefs should be after every history.}
Our observation is that we can do this dynamically while replicating the distribution over stopping times and preserving continuation incentives at all other times and histories. 


\subsection{Designer-optimal structures extract all surplus} 

A further observation is that the DM's surplus is zero across all designer-optimal structures. Hence, even if there are substantial gains from trade e.g., the designer values the DM's attention more than the DM's cost of delay, designer-optimal structures always hold the DM down to her outside option.\footnote{Indeed, this is true for more general designer preferences over actions, states, and stopping times as long as the designer values attention.} Remarkably, we will later see (Section \ref{sec:time_con}) that this remains true when the designer has no intertemporal commitment.
\begin{proposition}\label{prop:no_surplus} If the designer's value for DM's stopping time is strictly increasing, every designer-optimal structure extracts all surplus from the DM i.e., 
\[
I^* \in \text{argmax}_{I \in \mathcal{I}} \Ex^I[f(\tau)] \quad \text{implies} \quad 
\underbrace{\max_{\tau,a_{\tau}}  \Ex^{I^*}[v(a_{\tau},\theta,\tau)]}_{\text{DM's utility under $I^*$}}  - \underbrace{\max_{a \in A, t \in \mathcal{T}}\Ex_{\theta \sim \mu_0}[v(a,\theta,t)]  
}_{\text{DM's utility under no info}}= 0. 
\]
\end{proposition}

\subsection{Reduction to stopping times and belief paths} \ref{thrm:reduction} converts the problem of optimizing over dynamic information structures into simply choosing a stopping time $\tau$ and belief path $(\mu^C_t)_t$ subject to the following constraints they impose on each other: 

\begin{lemma} \label{prop: IC+boundary-condition} The following are equivalent:
\begin{enumerate}
\setlength\itemsep{0em}
    \item There exists a full-revelation structure with deterministic continuation belief paths $I \in \mathcal{I}^{*}$ which induces stopping time $\tau$ and belief path $(\mu_t^C)_{t\in\mathcal{T}}$.
    \item The following conditions are fulfilled:
    \begin{align*}
        &\phi(\mu_t^C) \geq \Ex^I[c\tau \mid \tau > t] - ct \text{ for every $t\in\mathcal{T}$}  \tag{Obedience}  \label{eq:obedience}   \\
        &\Pr^I(\tau > t+1)\mu_{t+1}^C(\theta) \leq \Pr^I(\tau > t)\mu_t^C(\theta) \text{ for every $t \in \mathcal{T}$ and $\theta \in \Theta$} \tag{Boundary} \label{eq:boundary}
    \end{align*}
\end{enumerate}
\end{lemma}

\ref{eq:obedience} is the usual constraint that at each time $t$, the DM must find it optimal to continue paying attention, noting that the DM's outside option from stopping to act is given by her continuation belief $\mu^C_t$. \ref{eq:boundary} imposes a constraint on the degree to which beliefs can move between periods and is directly implied by the martingale condition at time $t$. Say that the stopping time distribution-belief path pair $(d,\bm{\mu^C})$ is \emph{feasible} if it fulfils the obedience and boundary conditions. 

Moreover, from Theorem \ref{thrm:reduction}, to implement any feasible distribution $d$, it is \emph{sufficient} to restrict ourselves to increasing and maximal belief paths $\bm{\mu}^C$. It will turn out that these conditions are also \emph{necessary} to achieve the convex-order frontier among the set of feasible stopping times.

\section{Optimal Attention Capture}\label{sec:stopping_time}
From Theorem \ref{thrm:reduction} and Lemma \ref{prop: IC+boundary-condition} we move optimizing over dynamic information structures $\max_{I \in \mathcal{I}} \Ex^I[f(\tau)]$ to  
\begin{align*}        &\max_{\substack{\big(d,\bm{\mu^C}\big) 
            \\ 
            \in \Delta(\mathcal{T}) \times (\Delta(\Theta))^\mathcal{T}}} \Ex_{\tau \sim d}\Big[f(\tau)\Big] 
            \\
            & \quad \quad \quad \text{s.t. \ref{eq:obedience} and \ref{eq:boundary}}.
        \end{align*}

For expositional simplicity we focus on the case with nonlinear value of attention $f$ and linear waiting costs (constant per-unit time). This can equivalently interpreted as an analysis of the \emph{relative} shapes of the designer's value to the DM's cost of attention. That is, our results can be exactly translated to the environment in which the designer's value is $f: \mathcal{T} \to \mathbb{R}$, the DM's cost is $c: \mathcal{T} \to \mathbb{R}$, and the designer maximizes $f \circ c^{-1}$.

We first characterize the convex-order frontier of feasible stopping times, as well as---when beliefs start in the basin of uncertainty---the extreme points of feasible stopping times. Then, solutions to the designer's problem follow readily. 

\subsection{Characterization of convex-order frontier and extreme points} We recall some basic definitions. For $d, d' \in \Delta(\mathcal{T})$, say that $d$ dominates $d'$ in the convex order (denoted $d \succeq_{cx} d'$) if for any convex function $f: \mathcal{T} \to \mathbb{R}$, $\Ex_{\tau\sim d}[f(\tau)] \geq \Ex_{\tau\sim d'}[f(\tau)]$. Note that this implies mean-preservation. Say that $d$ strictly dominates $d'$ in the convex order ($d \succ_{cx} d'$) if the inequality is strict for some convex function $f$. We begin with two key definitions. 

\begin{definition}[Indifferent, increasing, and maximal (IIM) distributions] Let 
\[
\mathcal{D}^{IIM} :=\left\{ d \in \Delta(\mathcal{T}): \substack{\text{\normalsize (i) $\exists \bm{\mu}^C$ s.t. $(d,\bm{\mu}^C)$ is feasible, $\bm{\mu}^C$ increasing and maximal} \\
\text{\normalsize \ref{eq:obedience} binds for all $t \geq 1$;} \\
\text{\normalsize (ii) $(d,\bm{\mu}^C)$ feasible $\implies$ $\bm{\mu}^C$ increasing and maximal}}
\right\}
\]
Part (i) states that every distribution $d \in \mathcal{D}^{IMM}$ can be feasibly supported by an increasing and maximal belief path $\bm{\mu}^C$ which keeps the DM's \ref{eq:obedience} constraint binding for all times $\geq 1$. Part (ii) states that increasingness and maximality are furthermore necessary to support distributions in $\mathcal{D}^{IIM}$. 

\end{definition}

\begin{definition}[Block structure] The feasible pair $(d,\bm{\mu^C})$ is a block structure if there exists a sequence of times $(t_i)_{i=1}^n$ where $t_1 < t_2 < t_3 \ldots$ and, if $n < +\infty$, a pair of terminal times $TER := \{t_{n+1},t_{n+2}\}$ where $t_n < t_{n+1} \leq t_{n+2}$ such that 
    \[
    \text{supp } \tau = IND(d,\bm{\mu}^C) \cup TER = \{t_i\}_{i=1}^{n+2},
    \]
    where
\[
IND(d,\bm{\mu}^C) := \Big\{t \in \mathcal{T}: \text{\ref{eq:obedience} binds at time $t$} \Big\}
\]
are the set of times at which the DM is indifferent. 
    Let $\mathcal{I}^{BLOCK}$ denote the set of block structures. 
\end{definition}

Block structures are such that the support of the stopping time $\text{supp } d$ coincides with indifference times with the possible exception of a pair of terminal times. We are now ready to state our chracterization of the convex-order frontier and extreme points. 

\begin{theorem}\label{thrm:convex_ext}
    Let $\mathcal{D}_s := \{d \in \mathcal{D}(\mathcal{I}): \Ex_{\tau \sim d}[\tau] = s\}$ be the set of feasible stopping times with mean $s \in [0, \phi(\mu_0)/c]$. 
    \begin{itemize}
        \item[(i)] \textbf{Characterization of convex order.} For any $d \in \mathcal{D}_s$, there exists an indifferent, increasing, and maximal distribution $d^{IIM} \in \mathcal{D}^{IIM}$ for which
        \[
        d^{IIM} \succeq_{CX}  d \succeq_{CX} d^{DET} 
        \]
        where $\mathbb{P}_{\tau \sim d^{DET}}(\lfloor s\rfloor \leq \tau \leq \lceil s\rceil) = 1$. This implies if $d \notin \mathcal{D}^{IIM}$ then it is not on the convex-order frontier i.e., the first relation is strict. 
        \item[(ii)] \textbf{Characterization of extreme points.} If $\mu_0 \in \Phi^*$ or $|\Theta| = |A| = 2$, block structures induce extreme points of feasible stopping times: 
    \[\mathcal{D}\big(\mathcal{I}^{BLOCK}\big) \cap \mathcal{D}_s = \textrm{Ext} \big(\mathcal{D}_s \big).\]
    \end{itemize}
\end{theorem}

\paragraph{Discussion of Theorem \ref{thrm:convex_ext}.} Theorem \ref{thrm:convex_ext} (i) gives a characterization of the stopping times and continuation belief paths which attain the convex-order frontier. In Theorem \ref{thrm:reduction} we previously slowed that those which were increasing and maximal were always \emph{sufficient} to obtain any feasible distribution over stopping times. Theorem \ref{thrm:convex_ext} (i) shows that such properties are also \emph{necessary} to attain the convex frontier. 

\begin{figure}[h]
\centering
\caption{Illustration of convex-order frontier}
    \subfloat[Belief paths on the convex-order frontier]{\includegraphics[width=0.49\textwidth]{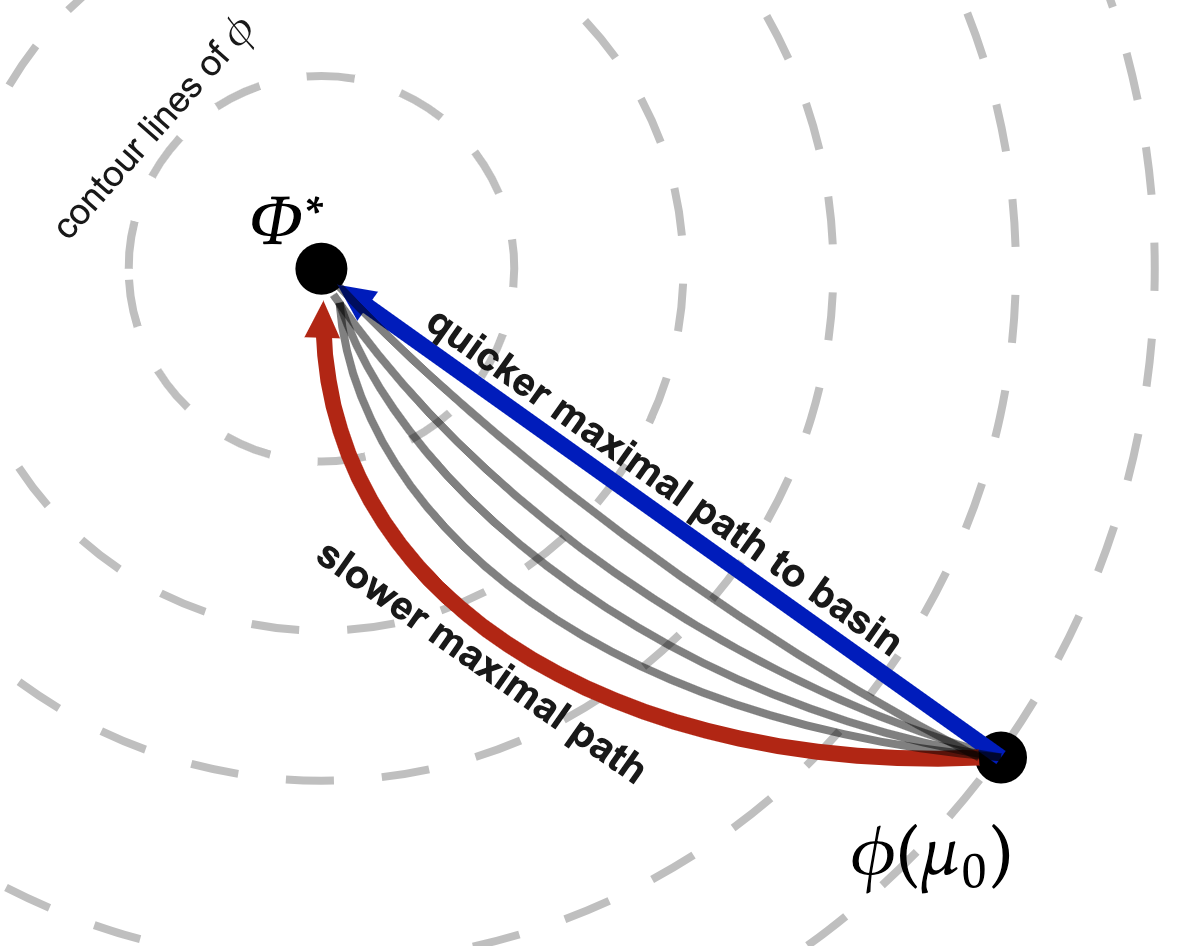}}
    \subfloat[CDFs comprising convex-order frontier]{\includegraphics[width=0.49\textwidth]{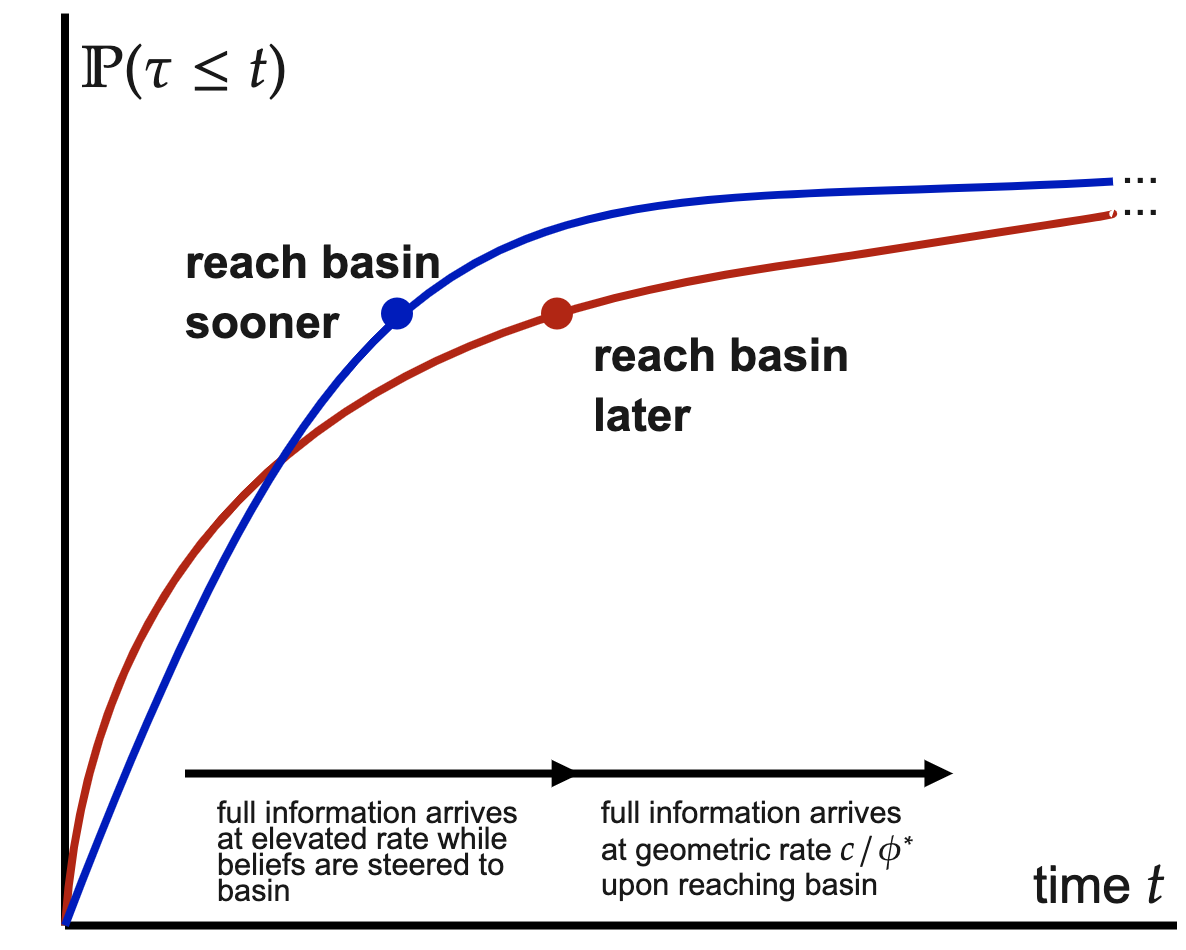}}
    \label{fig:cx_order}
\end{figure}

Distributions which attain the convex-order frontier are those which keep the DM indifferent between continuing to pay attention and stopping for all times $t \geq 1$. When the DM's beliefs start inside the basin of uncertainty $(\mu_0 \in \Phi^*)$ or $|\Theta| = 2$, there is a feasible stopping time which dominates all others in the convex-order frontier; if not, the convex-order frontier will typically comprise a family of distributions because there can be multiple increasing and maximal belief paths which keep the DM indifferent. Figure \ref{fig:cx_order} illustrates indifferent, increasing, and maximal distributions associated with different belief paths for $|\Theta| = 3$. Starting from the prior in the bottom right, the DM initially has value $\phi(\mu_0)$ for full information; as her beliefs are progressively steered towards the basin, she increasingly values information more. First consider the blue line with depicts a ``direct'' increasing path from $\mu_0$ to $\Phi^*$ along which the DM's beliefs are steered.\footnote{Loosely, this means that the path is everywhere perpendicular to the contour lines of $\phi$ (as with the blue line in Figure \ref{fig:cx_order} (a).} By Theorem \ref{thrm:convex_ext} (i), such the DM's belief paths must also be maximal so, together with DM indifference, the ``speed'' at which the DM's beliefs move along this blue line are pinned down by the boundary condition. The CDF of the induced distribution over stopping times is depicted by the blue line in Panel (b): the DM first receives full information at an elevated rate. This is driven by two distinct forces: first, while the DM's beliefs are outside the basin, she values information less. Thus, to entice her to pay attention, information must arrive at an elevated rate; second, the blue path steers beliefs ``directly'' to the basin but, to do so, the designer must offer an elevated probability of learning the state such as to generate the requisite movement in continuation beliefs.

Now consider the red path in panel (a) of Figure \ref{fig:cx_order} which depicts an ``indirect'' increasing path from $\mu_0$ to $\Phi^*$. Once again, from Theorem \ref{thrm:convex_ext} (i), to support a distribution on the convex frontier, such paths must be maximal so the ``speed'' at which the DM's belief move along this red line is pinned down and corresponds to a \emph{slower} path to the basin. The red line in panel (b) of Figure \ref{fig:cx_order} depicts the CDF of the indifferent, indifferent, and maximal stopping time associated with the red line. Compared to the distribution associated with the blue line, slower paths to the basin allow for {lower probability of early stopping} since the requisite movement in continuation beliefs is lower. However, this also means the basin at which the DM's value for information is reached later, which delays the time at which attention can be maximally extracted.

\begin{figure}[h!]
    \centering
      \captionsetup{width=\linewidth}
    \caption{Distributions induced by block structures}
    {\includegraphics[width=0.7\textwidth]{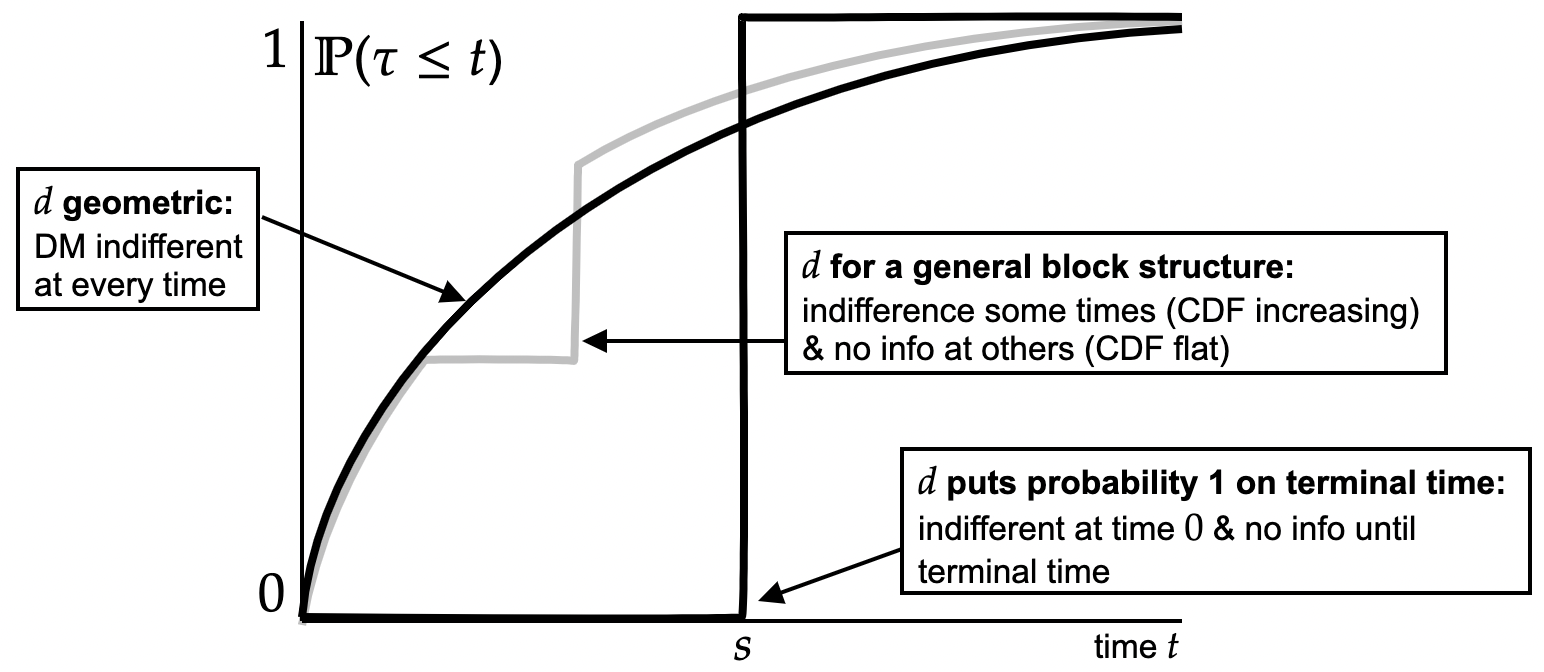}} 
    \label{fig:block_structure_CDF}
\end{figure}

Theorem \ref{thrm:convex_ext} (ii) goes beyond the convex-order frontier to characterize extreme points of feasible stopping times when the DM's prior starts in the basin ($\mu_0 \in \Phi^*$) or the state and action spaces are binary: block structures exactly implement these extreme points. The CDFs of various extreme points of feasible distributions with the same mean $s$ is illustrated in Figure \ref{fig:block_structure_CDF}.\footnote{Note that under the conditions of Theorem \ref{thrm:convex_ext} (ii) ($\mu_0 \in \Phi^*$ or $|\Theta| = |A| = 2$) there is a unique distribution (geometric) which attains the convex-order frontier.}

Block structures have the property of ``indifference or nothing'' before the terminal times: if the DM has a chance to stop, she is kept indifferent between continuing and stopping. This loosely resembles an important characterization of extreme points of majorization by \cite*{kleiner2021extreme}.\footnote{Our characterization is also distinct from the extreme points of first-order stochastic dominance \citep{yang2023economics} though this should be more apparent.} However, the constraints in our setting are strictly stronger than those imposed by mean-preserving spreads and mean-preserving contractions. This is because the set of feasible stopping times must obey \emph{conditional} constraints so that at each time $t$, the conditional expected additional value outweights the conditional expected additional cost. By contrast, majorization is an \emph{unconditional} constraint.

\begin{figure}[h]
\centering
\caption{Feasibility of stopping times is more stringent than majorization}
    \subfloat[$MPS$ is strictly weaker]{\includegraphics[width=0.42\textwidth]{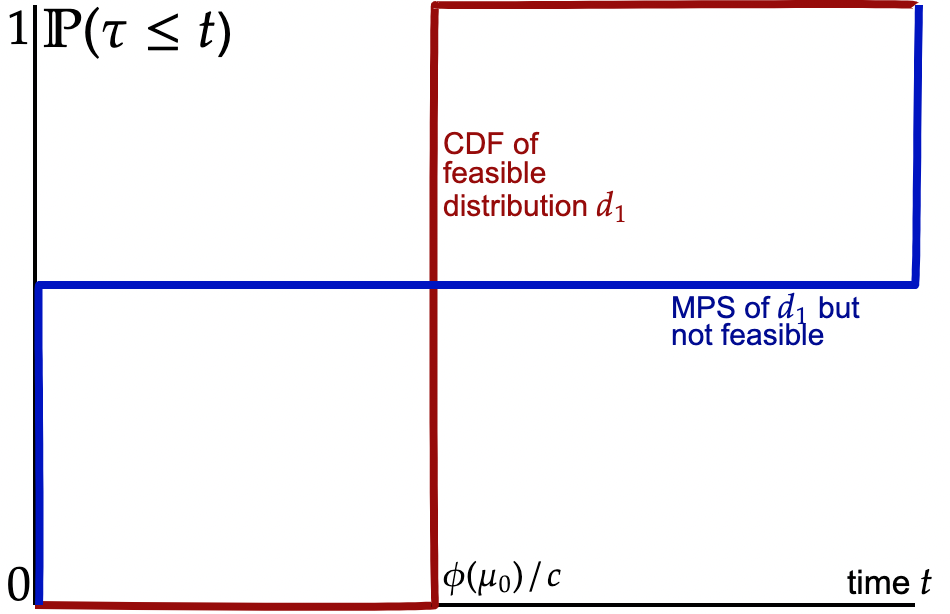}}
    \subfloat[$MPC$ is strictly weaker]{\includegraphics[width=0.42\textwidth]{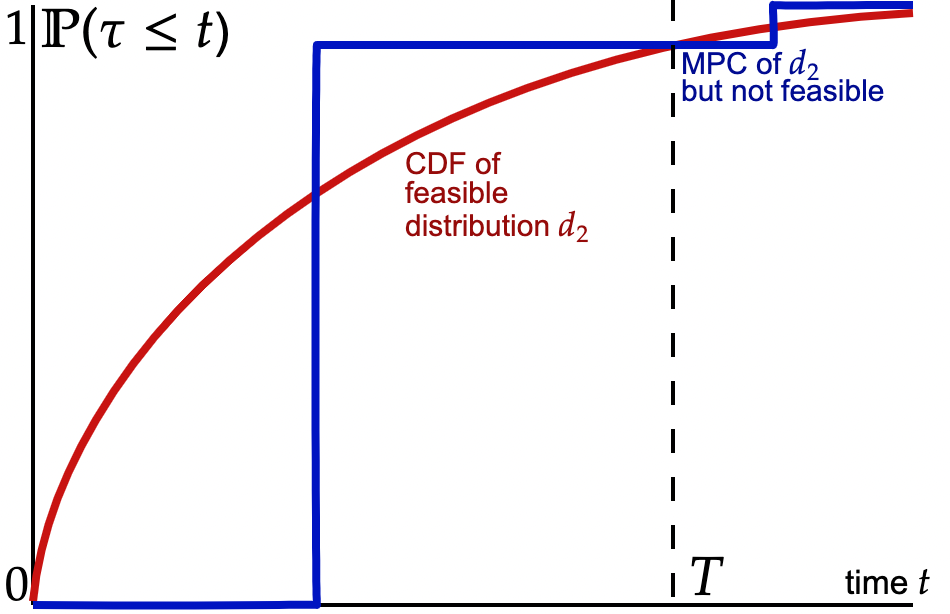}}
    \label{fig:majorization_comparison}
\end{figure}

This is illustrated in Figure \ref{fig:majorization_comparison}. The red line in Panel (a) illustrates the CDF of a feasible stopping time $d_1$ which concentrates the DM's stopping probability at time $t^* =\phi(\mu_0)/c$ (assuming this is an integer) so that, conditional on full information arriving then, the DM's time-$0$ obedience constraint is tight. The blue line illustrates a mean-preserving spread of $d_1$ which puts probability $1/2$ on time $1$, and probability $1/2$ on time $2(t^* - 1)$. This continues to respect the DM's time-$0$ incentives, but violates her interim incentives: if full information does not arrive at time $1$, then the additional waiting costs exceeds her interim value of information.\footnote{This is true whenever $\phi(\mu_0)/c > 3$; in continuous time the mean-preserving spread would put mass on $0$ and $2t^*$ so that the interim obedience constraint is always violated.} Now consider Panel (b) which illustrates the CDF of a feasible stopping time $d_2$ so that full information arrives at a geometric rate of $c/\phi(\mu_0)$. Now consider the following mean-preserving contraction: we contract the distribution before $T$ into a single atom, and the distribution after $T$ into another. This is depicted by the blue line. Now observe that the gap between these contracted atoms can be made arbitrarily large by increasing $T$; but once again this violates the DM's interim \ref{eq:obedience} constraint since, if no information arrives at the time of the first atom, the DM does not find waiting further worthwhile. 

\paragraph{The structure of optimal attention capture.} Theorem \ref{thrm:convex_ext} readily delivers explicit solutions for the problem of optimal attention capture which we showcase via the following corollaries:  

\begin{corollary}[Optimal attention capture for convex/concave] \label{cor:convexconcave} Suppose $f$ is 
\begin{itemize}
    \item[(i)] \textbf{convex}, then the optimal dynamic information structure capture is given by the stopping time-belief path pair $(d,\bm{\mu}^C)$ such that DM is indifferent for all times $t \geq 0$ and $\bm{\mu}^C$ is increasing and maximal. Moreover, there exists some time $T < +\infty$ such that continuation beliefs reach the basin ($\mu^C_T \in \Phi^*$) after which then full information arrives at a geometric rate of $c / \phi^*$: 
    \[
    \mathbb{P}_{\tau \sim d}(\tau = t+1 | \tau > T) = c / \phi^*; 
    \]
    \item[(ii)] \textbf{concave}, then the optimal dynamic information structure is given by the stopping time-belief path pair $(d,\bm{\mu}^C)$ such that DM is indifferent at time $t = 0$ and $\mathbb{P}_{\tau \sim d}(\lfloor \phi(\mu_0)/c \rfloor \leq \tau \leq \lceil \phi(\mu_0)/c \rceil) = 1$. 
\end{itemize}
\end{corollary}

Corollary \ref{cor:convexconcave} highlights how differences in \emph{relative time preferences} matter for the structure of attention capture.\footnote{Since this follows from our characterization of the convex-order frontier, the distribution of stopping times is unique whenever convexity or concavity is strict. However, the accompanying belief path need not be unique for $|\Theta| > 2$.} In particular, suppose that the DM's cost of waiting is given by $c(t)$. Then, if $f$ is \emph{more convex} than $c$ (i.e., $f \circ c^{-1}$ is convex), optimal attention capture must keep the DM indifferent at each point in time and, furthermore, the attendant belief path must be increasing and maximal. Conversely, if $f$ is \emph{less convex} than $c$ (i.e., $f \circ c^{-1}$ is concave), optimal attention capture concentrates the distribution over stopping times around $\phi(\mu_0)/c$, the highest feasible expected stopping time.

\begin{figure}[h!]
    \centering
      \captionsetup{width=\linewidth}
    \caption{Optimal belief paths for $\mu_0 \notin \Phi^*$}
    \begin{quote}
    \vspace{-1.5em}
    \centering 
    \footnotesize Note: Green: $u(a,\theta) \propto  -|a-\theta|$; blue: $u(a,\theta) \propto -(a-\theta)^2$; red: $u(a,\theta)  \propto (a-\theta)^4$.
\end{quote}
    {\includegraphics[width=0.7\textwidth]{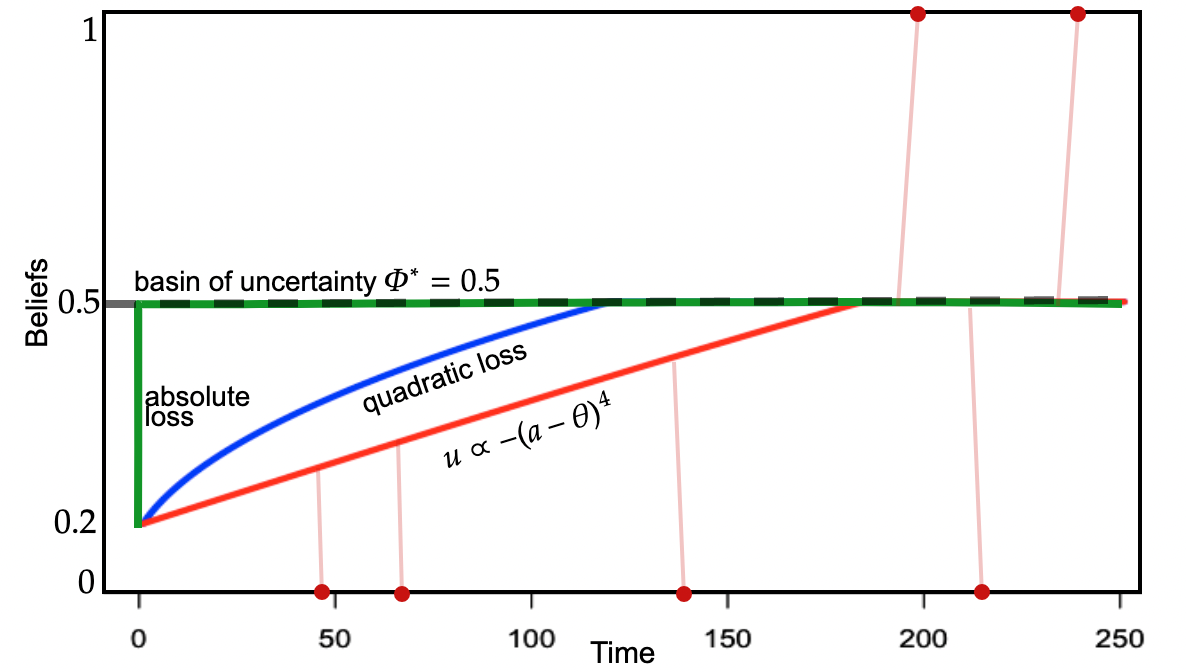}} 
    \label{fig:belief_path_converging}
\end{figure}

For binary states, indifference, increasingness, and maximality jointly pin down a unique dynamic information structure. Suppose, for instance that $\Theta = \{0,1\}$. We associate beliefs with the probability that $\theta = 1$ and suppose that $\mu_0 < \Phi^*$ so the prior starts outside the basin of uncertainty. From Corollary \ref{cor:convexconcave} (i), the optimal structure when $f$ is convex provides full information \emph{asymmetrically} by only providing conclusive news that $\theta = 0$.\footnote{More precisely, the belief path $(\mu^C_t)_t$ is pinned down by the relation $\phi(\mu^C_{t+1}) / \mu^C_{t+1} = \big(\phi(\mu^C_{t}) - c\big) / \mu^C_{t}$ whenever $\mu^C_{t+1} < \min \Phi^*$, and $\mu^C_{t+1} = \min \Phi^*$ otherwise.}  This is depicted by the red lines in Figure \ref{fig:belief_path_converging} where the thick lines corresponds to continuation beliefs, and the ``jumps'' correspond to the (random) arrival of full information. This bad news process steers the DM's beliefs toward the basin; upon reaching it, the optimal dynamic information structure provides full information \emph{symmetrically} across both states so that, upon not learning the state, the DM's beliefs are unchanged. 

The thick green, blue, and red lines in Figure \ref{fig:belief_path_converging} illustrate the optimal continuation belief paths for different DM utility functions when $A$ takes values between $0$ and $1$.\footnote{Although we assumed $A$ is finite, this can be approximated with a finite grid.  Utilities are normalized so value of full information in the basin is equal.}
When the DM's indirect utility is more concave, slower paths to the basin are better and vice versa.  A qualitatively similar binary-state case with exponential discounting has been studied in subsequent work \citep*{saeedi2024getting} via weak duality which requires guessing a sequence of multipliers. Beyond generalizing to many states, Corollary \ref{cor:convexconcave} offers an arguably more transparent characterization in the language of convex-orders and extreme-points.

\begin{corollary}[Block structures are optimal after reaching the basin]  \label{cor:blockafterbasin} 
For any increasing value of attention $f$, suppose that $(d,\bm{\mu}^C)$ is optimal for attention capture. If there exists some time $T < +\infty$ such that $\mu^C_T \in \Phi^*$ then the dynamic information structure starting from $T$ is a block structure: 
\[
\Big(d|_{\tau \geq T}, (\mu^C_{t-T})_{t \geq T}\Big) \in \mathcal{I}^{BLOCK}.
\]
where $d|_{\tau \geq T}(t) = \frac{d(t+T)}{\sum_{s \geq T}d(s)}$ for times $t > 1$ is the conditional distribution of $d$ from $T$ onwards renormalized and translated to start at time $0$. 
\end{corollary}

Corollary \ref{cor:blockafterbasin} states that conditional on reaching the basin of uncertainty, the optimal \emph{continuation} dynamic information structure is a block structure. The special case where beliefs start in the basin corresponds to $T = 0$ in which case optimal dynamic information structures are simply block structures. 

The reasoning behind Corollary \ref{cor:blockafterbasin} builds on the observation that the DM's attention costs are sunk upon reaching the basin. Hence, the dynamic information structure after then can be altered without affecting the DM's obedience constraints at earlier periods as long as this alteration preserves the conditional mean. Then (i) the law of total expectation decomposing the designer's payoff into the events that the DM stops before and after reaching the basin; (ii) the observation the designer's payoff is a continuous linear functional on $\Delta(\mathcal{T})$; and (iii) Bauer's maximum principle implies that the continuation information structure is attained an an extreme point over the set of feasible stopping times. From Theorem \ref{thrm:convex_ext} (ii), this is precisely given by block structures. 

\begin{corollary} [Block structures are optimal for binary states and actions] \label{cor:binarybinary}
Suppose that $|\Theta| = |A| = 2$. Then, for any increasing value of attention $f$, an optimal dynamic information structure is given by $(d, \bm{\mu}^C) \in \mathcal{I}^{BLOCK}$ where $\bm{\mu}^C$ jumps directly to the basin of uncertainty: 
\[
\mu^C_t = \begin{cases}
    \mu_0 \quad &\text{for $t = 0$}\\
    \mu^* \in \Phi^* &\text{for $t \geq 1$.}
\end{cases}
\]
\end{corollary}

Corollary \ref{cor:binarybinary} implies that for the simplest case with binary states and binary decisions, the belief paths accompanying optimal dynamic information structures are quite simple and consist of an immediate jump to the basin of uncertainty.

\paragraph{Attention capture with more exotic time preferences.}

We have thus far developed a fairly general analysis of optimal attention capture though the perspective of convex-orders and extreme points. This has allowed us to tackle settings in which the designer's value function $f$ is more convex (correspond to less time-risk aversion) or more concave (less)  than the DM's cost function $c$. While this is natural, there are economic environments which do not fall into either category.

\begin{definition}
    Say that $f$ is a step function if there exists some time $T$ and some increasing function $g$ such that 
    \[
    f(t) = \mathbb{I}\Big(t \geq T\Big) \cdot g(t). 
    \]
\end{definition}

\begin{figure}[h]
\centering
\caption{Illustration of step functions}
    \subfloat[$f$ step function, $c$ linear]{\includegraphics[width=0.45\textwidth]{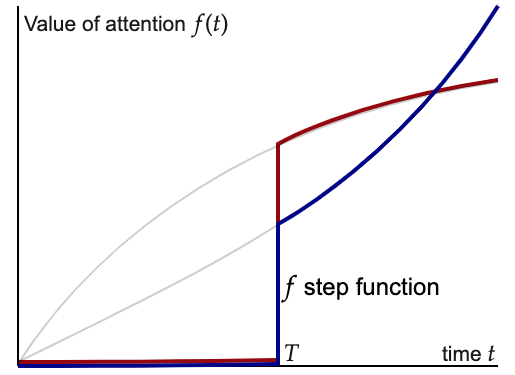}}
    \subfloat[$f$ linear, $c$ ``inverse step'']{\includegraphics[width=0.45\textwidth]{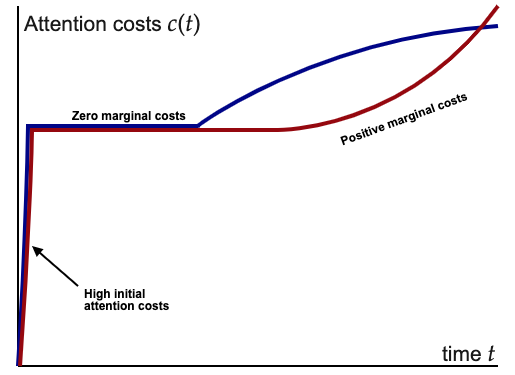}}
    \label{fig:step}
\end{figure}

Step functions reflect environments in which the designer might only begin to value attention beyond a certain threshold of engagement. Two such functions are depicted in panel (a) of Figure \ref{fig:step}.  Alternatively, step functions approximates environments in with the designer's value of attention is linear but the DM's per-unit time cost of attention is an ``inverse step function'': it is initially very steep (which might represent a ``fixed cost'' to attention) before the cost of attention falls temporarily to zero (e.g., the DM voluntarily engages with the advertisement) before eventually rising again.\footnote{Our results will hold for $f$ ``sufficiently flat'' so that the marginal value of attention is small but positive before time $T$. This allows the inverse $f^{-1}$ to be well-defined and, in panel (b) of Figure \ref{fig:step}, admit the interpretation of high marginal costs of attention by setting $c(t) := f^{-1}(t)$. 
} 

\begin{definition}
    Say that $f$ is $S-$shaped if there exists some time $T$ such that $f(t) - f(t-1) \lessgtr f(t+1) - f(t)$ if $t \lessgtr T$.
\end{definition}

`S-shaped' functions reflect environments in which users are initially unresponsive to advertising but, after crossing a threshold (consumer `wear in'), begin to respond strongly; at some point they become saturated and their demand once again tapers off (`wear out').\footnote{S-shaped response functions have been influential within economics, marketing, and consumer psychology; see \cite{fennis2015psychology} and references therein for more details. In the literature on static communication, S-shaped value functions also feature prominently (see, e.g., \cite{lipnowski2020cheap}).} An equivalent interpretation is where the designer's valuation of attention is linear but the DM's marginal time cost is initially very high but eventually lowers then rises again after waiting beyond a certain point.\footnote{This once again follows again from the observation that the constant cost $c$ is a normalization so the results holds for $f \circ c^{-1}$ S-shaped.}
 
\begin{proposition}[Attention capture for step and S-shaped functions] \label{prop:step_and_S-shaped} If $f$ is a
\begin{itemize}
    \item[(i)] \textbf{step function}, then for the step location $T > \phi^*/c$, every optimal dynamic information structure $(d, \bm{\mu}^C)$ must be such that 
    \begin{itemize}
        \item[(i)]  $\bm{\mu}^C$ is increasing and maximal
        \item[(ii)] \ref{eq:obedience} binds for all $t < T-\phi^*/c$. 
    \end{itemize}
    Moreover, if $T$ (independent of the function) is sufficiently large and $\{\mu \in \Delta(\Theta): \phi(\mu) \geq \phi(\mu_0)\} \subset int \Delta(\Theta)$, then continuation beliefs $\bm{\mu}^C$ reaches the basin before time $T-\phi^*/c.$
    
    \item[(ii)] \textbf{S-shaped function}  and $\mu_0 \in \Phi^*$ or $|\Theta| = |A| = 2$, then there exists some deterministic time $t^*$ such that the optimal information structure $(d,\bm{\mu^C}_t)$ is such that  
    \begin{itemize} 
            \item \textbf{Stopping time:} $d$ fulfils 
            \begin{align*}
                &\mathbb{P}_{\tau \sim d}(\tau = t+1 | \tau > t) = \begin{cases}
          \frac{c}{\phi^*} &\text{if } 1 \leq t <t^*, \\
          \frac{c}{\phi(\mu_0)} &\text{if } t = 0
          \end{cases}   \\
          &\mathbb{P}_{\tau \sim d}\Big(\tau = t^* + \frac{\phi^*}{c} \Big | \tau > t^*\Big) = 1
            \end{align*}
            \item \textbf{Belief path:} $\mu_t \in \Phi^*$ for all $t \geq 1$.
        \end{itemize}
\end{itemize}
\end{proposition}

\begin{figure}[h]
\centering
\caption{Optimal for $f$ S-shaped}
    \subfloat[Belief path and CDF]{\includegraphics[width=0.49\textwidth]{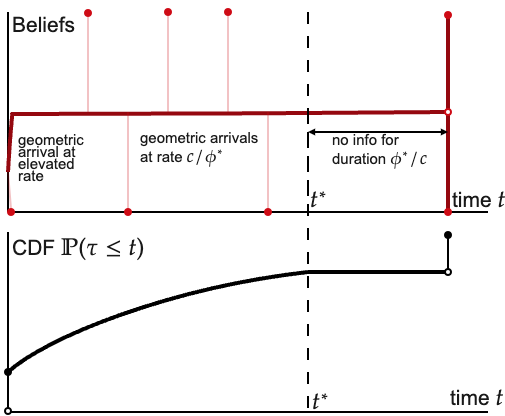}}
    \subfloat[Conditional concavification
    ]{\includegraphics[width=0.49\textwidth]{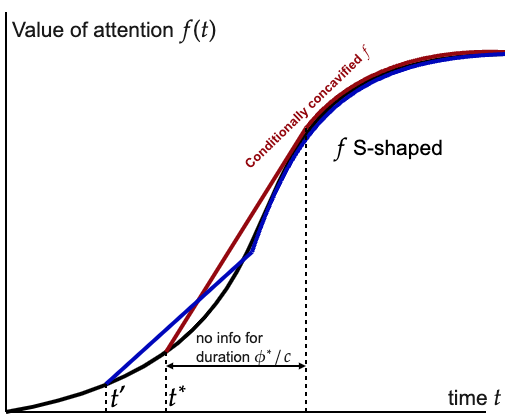}}
    \label{fig:sshape}
\end{figure}

When $f$ is S-shaped (part (ii)), the optimal dynamic information structure comprises an initial jump to the basin following which information arrives at a geometric rate of $c/\phi^*$ up to a deterministic time $t^*$. Conditional on reaching time $t^*$, the DM receives no information until time $t^* + \phi^*/c$ at which point the DM receives full information for sure. The belief paths and CDF associated with the optimal information structure are depicted in panel (a) of Figure \ref{fig:sshape}. The time $t^*$ is in turn pinned down as the 
smallest time at which the \emph{conditional} concavification of the value function $f$---the concavication of $f$ only considering times $t \geq t^*$---depicted as the red curve in panel (b) of Figure \ref{fig:sshape} is such that the first time tangent to $f$ is less than $\frac{\phi^*}{c}$ away from $t^*$. This ensures that the DM's stopping time conditional paying until until time $t^*$ concentrates on $t^* + \phi^* / c$ which exploits the ``steeply increasing'' portion of $f$. To see why this is optimal, consider, instead, an alternative dynamic information structure which delivers full information at a geometric rate up until time $t' < t^*$. Observe that at time $t'$, the designer is unable to induce the DM to wait until time $t^* + \phi^*/c$ with probability $1$ since since information is insufficiently valuable. Instead, the designer could ensure the DM waits until time $t' + \phi^*/c$. This, however, will be suboptimal because at time $t'+\phi^*/c$, the marginal value of attention is still high. Thus, by delivering full information at a geometric rate for longer (until $t^*$), the designer is able to fully exploit the convex-concave curvature of $f$. 

When $f$ is a step-function (part (i)), optimal dynamic information maximizes the probability that the DM pays attention until $T$ (at which point the designer's value jumps). To do so, the designer delivers full information at a stochastic rate to keep the DM indifferent while her continuation beliefs are steered towards the basin. This also suggests that a durable property of optimal dynamic information (across the S-shaped, convex, and step-function cases) is that, at least for initial periods, the DM is kept indifferent while her continuation beliefs are steered toward the basin in an increasing and maximal fashion. 

\section{Time-consistency}
\label{sec:time_con}
Our analysis thus far proceeded under the assumption that the designer has full intertemporal commitment. It turns out that this is unnecessary.

\begin{definition} $I$ is sequentially optimal for designer preference $f: \mathcal{T} \to \mathbb{R}$ if for every history $H_t$ which realizes with positive probability
    \[
        \max_{I' \in \mathcal{I}|H_t} \Ex^{I'}\Big[f\big(\tau(I')\big) \big|H_t\Big] = \Ex^{I}\Big[f\big(\tau(I)\big)\big|H_t\Big] 
    \]
    where $\mathcal{I}|H_t$ is the set of dynamic information structures where $H_t$ realizes with positive probability.
    Call the set of sequentially-optimal structures $\mathcal{I}^{SEQ} (f)$. 
\end{definition}

Sequential optimality is demanding, and requires that for every positive-probability history, the designer has no incentive to deviate to a different continuation information structure. It also implies optimality by choosing $H_t$ as the empty time-$0$ history. Such structures, if they exist, eliminate the need for intertemporal commitment.\footnote{That is, they ensure the designer's "Stackelberg payoff" under full commitment---the designer first chooses $I$ then facing $I$, DM chooses an optimal stopping time and action---coincides with the designer's payoff in a subgame perfect equilibrium of the stochastic game in which at each history, the designer chooses a new static information structure and, facing the chosen information structure, the DM decides between stopping to act (game ends and payoffs realized) or waiting to observe the generated message (game stochastically transitions to a new history).}

\begin{theorem}\label{thrm:time_con}
    Sequentially optimal dynamic information structures exist i.e.,  $\mathcal{I}^{SEQ}(f) \neq \emptyset$.
\end{theorem}

The key idea underlying Theorem \ref{thrm:time_con} is the observation that the designer has some freedom to manipulate the DM's beliefs at non-stopping histories while still preserving the induced joint distribution. By manipulating the DM's beliefs---for instance, by making her more certain of the state---the designer also increases the DM's value of stopping to act. This, in turn, ensures that the designer does in fact find it optimal to follow through with promised information at future histories. The modification relies on the fact that information is irreversible: as soon as the DM's continuation beliefs have shifted, the designer is subject to a martingale constraint with respect to these new continuation beliefs. Indeed, in the Appendix we show a more general version of Theorem \ref{thrm:time_con} which applies to any DM preference which dislikes delay.\footnote{That is, $t > t' \implies v(a,\theta,t) < v(a,\theta,t')$ for any $(a,\theta)$.} 

\begin{figure}[!h]  
\centering
\captionsetup{width=1\linewidth}
    \caption{Making attention capture sequentially optimal} \includegraphics[width=0.7\textwidth]{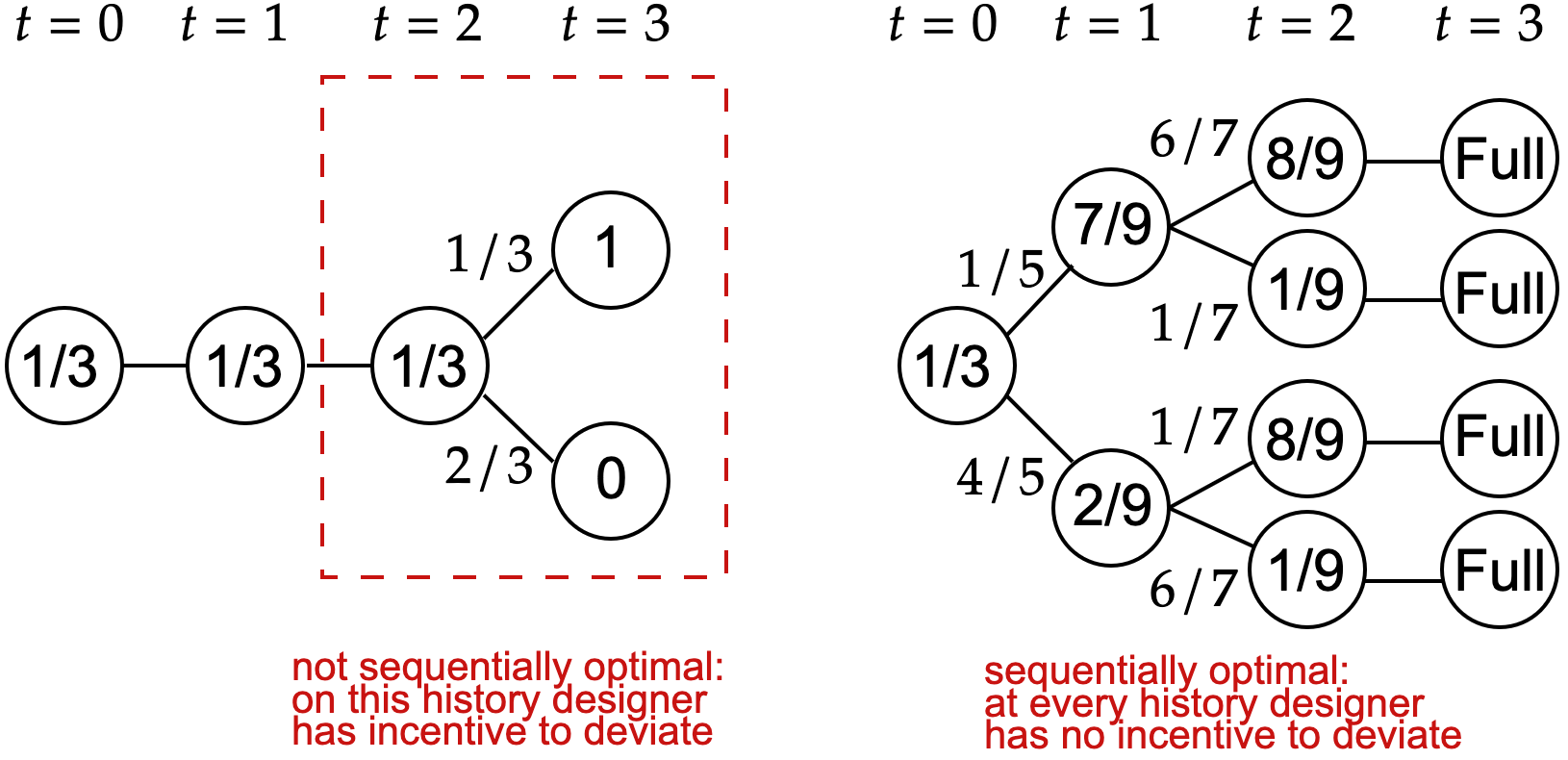}
     \label{fig:time_consis_illust}
\end{figure}

\paragraph{Illustration of Theorem \ref{thrm:time_con}.}
The proof of Theorem \ref{thrm:time_con} is deferred to Appendix \ref{appendix: time-consistent-modification} and proceeds by performing surgery on the tree representing the optimal dynamic information structure. We illustrate the main ideas through the following example. Suppose there is a single decision maker (DM) whose payoff from choosing action $a \in \{0,1\}$ when the state is $\theta \in \{0,1\}$ at time $\tau$ is quadratic-loss: 
\[
v(a,\theta,\tau) = -(a-\theta)^2 - c\cdot \tau \quad \text{with} \quad c = 1/9.
\] DM has prior $\mu_0 = \mathbb{P}(\theta = 1) = 1/3$ and the designer's payoff is $f(\tau)$ where $f$ is concave and strictly increasing. From Corollary \ref{cor:convexconcave} (ii), a designer-optimal distribution of stopping beliefs puts probability $1$ on $\tau = 3$ to deliver a payoff of $f(3)$.
Under commitment, this can be implemented via an information structure which gives the DM no information up to time $3$, upon which the DM learns the state perfectly. This is illustrated on the left of Figure \ref{fig:time_consis_illust}.

Note, however, that this not sequentially optimal: at time $t = 3$ 
 the designer does better by instead not following through with their promise and might instead try and use information to further extract attention. Indeed, \cite{knoepfle2020dynamic} and \cite{hebert2022engagement} study the case with linear value of attention and, noting this time-inconsistency, employ an alternate structure at which full information arrives at a Poisson rate. 
 However, such distributions are not optimal when $f$ is strictly concave: the uniquely optimal stopping time puts probability $1$ on $\{\tau = 3\}$. How can the designer implement this time-consistently? 

Consider the modification depicted in Figure \ref{fig:time_consis_illust} which delivers information \emph{gradually}: instead of inducing the unique beliefs $1/3$ at time $t = 1$, the designer instead induces the beliefs $7/9$ and $2/9$. At time $t = 2$, the designer induces the belief $8/9$ and $1/9$. Finally, at $t = 3$, the designer gives the DM full information. Observe that under this information structure, the DM is indifferent between continuing and stopping at every continuation history. For instance, at both histories with belief $8/9$, the DM is indifferent between stopping at time $t = 2$ and taking action $1$, yielding an expected payoff of $-1/9$, or paying an extra cost of $c = 1/9$ to learn the state at time $t = 3$. 

We have modified the information structure to induce the same distribution over stopping times, so this structure remains optimal. It remains to sketch why it is also sequentially optimal. Suppose, towards a contradiction, that it is not. Then there must be some continuation history $H_t$ such that the designer strictly prefers to deviate to an alternate continuation information structure $I'|H_t$. But observe that under the current structure, the DM was indifferent between continuing and stopping at $H_t$. Hence, under $I'|H_t$, the DM's continuation incentives must weakly improve. But this in turn means that continuation incentives at earlier histories leading to $H_t$ also weakly improve. Hence, the new information structure with $I'|H_t$ replacing $I|H_t$ remains obedient. Since $H_t$ realizes with positive probability, this is a strict improvement for the designer, contradicting the optimality of the original structure.

\section{Noninstrumental value of information}
\label{sec:noninstrumental}

\paragraph{Moving from sequentially optimality to suspense} We will now show a tight connection between dynamic information structures which are \emph{sequentially optimal} as analyzed in the previous section, and optimal attention capture when information is valued noninstrumentally (via suspense). On first reflection this connection might seem strange: the latter is a property of the designer's interim incentives while the former is a property of the DM payoffs. It will turn out that in both environments, whether \emph{interim continuation beliefs} are random or deterministic plays a key role. We start by outlining how DM might values suspense before analyzing optimal attention capture under suspense utility.

\begin{definition}[Suspense] Following \cite*{ely2015suspense}, let $\phi_s: \Delta(\Theta) \to \mathbb{R}$ is a strictly concave functional such that $\phi_s(\delta_{\theta}) = 0$ for each $\theta$. Call $\phi_s$ the \emph{suspense potential} for reasons which will be soon apparent. For the information structure $I$ and the (random) belief realizations $(\mu_t')_{t \leq \tau}$ up to stopping time $\tau > 0$, DM's utility under suspense is
\[
        v^{susp}\Big((\mu_t')_{t \leq \tau}, I,\tau\Big) = \sum_{t \leq \tau-1} g \bigg( \Ex^I \Big[\phi_s(\mu_t') - \phi_s(\mu_{t+1})    \Bigl\vert H_t := (\mu_s')_{s \leq t}
        \Big] 
\bigg)
 - c \cdot \tau.
\]
where $g$ is a strictly increasing and strictly concave \emph{aggregator} and if $\tau = 0$ then $v^{susp} = 0$.\footnote{Note that our definition of suspense is such that, if DM pays attention up to $\tau$, she gets the sum of her flow suspense utility for periods $t = 0,1,\ldots \tau - 1$ but not for $\tau$. As a positive assumption, this implies the DM does not value suspense for information she knows she will not receive. Nonetheless, our results remain unchanged (with time translated by $1$ period) if, stopping at $\tau$, the DM gets the sum of her flow suspense utility for periods $t = 0,1,\ldots \tau$.} 
\end{definition}
We emphasize that suspense utility is a function of all three arguments: the belief path $(\mu'_t)_t$ (which might be random), the dynamic information structure $I$, and the stopping time $\tau$. At the time-$t$ history of realized continuation beliefs $(\mu'_s)_{s \leq t}$, the DM assesses how much suspense she expects to receive from paying further attention which depends on $I$ since the dynamic information structure governs the law of future beliefs. These two ingredients determine the DM's flow suspense utility; the stopping time $\tau$ governs the boundary at which the DM stops paying attention.  

\makeatletter
\newcommand{\specialcell}[1]{\ifmeasuring@#1\else\omit$\displaystyle#1$\ignorespaces\fi}
\makeatother

\begin{example}[Squared variation] Define $\phi_s(\mu) = \sum_{\theta} (1 - \mu(\theta)) \mu(\theta)$ as the expected squared variation from learning the state at belief $\mu$. Then at history $H_t = (\mu_s)_{s \leq t}$, the DM's suspense utility for the period is
\begin{align*}
    \Ex^I \Big[\phi_s(\mu_{t}) - \phi_s(\mu_{t+1})  \Bigl\vert H_t 
        \Big] 
        &= 
        \Ex^I \Big[{\sum_{\theta} (\mu_{t+1}(\theta) - \mu_t(\theta))^2}
        \Bigl\vert H_t 
        \Big]
\end{align*}
as in the main text of \cite*{ely2015suspense}.\footnote{To see this, write out the term within the expectation on the LHS via the definition of $\phi_s$ and use the the martingale property of beliefs: $\Ex^I[\mu_{t+1}(\theta)|H_t] = \mu_{t}(\theta) = \Ex^I[\mu_{t}(\theta)]$ and rearrange. We follow \citep*{ely2015suspense} in assuming flow suspense utility is aggregated additively across time via a concave function.} 
$\hfill \diamondsuit$
\end{example}

Facing $I$, the DM chooses an optimal stopping time $\tau$ adapted to the filtration generated by belief realizations. The designer's problem is, as before, $\sup_{I} \Ex^I[f(\tau(I))]$ where $\tau(I)$ is the DM's stopping time under suspense utility.

The DM's stopping time $\tau$ must fulfil a sequence of dynamic constraints analogous those in Lemma \ref{prop: IC+boundary-condition}: for each time $t < \tau$ and history $H_t = (\mu_s)_{s \leq t}$ which realizes with positive probability, her expected additional suspense utility up to her optimal stopping time $\tau$ must be greater than her expected additional flow cost: 

\begin{align*}
    \Ex \bigg[\sum_{s= t}^{\tau - 1}  g\bigg(\underbrace{\Ex_s \Big[ \phi_{s}(\mu_{s}) - \phi_s(\mu_{s+1})  \mid H_s \Big]}_{\text{time-$s$ suspense}}\bigg) \Big \lvert  H_t\bigg] \geq \Ex[c(\tau-t) \mid H_t],
\end{align*}
noting that the expected additional utility at time-$t$ from suspense until stopping at time $\tau$ depends on both the information structure $I$ as well as realization of (potentially random) interim beliefs between $t$ and $\tau$. 

Recall we defined the \emph{suspense potential} $\phi_s: \Delta(\Theta) \to \mathbb{R}$ as a primitive of the DM's preference for suspense. Further define $\Phi^*_s := \max_{\mu} \phi_s(\mu)$ as the basin of suspense where the DM's suspense potential is maximized. Say the sequence of beliefs $(\mu_t)_t$ is increasing in suspense potential if $(\phi_s(\mu_t))_t$ is increasing. 

Our next result develops a tight connection between optimal attention capture under suspense, and sequentially optimal attention capture when information is instrumentally valuable. 

\begin{theorem} \label{thrm:suspenseoptimal} For any designer value function $f$, 
\[
\boxed{\substack{\text{\normalsize $I$ is optimal under suspense utility} \\ \text{\normalsize with suspense potential $\phi_s$,}\\
\text{\normalsize cost $c$ and aggregator $g$}}}
\iff 
\boxed{\substack{\text{\normalsize $I$ is sequentially-optimal under instrumental value} \\ \text{\normalsize for decision problem with $\phi = \phi_s$}\\
\text{\normalsize and cost $g^{-1}(c)$}}}
\]
In particular, if $f$ is 
\begin{enumerate} [nosep]
    \item[(i)] \textbf{concave} and $\phi_s(\mu_0)/g^{-1}(c)$ is an integer, the designer-optimal information structure $I$ induces a stochastic belief path which reduces suspense potential deterministically: for all $t < \tau$,
    $\phi(\mu_{t}) - \phi(\mu_{t+1}) = g^{-1}(c)$ a.s. and  $\mathbb{P}^I\big(\tau = \frac{\phi_s(\mu_0)}{g^{-1}(c)}\big) = 1$;\footnote{Our assumption that $\phi_s(\mu_0)/g^{-1}(c)$ is an integer is purely an artifact of working in discrete time; if it were not, the optimal time would be spread over two consecutive periods.}
    \item[(ii)] \textbf{convex}, the designer-optimal information structure $I$ induces a deterministic and increasing belief path which reduces suspense potential stochastically: there exists a deterministic and increasing path $(\mu^C_t)_t$ such that for all $t \in \mathcal{T}$, 
    \[ \text{supp }\, I_{t+1}(\cdot|H_t = (\mu^C_s)_{s \leq t}) \subseteq \{\mu^C_{t}\} \cup \{\delta_{\theta}: \theta \in \Theta\}
    \]
    and $\mathbb{P}^I(\mu_{t+1} \in \{\delta_{\theta}: \theta \in \Theta\} | \tau \geq t) = g^{-1}(c)/\phi_s(\mu^C_t)$. 
\end{enumerate}
\end{theorem}

\begin{figure}[!h]  
\centering
\captionsetup{width=1\linewidth}
    \caption{Optimal attention capture when DM values suspense} \includegraphics[width=0.9\textwidth]{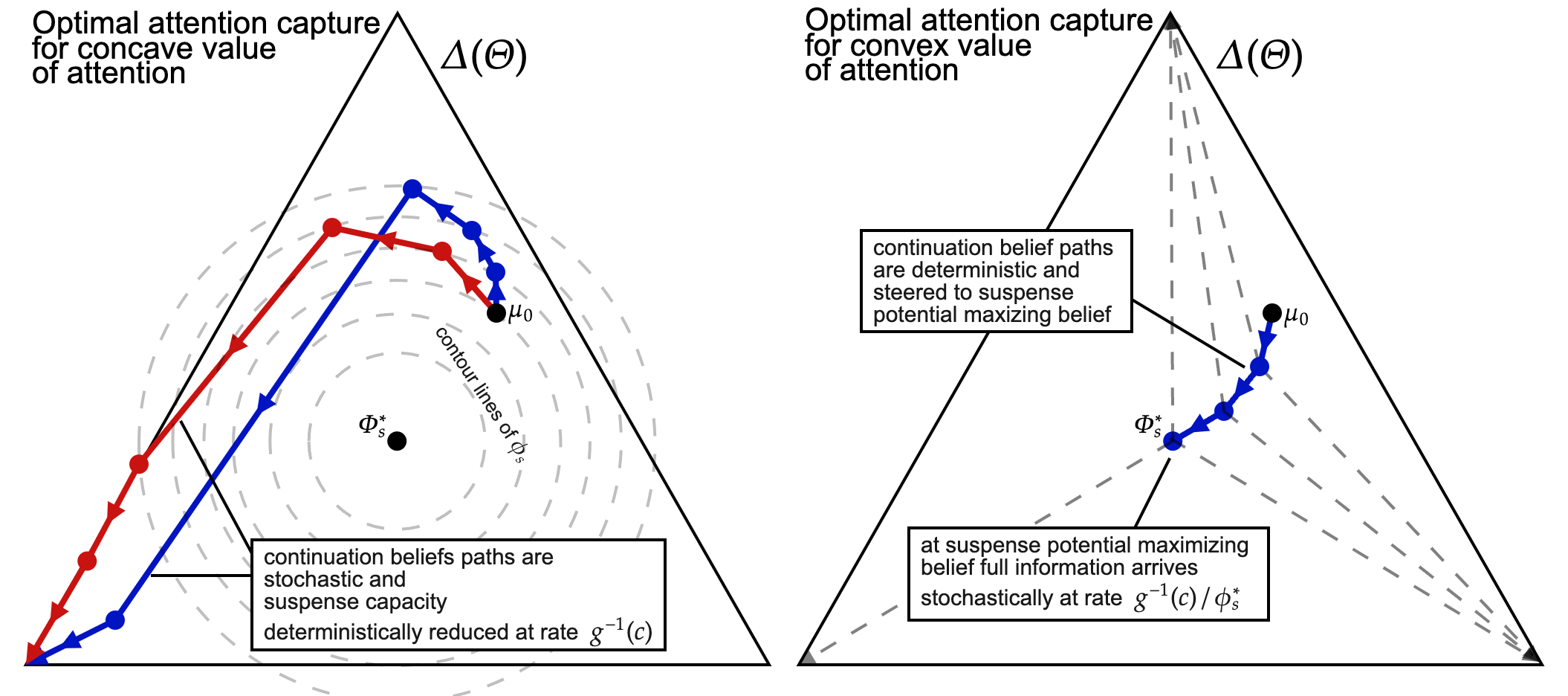}
     \label{fig:suspense_optimal}
\end{figure}

Theorem \ref{thrm:suspenseoptimal} establishes a bijection between optimal attention capture under suspense and sequentially optimal attention capture when information is valued instrumentally we developed in Sections \ref{sec:stopping_time} and \ref{sec:time_con}. It further illustrates how relative time preferences between the designer and DM shapes not only whether full information should arrive at random or deterministic times (analyzed in Section \ref{sec:stopping_time}), but also whether {interim beliefs} are deterministic/collapsed or random/spread out. At the heart of Theorem \ref{thrm:suspenseoptimal} is the observation that there are two distinct ways to portion out suspense over time: 

 When $f$ is concave, suspense should be portioned out \emph{deterministically} over time with \emph{stochastic continuation belief paths} so that the DM receives $g^{-1}(c)$ per-period in flow suspense utility. The left panel of Figure \ref{fig:suspense_optimal} shows two realizations of belief paths. Given $\mu_t$, the continuation belief $\mu_{t+1}$ is supported on the set of beliefs where suspense is $\phi_s(\mu_{t}) - g^{-1}(c)$. This then induces a \emph{deterministic} stopping time which is optimal since $f$ is concave. Moreover, since the DM's per-period utility exactly offsets her cost, she is kept indifferent between continuing and stopping. Indeed, this is also true of sequentially optimal attention capture for concave $f$ when information is valued instrumentally. Further observe that this also coincides with the utility-maximizing dynamic information structure of \cite*{ely2015suspense} for the fixed time horizon $\phi_s(\mu_0) / g^{-1}(c)$. 

When $f$ is convex, suspense should be portioned out \emph{stochastically} over time with \emph{deterministic continuation belief paths} so that the DM receives $g^{-1}(c)$ in per-period flow utility. The right panel of Figure \ref{fig:suspense_optimal} shows the belief path which is steered deterministically towards the basin $\Phi^*_s$ where suspense potential is maximized. Thus, for each belief $\mu_t$, the optimal information structure delivers full information in the next period with probability $\frac{g^{-1}(c)}{ \phi_s(\mu_t)}$ such that her flow suspense utility exactly offsets the cost of waiting. Note, however, that unlike the case where information is instrumentally valuable, suspense utility depends on the entire path of realized beliefs. 

Why is there a connection between optimal attention capture under suspense and sequentially optimal attention capture under instrumental information? Recall that sequentially optimal dynamic information  progressively delivers more reservation utility to the DM such that she is kept indifferent between stopping to act and continuing to pay attention. This equates the additional \emph{final} value of information $\phi(\mu_t)$ with her expected additional cost conditional on already paying attention up to $t$. In the suspense case, we equate the additional \emph{flow} value of suspense $\Ex[\sum_{s \geq t}^{\tau - 1} g(\phi_s(\mu_{s}) - \phi_s(\mu_{s+1})) \mid H_t]$ with the DM's expected additional cost. Although these settings resemble each other, they are not isomorphic.\footnote{In particular, information structures which fulfil dynamic obedience constraints when information is valued instrumentally (for a decision problem where $\phi = \phi_s$ and cost is $g^{-1}(c)$ per unit time as in Theorem \ref{thrm:suspenseoptimal}) may not fulfil dynamic obedience constraints under suspense.
} Nonetheless, Theorem \ref{thrm:suspenseoptimal} shows that \emph{at the sender's optimum}, there is a tight connection: in Appendix \ref{appendix: time-consistent-modification} we construct a relaxed problem for optimal attention capture under suspense, and show that the sequentially optimal structures for a modified problem attains the bound. 

\section{Attention capture with persuasion motives} \label{sec:bangbang}
We finally turn to the problem of optimal dynamic information structures when the designer has preferences over both the DM's action and stopping time. For instance, a retail platform might generate revenue from both commissions (purchase decision) and advertisements (browsing duration); a law firm on a hybrid fee model might charge for both winning the case (contingent fees) as well as for their time (hourly fees); and so on. Within the binary environment $(\Theta = A = \{0,1\})$, say that the designer's preferences $\hat f: A \times \mathcal{T} \to \mathbb{R}$ are \emph{additively separable across actions and times} if \[\hat f(a,\tau) = a + f(\tau)\]
for some strictly increasing function $f(\tau)$.\footnote{Subsequent work \citep*{koh2024persuasion} studies implementation of joint distributions over actions and stopping times. However, we have not found a way to employ those duality approaches to the additively separable case studied here. This suggests that the perturbation techniques developed here are complementary.} 
Implicit in this formulation is that the designer has state-independent preferences.

Relative to pure attention capture, the DM's continuation value now depends on both \emph{when} she expects to stop, as well as \emph{what} beliefs she expects to hold when she does. If, for instance, after time $t$ the continuation information structure assigns high probability to stopping at a belief at which DM is indifferent between actions, this depresses incentives to continue paying attention since information has no value then. 

Indeed, this tradeoff is at the heart of the designer's problem: in order to increase the probability that DM takes the designer's preferred action, the designer must garble information which, in turn, depresses the DM's incentives to pay attention.\footnote{Several recent papers study persuasion in the presence of a rationally inattentive agent \cite*{lipnowski2020attention, bloedel2021persuasion}. Our setting is complementary to theirs: it is richer along the time dimension since the designer chooses dynamic information structures, but simpler in the sense that the receiver simply decides when to stop paying attention, not what to pay attention to.} 
Hence, dynamic information structures can be quite complicated and intersperse periods of persuasion (at which DM stops with intermediate beliefs and takes action $1$) with periods of attention capture (at which DM stops upon learning the state). Nonetheless, we will show that when attention and persuasion are valued additively, optimal information structures are surprisingly simple. 

\begin{definition}
    A dynamic information structure is one-shot persuasion at time $T$ if it reveals no information for times $t < T$ and at time $T$ coincides with the designer-optimal static information structure subject to the constraint that the DM's expected surplus is greater than $c\cdot T$.\footnote{That is, the static information structure at time $T$ denoted $\pi^* = I(\cdot|H_{T-1}) \in \Delta(\Delta(\Theta))$ maximizes $\Ex_{\mu \sim \pi}[h(\mu,T)]$ among the set 
    \[
    \Big\{ \pi \in \Delta(\Delta(\Theta)): \int \Big(\max_a \Ex_{\mu \sim \pi}[u(a,\theta)] - \max_a  \Ex_{\mu_0}[u(a,\theta)]\Big) d \pi(\mu) \geq c \cdot T, \int \mu d\pi(\mu) = \mu_0\Big\},
    \] where $h(\mu,t)$ is the designer's indirect utility function when the DM stops at belief $\mu$ at time $t$, breaking ties in actions in favor of the designer.}
\end{definition}

\begin{theorem} \label{thrm:attention_persuasion_separable}
    Suppose we are in the binary environment and designer's preferences $\hat f = a + f$ is additively separable across actions and times. Then an optimal dynamic information structure either: 
    \begin{itemize}
    \setlength\itemsep{0em}
        \item[(i)] (Pure attention capture) is equivalent to the designer-optimal when $\hat f(a,\tau) = f(\tau)$; or 
        \item[(ii)] (One-shot persuasion) is one-shot persuasion at some time $T > 0$.
    \end{itemize}
\end{theorem}

Theorem  \ref{thrm:attention_persuasion_separable} states that optimal designs either focuses on extracting attention such that the DM stops only upon learning the state. In this case, the probability that DM takes $1$ is $\mu_0$ so the designer's value function is 
\[
    \Ex^I[\hat f(a,\tau)] = \mu_0 + \Ex^I[f(\tau)]
\]
so that the optimal dynamic information structure coincides with that if the designer's value function were simply $h(\tau)$ as studied in Section \ref{sec:stopping_time}. Alternatively, the designer only provides information at a fixed and deterministic time $T$ such that their payoff is 
\[
    \Ex^I [\hat f(a,\tau)] =  \Ex^I[a_{\tau}] + f(T) 
\]
which persuades the DM into taking action $1$ when it is in fact state $0$ at a frequency which leaves the DM enough value so she finds it worthwhile to pay attention up to $T$. Thus, Theorem \ref{thrm:attention_persuasion_separable} gives a "bang-bang" characterization of optimal dynamic information structures for arbitrary value of attention.  In particular, this implies that, perhaps surprisingly, the designer never finds it optimal to exploit potential randomization gains (which, as we have seen, is optimal with convex value of attention) while persuading the DM by garbling stopping beliefs.


\paragraph{Proof idea of Theorem \ref{thrm:attention_persuasion_separable}.} The proof is involved and deferred to Appendix \ref{appendix:stopping_time}; we sketch the basic ideas here. First observe that it is without loss to consider \emph{extremal stopping beliefs}---that is, stopping beliefs in $\{0,\bar \mu, 1\}$ where the DM is indifferent between actions at belief $\bar \mu$ and we assume tiebreaking in favor of action $1$.\footnote{Extremal beliefs have been used prominently in static information design (e.g., \cite*{bergemann2015limits}). In Appendix \ref{appendix:bangbang} develop the observation that any feasible joint distribution over actions and times can be implemented with (dynamic) extremal stopping beliefs.} 

\begin{figure}[h!]
\centering
\caption{Switching and Pasting Lemma} 
\begin{quote}
    \vspace{-1.5em}
    \centering 
    \footnotesize Note: For simplicity assume continuation beliefs $> \bar \mu$.
\end{quote}
    {\includegraphics[width=0.8\textwidth]{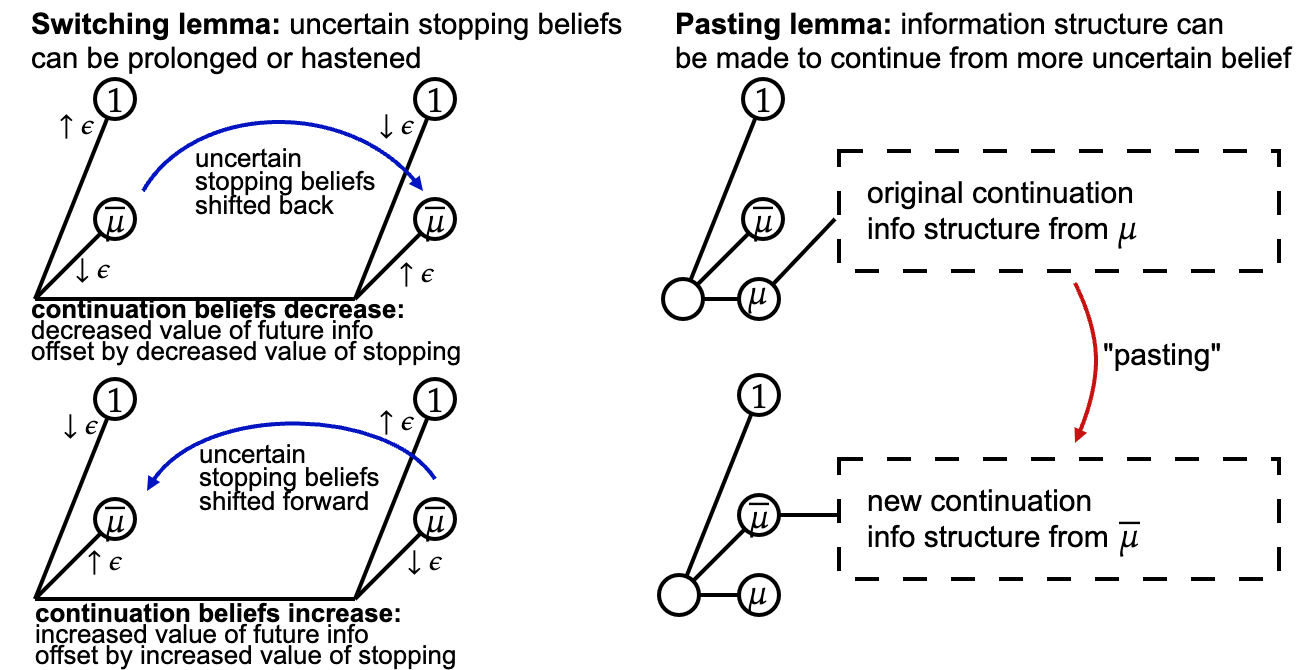}}
    \label{fig:switching_pasting}
\end{figure}

We now develop two key ideas. The first idea is a \emph{switching lemma} which states that the designer can either hasten or postpone the arrival of the uncertain stopping belief $\bar \mu$ while preserving dynamic obedience constraints. This allows the designer to vary the correlation between stopping beliefs and stopping times while preserving marginals as well as continuation incentives. It is illustrated on the left of Figure \ref{fig:switching_pasting}. 

The second idea is a \emph{pasting lemma} which states that if the DM stops with positive probability at time $t$ at belief $\bar \mu$ but continues with belief $\mu > \bar \mu$, then the designer can do strictly better by having the DM continue at a more uncertain belief (e.g., from $\bar \mu$). This is done by "pasting" a suitable modification of the information structure continuing from $\mu$ to instead continue from $\bar \mu$ and is illustrated on the right of Figure \ref{fig:switching_pasting}.\footnote{The pasting lemma is loosely related to the notion of "stop-go" pairs in the mathematics literature on optimal Skorohod embeddings as recently used in \cite*{beiglbock2017optimal}. There, they use path-swapping arguments by continuing on one path, and stopping on another to obtain necessary conditions on optimal stopping for a given stochastic process. Here, we take as given that the DM is optimally stopping and instead modify the belief process to continue on $\bar \mu$ instead of $\mu$, and stop on $\mu$ instead of $\bar \mu$ to obtain necessary conditions on the designer's optimality.}

With these ideas in hand, suppose that the optimal structure is such that the designer provides the possibility of full information over multiple periods but the probability of stopping at the indifferent belief $\bar \mu$ is positive. Employing the switching lemma, we can shift the probability of stopping at different beliefs $\{0,\bar \mu, 1\}$ around in time while weakly increasing the designer's payoff and preserving the DM's dynamic ICs. We do so until there exists some time $t$ at which the DM has positive probability of either: (i) stopping at belief $\{0,\bar \mu,  1\}$; or (ii) continuing with belief $\mu > \bar \mu$. Then, by applying the pasting lemma, we can modify the information structure so that DM (i) stops at belief $\bar \mu$; and (ii) continues at belief $\mu$ according to some modified information structure. This strictly dominates the original structure, a contradiction.

\section{Summary}\label{sec:concluding}
We have developed a unified analysis of the form and limits of attention capture. The reduction principle (Theorem \ref{thrm:reduction}) showed that deterministic continuation belief paths which maximally make DM increasingly uncertain play a crucial role in shaping DM's continuation incentives. We then characterized the convex-order frontier and extreme points of feasible stopping times (Theorem \ref{thrm:convex_ext}) which delivered the form of optimal attention capture for a wide class of environments (Corollaries \ref{cor:convexconcave}-\ref{cor:binarybinary}). We also showed that intertemporal commitment is unnecessary and provided an explicit procedure for making optimal structures sequentially optimal (Theorem \ref{thrm:time_con}) which turned out to have a tight and surprising connection to optimal attention capture for a DM who values information noninstrumentally (Theorem \ref{thrm:suspenseoptimal}). Finally, we considered designer who additionally valued persuasion (Theorem \ref{thrm:attention_persuasion_separable}) and showed that attention capture remains an important force.

\setstretch{0.65}
\small
\setlength{\bibsep}{0pt}
\bibliography{BAM}


\newgeometry{a4paper, left=2.5cm,right=2.5cm,top=2cm,bottom=2cm, includefoot,heightrounded}

\setstretch{1.25}
\begin{center}
\vspace{-1em}
    \large{\scshape{\textbf{Appendix to `Attention Capture'}}} 
\end{center}

\normalsize  
\paragraph{Outline of Appendix.} Appendix \ref{appendix:reduction_timecon} proves the reduction principle (Theorem \ref{thrm:reduction}) and develops generalized martingale and obedience constraints. Appendix \ref{appendix:stopping_time} proves results on optimal dynamic information structures. Appendix \ref{appendix: time-consistent-modification} proves that every optimal structure can be made sequentially optimal (Theorem \ref{thrm:time_con}) and develops a bijection between sequentially optimal structures and optimal structures under suspense utility (Theorem \ref{thrm:suspenseoptimal}). Appendix \ref{appendix:bangbang} proves the bang-bang result (Theorem \ref{thrm:attention_persuasion_separable}) when the designer has persuasion and attention motives.

\paragraph{Augmented dynamic information structures.}
In the appendix we will augment our description of dynamic information structures. Let $\overline{\Delta \Theta} := \Delta(\Theta) \times M$ be the space of belief-message pairs, where $M$ is an arbitrary message space. We now redefine several objects. $I \in \Delta(\prod_{t=1}^{+\infty} \overline{\Delta(\Theta)})$ as a joint distribution over paths of belief-message pairs. $H_t := (\mu_s,m_s)_{s \leq t}$ is a time-$t$ history. Write $I_{t+1}(\cdot|H_t) \in \Delta(\overline{\Delta(\Theta)})$ to denote the conditional distribution over the time-$t+1$ belief-message pair following history $H_t$. Say that $H_t$ is a positive probability history if $\{(\mu_s,m_s)_{s}: (\mu_s,m_s)_{s \leq t} = H_t\} \subseteq \text{supp\, } I$. $I$ is a dynamic information structure if the belief component is a martingale with respect to the natural filtration generated by its histories.

We emphasize that all results in the main text hold as stated for dynamic information structures as belief martingales. This augmented definition helps describe modifications by allowing us to distinguish histories with the same belief path, but different continuation information structures, but one can verify that all results do not rely on this augmentation. For instance, at each optimal augmented dynamic information structure $I \in \Delta(\prod_{t=1}^{+\infty} \overline{\Delta(\Theta)})$, the belief paths for histories where the DM optimally stops are disjoint from belief paths for histories where the DM optimally continues.\footnote{Hence $I$ can be canonically projected down onto the space of belief martingales analyzed in the main text by setting $m = \emptyset$ for each $m$. The distinction between dynamic information structures as conveying information about the state (via beliefs) and about continuation information (via messages) is well-understood \citep{greenshtein1996comparison}.}

\titleformat{\section}
		{\bfseries\center\scshape}     
         {Appendix \thesection:}
        {0.5em}
        {}
        []

\appendix 


\section{Proof of reduction principle}\label{appendix:reduction_timecon}

\subsection{Preliminaries} 

\textbf{Sequential formulation of DM's problem.} DM's problem can be equivalently formulated as follows: for any history $H_t $, the DM solves the following optimization problem:
\begin{align*}
    U^I(H_t)  \coloneqq \sup_{\substack{\tau,a_{\tau}: \\ \tau \geq t}} \mathbb{E}^I[v(a_{\tau},\theta,\tau) \mid H_t],
\end{align*}
where $\tau$ is a stopping time, $a_{\tau}$ is a stochastic action under the natural filtration, and $\mathbb{E}^I[\cdot |H_t]$ is the conditional expectation under information structure $I \in \mathcal{I}$ after history  $H_t.$ Since we break indifferences in favor of not stopping, given history $H_t$, the DM will stop paying attention if and only if $\max_{a \in A} \Ex^I[v(a,\theta, t)|H_t]$, the expected payoff from stopping to act immediately, is strictly greater than the continuation payoff $\mathbb{E}^I[U^I(H_{t+1}) \mid H_t]$. 

\paragraph{General DM and designer preferences.} Define the indirect utility function $v^*: \Delta(\Theta) \times \mathcal{T} \to \mathbb{R}$ and DM's best response correspondence $a^*:\Delta(\Theta) \times \mathcal{T} \rightrightarrows A$ at time $t \in \mathcal{T}$ as follows: 
\begin{align*}
    v^*(\mu,t) = \max_{a\in A}\mathbb{E}_{\theta \sim \mu} v(a,\theta,t), \quad \quad a^*(\mu,t) = \text{arg max}_{a\in A}\mathbb{E}_{\theta \sim \mu} v(a,\theta,t).
\end{align*}  
Let $\hat{f}: A \times \mathcal{T} \to \mathbb{R}$ be the designer's value function. Define 
\begin{align*}
h(\mu^S,t) \coloneqq  \Ex_{\theta \sim \mu^S} \Big[ \max_{a \in a^*(\mu^S,t)} \hat{f}(a,t)\Big]
\end{align*}
as the designer's indirect utility function when the DM stops at belief $\mu^S$ at time $t$ and breaking ties in favor of the designer. The designer's maximization problem is equivalent to 
\[\sup_{{p} \in \mathcal{P}} \Ex_{(\mu^S,\tau) \sim p} [h(\mu^S,\tau)].\]
where $\mathcal{P}$ is the set of feasible joint distributions over beliefs and stopping times. To ensure the solution to the designer's optimization problem, we impose the following assumption which was described informally in the main text.
\begin{assumption} \label{assumption: metric_exist}
There exists a metric $d(\cdot,\cdot)$ over the set $\Delta(\Delta(\Theta) \times \mathcal{T} )$ such that $\mathcal{P}$ is compact under metric $d$. Moreover $p \mapsto \Ex_{(\mu^S,\tau) \sim p} [h(\mu^S,\tau)]$ is continuous under $d$.
\end{assumption}

This implies the optimization problem attains its supremum i.e., $\max_{I \in \mathcal{I}} \Ex[f(a_\tau,\tau)]$ has a solution. This assumption holds, for instance, whenever the designer's function $f$ is bounded by some exponential function. More generally, this is fulfilled by standard functional forms in economics; we elaborate on this in Online Appendix \ref{onlineappendix: topology_space}. We will weaken the constant cost per-unit time assumption to:

\begin{assumption} [Impatience]
    The DM's utility function $v$ satisfies impatience if $v(a,\theta,t)$ is strictly decreasing in $t$ for all $a\in A$ and $\theta \in \Theta$.
\end{assumption}

We are ready to prove reduction principle (Theorem \ref{thrm:reduction}). The proof of Theorem \ref{thrm:reduction} consists of two steps:

\begin{itemize}
    \item \textbf{Step A:} Show that every feasible distribution over stopping times can be implemented with full-revelation and deterministic structures.
    \item \textbf{Step B:} Show that every feasible stopping time can be implemented with a maximal belief path via an iterative ironing procedure.
\end{itemize}
These two steps imply Theorem \ref{thrm:reduction}.

\subsection{Proof of Step A}
\begin{proof}[Proof of Step A] The following step show that every feasible distribution over stopping times can be implemented with full-revelation and deterministic structures. 

\noindent \underline{{\textbf{Step 1: Collapse non-optimal stopping histories ($I \to I'$).}}}

For each $t \in \mathcal{T}$, define a (deterministic) belief $\mu^C_t$  as $\mu^C_t \coloneqq   \Ex^I\big[\mu_t \big \lvert \tau(I) > t \big]$. We construct an information structure $I'$  so that the DM prefers to stop paying attention if and only if she sees a message $S$. At time $t$ with history $H^C_t \coloneqq (\mu^C_s,C)_{s \leq t}$ (the history comprising continuation messages for $t$ periods), define a static information structure $I'_{t+1}( \cdot \mid H^C_t) \in \Delta(\Delta(\Theta) \times M)$ as follows: 
\begin{enumerate} [nosep]
\setlength\itemsep{0em}
    \item[(i)] Stopping beliefs agrees with that under $I$: 
    \[I'_{t+1} \Big(\{(\mu,S) : \mu \in B \} \Big\lvert H^C_t\Big) = \Pr^I\Big(\mu_{t+1} \in B, \tau(I) = t+1 \Big\lvert  \tau(I) > t\Big)\]
    for every Borel set $B \subseteq \Delta(\Theta)$; 
    \item[(ii)]Continuation belief concentrates on $\mu^C_{t+1}$:
    \[I'_{t+1}\Big((\mu^C_{t+1},C) \Big\lvert H^C_t\Big) = \Pr^I\Big( \tau(I) > t+1 \Big\lvert  \tau(I) > t\Big).\]
\end{enumerate}
Moreover, whenever a history contains message $S$, an information structure $I'$ provides no information. It is easy to check that $I'$ is indeed an information structure (such that the martingale condition holds).

    Define $\bar \tau$ as the random time at which DM first sees message $S$ under $I'$. Observe
\[\Pr^{I'}(\mu_{s} \in B, \bar{\tau} = s \mid H^C_t) = \Pr^{I}(\mu_s \in B, \tau(I) = s \mid \tau(I) > t), \]
for every Borel set $B \subseteq \Delta(\Theta)$ and time $s>t$. Letting $\bar \mu$ be the random belief which the DM sees message $S$ under $I'$, we have $(\bar \mu, \bar \tau) \overset{d}{=} (\mu_{\tau(I)}, \tau(I))$ since the modification preserved the distribution of stopping beliefs for each time. It remains to show $I'$ is obedient so the time at which DM receives $S$ is indeed an optimal stopping time. 
\begin{lemma}\label{lemma:continuation_preserved}
Under $I'$, for any $t \in \mathcal{T}$, the DM continues paying attention at history $H^C_t$.
\end{lemma}
\begin{proof}[Proof of Lemma \ref{lemma:continuation_preserved}] We show that the DM does weakly better by continuing until seeing message $S$ since that is a well-defined stopping time. To this end, note that the DM's expected utility if she stops at history $H^C_t$ under $I^{'}$ is 
\begin{align*}
    v^*(\mu^C_t,t) &= v^*\Big(\Ex^I[\mu_t \mid \tau(I)>t],t \Big) \\
    &\leq \Ex^I[v^*(\mu_t,t) \mid \tau(I) > t] \tag{$v^*(\cdot,t)$ is convex} \\
    &\leq \Ex^{I}[v^*(\mu_{\tau(I)},\tau(I)) \mid \tau(I) > t] \tag{sequential optimality of $\tau(I)$}\\
    &= \Ex^{I'}[v^*(\bar \mu,\bar{\tau}) \mid H^C_t].
\end{align*}
 This implies DM does weakly better by continuing at history $H^C_t$. 
\end{proof}
From Lemma \ref{lemma:continuation_preserved}, under $I'$ the DM does not stop unless she sees message $S$. Hence, $(\mu_{I(\tau')},\tau(I')) \overset{d}{=} (\mu_{\bar \tau},\bar\tau )$ which implies $(\mu_{\tau(I')},\tau(I')) \overset{d}{=} (\mu_{\tau(I)},\tau(I))$.

\noindent \underline{{\textbf{Step 2: Modify stopping beliefs to be full-revelation.}}}

We now modify $I'$ so that $I''$ is full-revelation and deterministic which gives the second part of Theorem \ref{thrm:reduction}. Construct a new information structure $I''$ from $I'$ as follows: $\text{supp }(I''_{t+1} \mid H^C_t) = \{(\delta_\theta,S): \theta \in \Theta\} \cup \{(\mu^C_{t+1},C)\}$
\begin{enumerate}[nosep]
    \item[(i)] For every $\theta \in \Theta$, $I''_{t+1}((\delta_\theta,S) \mid H^C_t) = \int_{\mu \in \Delta(\Theta)} \mu(\theta) dI'_{t+1}((\mu,S) \mid H^C_t)$
    
    \item[(ii)]$I''_{t+1}((\mu^C_{t+1},C) \mid H^C_t) = \Pr^{I'}(\tau(I') > t+1 \mid \tau(I') > t)$.
\end{enumerate}
 Part (i) says $I''$ redistributes stopping messages with arbitrary posteriors belief $\mu$ from $I'$ into those with full beliefs. 
It is easy to check that $I'$ is a valid dynamic information structure (so the martingale condition holds). Denote the stopping time ${\bar \tau}$ as a random time at which the DM stops once she sees message $S$ under $I''$. Part (ii) then implies $\Pr^{I''}(H^C_s \mid H^C_t) = \Pr^{I'}(H^C_s \mid H^C_t)$ for every $s>t$ hence $\tau(I') \overset{d}{=} {\bar \tau}$. 

\begin{lemma}\label{lemma:first_modification_preserve_IC_second_step}
Under $I''$, for any $t \in \mathcal{T}$, the DM continues at history $H^C_t = (\mu^C_s,C)_{s \leq t}.$ 
\end{lemma}

\begin{proof}[Proof of Lemma \ref{lemma:first_modification_preserve_IC_second_step}] 
Note that DM's expected utility if she stops at history $H^C_t$ under $I''$ is
\begin{align*}
  v^*(\mu^C_t,t) &\leq \Ex^{I'}[v^*(\mu_{\tau(I')},\tau(I')) \mid H^C_t]  \tag{sequential optimality of $\tau'(I)$} \\
  &= \sum_{s=t+1}^\infty \bigg( \int_{\mu \in \Delta(\Theta)} v^*(\mu,s) dI'_{s}((\mu,S) \mid H^C_{s-1}) \bigg) \Pr^{I'}(\tau(I') >s-1 \mid \tau(I') > t ) \\
  &\leq \sum_{s=t+1}^\infty \bigg(\int_{\mu \in \Delta(\Theta)} \sum_{\theta \in \Theta} \mu(\theta) v^*(\delta_\theta,s) dI'_{s}((\mu,S) \mid H^C_{s-1}) \bigg) \Pr^{I'}(\tau(I') >s-1 \mid \tau(I') > t )\\
  &= \sum_{s=t+1}^\infty \bigg(\sum_{\theta \in \Theta} I''_{s}((\delta_\theta,S) \mid H^C_{s-1})v^*(\delta_\theta,s)\bigg) \Pr^{I''}({\bar \tau} >s-1 \mid {\bar \tau} > t )   \\
  &= \Ex^{I''}[v^*(\mu_{{\bar \tau}},{\bar \tau}) \mid H^C_t].
\end{align*}
Therefore, DM does weakly better by continuing to pay attention given history $H^C_t$.
\end{proof}
We showed in Step 1 that $d(I) = d(I')$. By Lemma \ref{lemma:first_modification_preserve_IC_second_step}, we know $\tau(I'') \overset{d}{=} {\bar \tau}$. Moreover, we showed $\tau(I') \overset{d}{=} \bar \tau$ which implies $\tau(I) \overset{d}{=} \tau(I'')$. Let $\mathcal{I}^{FULL}$ be the set of full-revelation structures with deterministic paths. Since $I'' \in \mathcal{I}^{FULL}$, $\mathcal{D}(\mathcal{I}^{FULL}) = \mathcal{D}(\mathcal{I})$, as desired
\end{proof}
\subsection{Proof of Step B} \label{appendix:increasingbeliefs}
Lemma \ref{prop: IC+boundary-condition} implies that feasible pairs of stopping time and a belief path suffice for information structures. We will construct a topological space of such belief paths, fixing a feasible stopping time.  

We first introduce some useful notation. Define the set of belief paths with $\mathcal{W} = (\Delta(\Theta))^\mathcal{T}$. Fixing a stopping time $\tau$, define $\mathcal{W}(\tau) \subset \mathcal{W}$ as the set of belief paths corresponding to stopping time $\tau$. 
\begin{definition} [Undominated belief path]
Fix stopping time $\tau$. A belief path $(\mu_t)_{t \in \mathcal{T}} \in \mathcal{W}(\tau)$ is an {undominated belief path} under $\tau$ if there is no $(\mu'_t)_{t \in \mathcal{T}} \in \mathcal{W}(\tau)$ such that $\phi(\mu'_t) \geq \phi(\mu_t)$ for every $t \in \mathcal{T}$ and the inequality is strict for some $t \in \mathcal{T}$.
\end{definition}
The following proposition guarantees the existence of an undominated belief path for any feasible stopping time $\tau$. This will be a useful object to prove Step B. Note that the proof of Lemma \ref{prop: IC+boundary-condition} relies on only the first half of Theorem \ref{thrm:reduction}.

\begin{lemma} \label{prop: existence_maximal}
For every feasible stopping time $\tau$, there exists an undominated belief path corresponding to $\tau$. 
\end{lemma}
\begin{proof}
    See Online Appendix \ref{onlineappendix: existence_undominated_belief}.
\end{proof}
Since undominated belief paths exist, we provide a stronger version of Step B.
\begin{lemma}\label{lem:increasing_extremal}
Fixing a feasible distribution $d({\tau}) \in \{\text{marg}_{\mathcal{T}} d: d \in \mathcal{D}\}$, every undominated belief path $(\mu^C_t)_{t \in \mathcal{T}} \in \mathcal{W}(\tau)$ must be increasing and maximal.

\end{lemma}
\begin{proof} [Proof of Lemma \ref{lem:increasing_extremal}]
    Fixing a feasible stopping time $\tau$ and its corresponding undominated belief path $(\mu^C_t)_{t \in \mathcal{T}}$. Consider any sequence of $(\mu^i_t)_{t\in\mathcal{T}} \in (\Delta(\Theta))^\mathcal{T}$. We first prove $(\mu^C_t)_{t \in \mathcal{T}}$ is increasing. We recursively construct another sequence of beliefs $(\mu^{i+1}_t)_{t\in\mathcal{T}} \in (\Delta(\Theta))^\mathcal{T}$ which satisfies the conditions in Lemma \ref{prop: IC+boundary-condition} for each $i \in \mathcal{T} \cup \{-1\}$ as follows: 
\begin{itemize}\setlength\itemsep{0em}
    \item[(i)] If $i=-1$, set $\mu^i_t = \mu^C_t$ for every $t \in \mathcal{T}$.
    \item[(ii)] If $i\geq 0$, given the sequence of $(\mu^i_t)_{t\in\mathcal{T}} \in (\Delta(\Theta))^\mathcal{T}$, define
    \[t_i = \begin{cases}
    \min\{t \in \mathcal{T}: t>i,  \phi(\mu^i_i) \leq \phi(\mu^i_t)\} & \, \text{if the minimum exists,}\\ 
    +\infty & \, \text{otherwise.}
    \end{cases}\]
\begin{itemize}[leftmargin=*] \setlength\itemsep{0em}
    \item[$\bullet$] If $t_i = +\infty$, set $\mu'_t = \begin{cases}
    \mu'_t, &\text{if } t < i \\ \mu^i_{i}, &\text{if } t \geq i 
    \end{cases}$ 
    i.e., the sequence $(\mu^{i+1}_t)_t$ follows the sequence $(\mu^i_t)_t$ up to period $i$, and remains constant at belief $\mu^i_i$ thereafter. 
    \item[$\bullet$] If $t_i < {+\infty}$, define $\theta_i \in  \text{argmax}_{\theta \in \Theta} \frac{\mu_{t_i}^i(\theta)}{\mu_{i}^i(\theta)}$ i.e., the state at which the ratio of the beliefs at time $i$ and time $t_i$---the time at which $\phi$ increases relative to that at time $i$. 
    
    It is clear that $\frac{\mu^i_{t_i}(\theta_i)}{\mu^i_{i}(\theta_i)} \geq 1$. Thus, there exists a sequence $\pi^i_{i},\dots,\pi^i_{t_i-1} \in \mathbb{R}$ such that $\pi^i_t \in \big[1,\frac{\Pr(\tau> t)}{\Pr(\tau > t+1)}\big]$ for every $t \in \{i,\dots,t_i-1\}$ and $\prod_{t=i}^{t_i-1} \pi^i_t = \frac{\mu^i_{t_i}(\theta_i)}{\mu^i_{i}(\theta_i)}$ because the second condition of Lemma  \ref{prop: IC+boundary-condition} for $\mu^i$ implies \[\frac{\mu^i_{t_i}(\theta_i)}{\mu^i_{i}(\theta_i)} \leq \frac{\Pr(\tau > i)}{\Pr(\tau > t_i)} = \prod_{t=i}^{t_i-1}\frac{\Pr(\tau > t)}{\Pr(\tau > t+1)}.\] 
    This sequence $(\pi_s^i)_{s=i}^{t_i-1}$ simply splits up the ratio of $\mu_{t_i}^i(\theta_i) /\mu^i_i(\theta_i)$ into $t_i-1 - i$ sub-ratios. We will now construct a new sequence of beliefs to bridge $\mu^i_{i}$  and $\mu^i_{t_i}$ while ensuring that $\phi$ increases over the interval. For any $t \in \{i,\dots,t_i\}$, set $\mu^{i+1}_{t} = \lambda_{i,t}\mu^i_i +  (1-\lambda_{i,t}) \mu^i_{t_i},$ i.e., a linear combination of the belief at $t_i$ and that at $i$ where the weights are given by 
\begin{align*}
    \lambda_{i,t} = \frac{1}{ \mu^i_{t_i}(\theta_i)-\mu^i_{i}(\theta_i) }\Big(  \mu^i_{t_i}(\theta_i) - \prod_{s=i}^{t-1}{\pi^i_s}\cdot \mu^i_{i}(\theta_i) \Big) \in [0,1]. 
\end{align*}
Moreover, if $t \notin\{i,\dots,t_i\}$, we set $\mu^{i+1}_t = \mu^i_t.$ From the construction of $\mu^{i+1}$, note that $\mu^{i+1}_i = \mu^{i}_i$ and $\mu^{i+1}_{t_i} = \mu^{i}_{t_i}$
\end{itemize}
\end{itemize}
We will now inductively show that a sequence $(\mu_t^i)_{t \in \mathcal{T}}$ satisfies the conditions in Lemma \ref{prop: IC+boundary-condition} for each $i \in \mathcal{T} \cup \{-1\}$:

\noindent \underline{\textbf{Base step.}}$(\mu^{-1}_t)_{t\in\mathcal{T}} = (\mu^C_t)_{t \in \mathcal{T}}$ satisfies the conditions in Lemma \ref{prop: IC+boundary-condition} because by $I$ was assumed to be full-revelation and deterministic.

\noindent \underline{\textbf{Inductive step.}}
Suppose $t \in \mathcal{T}$ such that $(\mu^{i}_t)_{t\in\mathcal{T}}$ satisfies the conditions in Lemma \ref{prop: IC+boundary-condition}. We consider two cases. 

\noindent  \underline{Case 1: $t_i = {+\infty}$.} We have that $\phi(\mu^i_i) > \phi(\mu^i_t)$ for every $t>i$. This implies $\phi(\mu^{i+1}_t) = \phi(\mu^i_i) > \phi(\mu^i_t) \geq \Ex[c(\tau) \mid \tau > t] - c(t)$ for every $t>i$. For every $t \leq i$, $\phi(\mu^{i+1}_{t}) = \phi(\mu^{i}_{t}) \geq \Ex[c(\tau) \mid \tau > t] - c(t).$ Therefore, the boundary constraint holds for every time $t \in \mathcal{T}$ for the sequence of beliefs $(\mu^{i+1}_t)_{t\in\mathcal{T}}$. Moreover, for every $t < i$, we know $\frac{\mu^{i+1}_{t+1}(\theta)}{\mu^{i+1}_{t}(\theta)} = \frac{\mu^{i}_{t+1}(\theta)}{\mu^{i}_{t}(\theta)} \leq \frac{\Pr(\tau > t)}{\Pr(\tau > t+1)}.$ For every $t \geq i$, we have $\frac{\mu^{i+1}_{t+1}(\theta)}{\mu^{i+1}_{t}(\theta)} = \frac{\mu^{i}_{i}(\theta)}{\mu^{i}_{i}(\theta)}= 1 \leq \frac{\Pr(\tau> t)}{\Pr(\tau > t+1)}$. Thus, the boundary constraint holds for every time $t \in \mathcal{T}$ for the sequence of beliefs $(\mu^{i+1}_t)_{t\in\mathcal{T}}$, as required. \\

\noindent \underline{Case 2: $t_i < +\infty$.} We now verify each of the constraints holds.
\begin{enumerate}
    \item[(i)] (Obedience constraint) The definition of $t_i$ implies $\phi(\mu_i) > \phi(\mu_t)$ for every $t\in\{i+1,\dots,t_i-1\}$ and $\phi(\mu_i) \leq \phi(\mu_{t_i})$. Because $\phi$ is concave and $\lambda_{i,t}\in[0,1]$, Jensen's inequality implies that for every $t \in \{i,\dots,t_i-1\},$
    \begin{align}
        \phi(\mu^{i+1}_t) &= \phi(\lambda_{i,t}\mu^i_i +  (1-\lambda_{i,t}) \mu^i_{t_i})\nonumber \\
        &\geq \lambda_{i,t} \phi(\mu^i_i) + (1-\lambda_{i,t}) \phi(\mu^i_{t_i}) \nonumber\\
        &\geq \phi(\mu^i_i) \geq \phi(\mu^i_t) \geq \Ex[c(\tau) \mid \tau>t]-c(t), \label{increasing}
    \end{align}
    where the last inequality follows from the first condition in Lemma \ref{prop: IC+boundary-condition} of $\mu^i$. As such, for every $t \notin \{i,\dots,t_i-1\}$, we have $\phi(\mu^{i+1}_t) = \phi(\mu^{i}_t) \geq \phi(\mu^i_i) \geq \phi(\mu^i_t) \geq \Ex[c(\tau) \mid \tau>t]-c(t)$. Therefore, the obedience constraint holds for every time $t \in \mathcal{T}$ for the sequence of beliefs $(\mu^{i+1}_t)_{t \in \mathcal{T}}.$
     \item[(ii)] (Boundary constraint) For every $t \in \{i,\dots,t_i-1\}$ and $\theta \in \Theta$, we have
    \begin{align*}
        \frac{\mu^{i+1}_{t+1}(\theta)}{\mu^{i+1}_t(\theta)} &= \frac{\lambda_{i,t+1}\mu^i_i(\theta) + (1-\lambda_{i,t+1})\mu^i_{t_i}(\theta)}{\lambda_{i,t}\mu^i_i(\theta) + (1-\lambda_{i,t})\mu^i_{t_i}(\theta)} \\
        &\leq \frac{\lambda_{i,t+1} + (1-\lambda_{i,t+1})\mu^i_{t_i}(\theta_i)/\mu^i_i(\theta_i)}{\lambda_{i,t} + (1-\lambda_{i,t})\mu^i_{t_i}(\theta_i)/\mu^i_i(\theta_i)} \hfill{\quad \quad \Big(\lambda_{i,t} \leq \lambda_{i,t+1} \text{ and } \frac{\mu^i_{t_i}(\theta)}{\mu^i_i(\theta)} \leq \frac{\mu^i_{t_i}(\theta_i)}{\mu^i_i(\theta_i)}\Big)} \\
        &= \frac{\prod_{s=i}^t \pi^i_s \cdot \mu^i_i(\theta_i)}{\prod_{s=i}^{t-1} \pi^i_s \cdot \mu^i_i(\theta_i)} = \pi^i_t \leq \frac{\Pr(\tau > t)}{\Pr(\tau > t+1)}.
    \end{align*}
    Moreover, for every $t \notin \{i,\dots,t_i-1\}$ and $\theta \in \Theta$, we have $\frac{\mu^{i+1}_{t+1}(\theta)}{\mu^{i+1}_t(\theta)} = \frac{\mu^{i}_{t+1}(\theta)}{\mu^{i}_t(\theta)} \leq \frac{\Pr(\tau \geq t)}{\Pr(\tau \geq t+1)}$ from the induction hypothesis. Therefore, the boundary constraint holds for every time $t \in \mathcal{T}$ and state $\theta \in \Theta$ for the sequence of beliefs $(\mu^{i+1}_t)_{t \in \mathcal{T}}.$
\end{enumerate}
We have shown that the sequence $(\mu_t^i)_{t \in \mathcal{T}}$ satisfies the conditions in Lemma \ref{prop: IC+boundary-condition} for each $i \in \mathcal{T} \cup \{-1\}$. We now complete the proof of Theorem \ref{thrm:reduction} (ii). Define a sequence $(\mu^*_t)_{t\in\mathcal{T}} \in (\Delta(\Theta))^\mathcal{T}$ such that $\mu^*_t = \mu^t_t$ for every $t \in \mathcal{T}$ i.e., taking the `diagonal' by choosing time $t$'s beliefs to be sequence $t$'s beliefs at time $t$.
\begin{itemize} \setlength\itemsep{0em}
    \item For the obedience constraint, $\phi(\mu^*_t)=\phi(\mu^t_t) \geq \Ex[c\tau \mid \tau>t] - ct.$
    \item For the boundary constraint, $\frac{\mu^*_{t+1}(\theta)}{\mu^*_{t}(\theta)} = \frac{\mu^{t+1}_{t+1}(\theta)}{\mu^{t}_{t}(\theta)} = \frac{\mu^{t+1}_{t+1}(\theta)}{\mu^{t+1}_{t}(\theta)} \leq \frac{\Pr(\tau > t)}{\Pr(\tau > t+1)}.$
    \item For the increasing property, from Equation \ref{increasing} $\phi(\mu^*_t) = \phi(\mu^t_t) \leq \phi(\mu^{t+1}_{t+1}) = \phi(\mu_{t+1}^*)$.
\end{itemize}
 Thus, Lemma \ref{prop: IC+boundary-condition} implies $(d({\tau}), (\mu^*_t)_t)$ is feasible. From the construction, $\phi(\mu^*_t) = \phi(\mu^t_t) \geq \phi(\mu^C_t).$ If $\phi(\mu^C_t)$ is not increasing for some $t$, then this modification is not trivial: there exists $t_0$ such that $\phi(\mu^*_{t_0}) > \phi(\mu^C_{t_0})$, which contradicts that $(\mu^C_t)_t$ is an undominated belief path of $d({\tau}).$ 
 
 Now we prove $(\mu^C_t)_{t \in \mathcal{T}}$ is extremal. Suppose towards a contradiction that there is a minimum $t_0 \in \mathcal{T}$ such that $\mu_{t_0+1} \notin \Phi^*$ and $\frac{\Pr(\tau>t_0+1)}{\Pr(\tau >t_0)} < \min_{\theta \in \Theta}\frac{\mu_{t_0}(\theta)}{\mu_{t_0+1}(\theta)}$. Pick any $\mu^* \in \Phi^*$. We pick $\lambda \in [0,1)$ as following. If $\min_{\theta \in \Theta} \frac{\mu_{t_0}(\theta)}{\mu^*(\theta)} \geq \frac{\Pr(\tau > t_0+1)}{\Pr(\tau>t_0)}$, choose $\lambda = 0$. Otherwise, choose $\lambda \in [0,1)$ such that
\begin{align*}
    \frac{\Pr(\tau>t_0+1)}{\Pr(\tau > t_0)} = \min_{\theta \in \Theta} \frac{\mu_{t_0}(\theta)}{\lambda \mu_{t_0+1}(\theta) + (1-\lambda)\mu^*(\theta) }.
\end{align*}
The existence of $\lambda$ follows from the intermediate value theorem. We define a new belief path $(\mu'_t)_{t \in \mathcal{T}}$ as follows:
\begin{align*}
    \mu'_t = \begin{cases}
     \mu_t, &\text{if $t \leq t_0$} \\
     \lambda \mu_t + (1-\lambda)\mu^*,& \text{ otherwise.}
    \end{cases}
\end{align*}
We will show that a pair of a belief path and a distribution of stopping time $((\mu'_t)_{t \in \mathcal{T}},d(\tau))$ is feasible. The obedience constraint is still the same for $t \in \{0,\dots,t_0\}$. For $t\geq t_0$, we have
\begin{align*}
    \phi(\mu'_t) = \phi(\lambda \mu_t + (1-\lambda)\mu^*) \geq \lambda \phi(\mu_t) + (1-\lambda)\phi^* \geq \phi(\mu_t) \geq \Ex[c(\tau) \mid \tau>t]-c(t),
\end{align*}
so the obedience constraint for every $t \in \mathcal{T}$. The boundary constraint is still the same for $t \in \{0,\dots,t_0-1\}$. For $t = t_0$, the boundary constraint holds because of the construction of $\lambda$. For $t>t_0$, we have
\begin{align*}
    \min_{\theta \in \Theta} \frac{\mu'_t(\theta)}{\mu'_{t+1}(\theta)} = \min_{\theta \in \Theta} \frac{\lambda\mu_t(\theta)+(1-\lambda)\mu^*(\theta)}{\lambda\mu_{t+1}(\theta)+(1-\lambda)\mu^*(\theta)} \geq \min_{\theta \in \Theta} \frac{\mu_t(\theta)}{\mu_{t+1}(\theta) }\geq \frac{\Pr(\tau>t+1)}{\Pr(\tau>t)},
\end{align*}
where the first inequality follows from the fact that $\theta$ that minimizes the LHS must satisfy $\mu_t(\theta) \leq \mu_{t+1}(\theta)$. This concludes that a pair of a belief path and a distribution of stopping time $((\mu'_t)_{t \in \mathcal{T}},d(\tau))$ is feasible. 

For every $t \leq t_0$, $\phi(\mu'_t) = \phi(\mu_t)$. Moreover, for every $t> t_0$, we have 
\begin{align*}
    \phi(\mu'_t) = \phi(\lambda \mu_t + (1-\lambda)\mu^*) \geq \lambda \phi(\mu_t) + (1-\lambda)\phi^* \geq \phi(\mu_t).
\end{align*}
Therefore, $\phi(\mu'_t) \geq \phi(\mu_t)$ for every $t \in \mathcal{T}$. Because $(\mu_t)_{t \in \mathcal{T}}$ is an undominated path coresponding to $\tau$, we must have $\phi(\mu'_t) = \phi(\mu_t)$ for every $t \in \mathcal{T}$, which implies that $\phi(\mu_t) = \phi^*$ for every $t> t_0$. This contradicts the fact that $\mu_{t_0+1} \notin \Phi^*$. Therefore, the belief path $(\mu_t)_{t \in \mathcal{T}}$ must have a property of extremal paths, as desired. 
\end{proof}
\subsection{Proof that designer-optimal structures leave DM with no surplus}
\begin{proof}[Proof of Proposition \ref{prop:no_surplus}]
    Suppose towards a contradiction there exists a designer-optimal structure $I^*$ that leaves the $DM$ with a positive surplus. Fix sufficiently small $\epsilon > 0$ and construct an information structure $I^{**}$ as follows: with probability $1-\epsilon$ the designer follows $I^*$, and with probability $\epsilon$ the designer reveals nothing with an augmented message $m_\emptyset \in M$ in the first period and then follows $I^*$ in later periods:
    \begin{enumerate}
        \item[(i)] At time $t>1$ for histories $H_t$ that do not contain  $(\mu_0,m_\emptyset)$, we have 
        $I^{**}_{t+1}((\mu,m) \mid H_t) = I^{*}_{t+1}((\mu,m) \mid H_t).$
        \item[(ii)] At time $t>1$ with history $H_t = \big((\mu_0,m_\emptyset),H_{t-1} \big)$, we have $I^{**}_{t+1}((\mu,m) \mid H_t) = I^{*}_{t}((\mu,m) \mid H_{t-1}).$
    \end{enumerate}

    We now claim that the distribution of the resultant stopping time induced by $I^{**}$ is $d(I^{**}) = (1-\epsilon) d(I^*) \oplus \epsilon (d(I^*) + 1).$
    
    To see this, we verify that the DM's obedience constraints are preserved. For any $H_{t-1}$ which realizes with positive probability on the structure $I^*$, note that DM prefers to continue paying attention at history $\big((\mu_0,m_\emptyset),H_{t-1} \big)$ under $I^{**}$ if and only if she does so at history $H_{t-1}$ under $I^*$.
   To see this, let $H_{t-1} = (\mu'_s,m'_s)_{s \leq t-1}$ and observe  
    \begin{align*}
        &v^*(\mu'_{t-1},t) \leq \Ex^{I^{**}}[v^*(\mu_{{\tau}(I^{**})},{\tau}(I^{**})) \mid \big((\mu_0,m_\emptyset),H_{t-1} \big)] \\
        \iff & v^*(\mu'_{t-1},t) \leq \Ex^{I^{*}}[v^*(\mu_{{\tau}(I^{*})},\tau(I^*)+1) \mid H_{t-1} ] \\
        \iff & v^*(\mu'_{t-1},t-1)-c \leq \Ex^{I^{*}}[v^*(\mu_{{\tau}(I^{*})},\tau(I^*)) \mid H_{t-1} ] -c \tag{$v$ has additively separable cost} \\
        \iff & v^*(\mu'_{t-1},t-1) \leq \Ex^{I^{*}}[v^*(\mu_{{\tau}(I^{*})},\tau(I^*)) \mid H_{t-1} ].
    \end{align*}
    Moreover, the obedience constraint at time $0$ under $I^{**}$ still holds for sufficient small $\epsilon > 0$ because $I^*$ and $I^{**}$ coincide with probability $1-\epsilon$ and the obedience constraint at time $0$ under $I^*$ is slack by assumption. These together imply $\tau(I^{**})$ is indeed is still the DM's optimal stopping time under $I^{**}$.
    However, the designer's value of $I^{**}$ is 
    \begin{align*}
        \Ex^{I^{**}}[h(\mu_{\tau(I^{**})},\tau(I^{**}))] &= (1-\epsilon) \Ex^{I^{*}}[h(\mu_{\tau(I^*)},\tau(I^*))] + \epsilon \Ex^{I^{*}}[h(\mu_{\tau(I^*)},\tau(I^*)+1)] \\
        &> \Ex^{I^{*}}[h(\mu_{\tau(I^*)},\tau(I^*))] 
    \end{align*}
    because $h(\mu,\cdot)$ is strictly increasing, which contradicts the optimality of $I^*$.
\end{proof}
\section{Proof of optimal attention capture}\label{appendix:stopping_time}
In this appendix we collect proofs which were omitted in Section \ref{sec:stopping_time}. 
From Lemma \ref{prop: IC+boundary-condition}, a stopping time $\tau$ is feasible if we can find a supporting belief path $(\mu^C_t)_{t \in \mathcal{T}}$ such that $(\tau, (\mu^C_t)_{t \in \mathcal{T}})$ fulfils the obedience and boundary constraint.

We introduce an important lemma which we will draw on extensively.
\begin{lemma}[Pivot Lemma] \label{lem: convex_indiff} Suppose the designer's optimal full-revelation information structure $I$ with deterministic belief path induces the stopping time $\tau$. Let $t_0 < t_1 \in \mathcal{T} \cup \{+\infty\}$ such that $t_0+1 \in \text{supp } \tau$, and  
    \[\Ex[f(\tau+1)-f(\tau) \mid \tau \in [t_0+1,t_1)] > f(t_0+1)-f(t_0).\]
    Then, the DM must be indifferent between continuing and stopping at time $t_0$ under $I$, implying $t_0 \in \text{supp } \tau$. This further implies for any $t_0 \in \mathcal{T}$ and any feasible distribution $d \in \mathcal{D}_s$ with mean $s$, if the DM strictly pefers to continue at time $t_0$ and $t_0 + 1 \in \text{supp } d$, then there exists $d' \in \mathcal{D}_s$ such that $d' \succ_{CX} d$. 
\end{lemma}

\begin{figure}[!h]  
\centering
\captionsetup{width=1\linewidth}
    \caption{Intuition for pivot lemma} \includegraphics[width=0.5\textwidth]{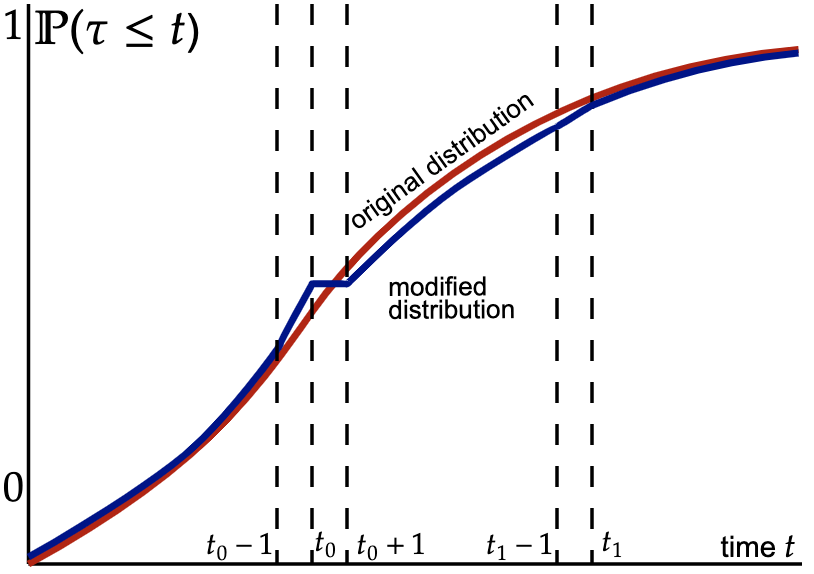}
     \label{fig:pivot}
\end{figure}

The proof is quite notation-intensive and deferred to Online Appendix \ref{onlineappendix: convex_frontier}. Nonetheless, the intuition is simple; we sketch it here. Suppose the DM strictly prefers to continue at time $t_0$ and the conjectured optimal distribution induces the stopping time with CDF given by the red line in Figure \ref{fig:pivot}.\footnote{We have depicted `smoothed' versions of the CDF.}

Now consider the following perturbation: at time $t_0$, we `uniformly' delay the information structure by a single unit of time with probability $\epsilon$, but increase the probability of stopping at time $t_0$ by the same probability. Then, conditional on facing this 1-unit delay beyond time $t_0$, we once again accelerate the arrival of full information at time $t_1$ so that the original distribution and modified distribution are identical after time $t_1$. This modified CDF is depicted as the blue line in Figure \ref{fig:pivot}. 

Note that this perturbation does not change the DM's continuation value for times $t<t_0$ since it preserves the mean. Moreover, if the delay occurs, for times $t>t_1$ the DM faces the same decision problem as she faces under the old information structure, so the obedience constraints on the delayed path still hold. If the delay occurs, for times $t \in (t_0, t_1]$, the DM faces approximately the same continuation information structure, with the difference that information at and after time $t_1$ is accelerated. Thus, this strictly improves the obedience constraints on the delayed path at times $t \in (t_0,t_1]$. Now consider the obedience constraint at time $t_0$. If it is slack, then this perturbation for sufficiently small $\epsilon>0$ induces a feasible stopping time yet strictly improves the designer's payoff, contradicting optimality.

\subsection{Proof of Theorem \ref{thrm:convex_ext}}
\begin{proof}[Proof of Theorem \ref{thrm:convex_ext} (i)]
    We begin with the existence of a convex frontier:
    \begin{lemma} \label{lem: convex_frontier_exist}
        Fix any $d \in \mathcal{D}_s$. There exists $d^* \in \mathcal{D}_s$ such that 
        \begin{enumerate}
            \item $d^* \succeq_{CX} d$, and
            \item there is no $d' \in \mathcal{D}_s$ such that $d' \succ_{CX} d^*.$
        \end{enumerate} 
    \end{lemma}
    \begin{proof}
        See Online Appendix \ref{onlineappendix: convex_frontier}.
    \end{proof}
    Pick $d^*$ that satisfies the above conditions. We will show that $d^* \in \mathcal{D}^{IIM}.$ Pick any belief path $\bm{\mu^C} \in \Delta(\Theta)^\mathcal{T}$ such that $(d^*,\bm{\mu^C})$ is feasible. By Lemma \ref{prop: existence_maximal}, there exists an undominated belief path $\bm{\mu^{*C}} \in \mathcal{W}(\tau^*)$ where $\tau^* \sim d^*$ such that $\phi(\mu^{*C}_t) \geq \phi(\mu^C_t)$ for every $t \in \mathcal{T}.$

    Let $T = \max \text{supp } \tau^*$, which can be $+\infty.$ Suppose that $T$ is finite. Lemma \ref{lem: convex_indiff} and the fact that there is no $d' \in \mathcal{D}_s$ such that $d' \succ_{CX} d^*$ imply the DM must be indifferent at time $T-1$ under $(d^*,\bm{\mu^{*C}})$. However, this is impossible because
    \begin{align*}
        \Ex[c(\tau^* - (T-1)) \mid \tau > T-1] = c < \phi(\mu_0) \leq \phi(\mu_t),
    \end{align*}
where the last inequality is from the result that a maximal belief path must be increasing. Thus, $\max \text{supp } \tau^* = +\infty.$ We show that the DM must be indifferent at every time $t \in \mathcal{T}$ under $(d^*,\bm{\mu^{*C}})$. For each time $t\ \in \mathcal{T}$, there exists time $t'>t$ such that $t' \in \text{supp } \tau^*$. Applying Lemma \ref{lem: convex_indiff} again implies the DM must be indifferent at time $t'-1$ and $t'-1 \in \text{supp } \tau^*.$ Applying this argument (backward) inductively, we obtain the DM must be indifferent at time $t<t'$, as desired.

Since the DM must be indifferent at every time $t \in \mathcal{T}$ under $(d^*,\bm{\mu^{*C}})$, we have 
\[\Ex[c(\tau^* -t) \mid \tau^*>t] = \phi(\mu^{*C}_t) \geq \phi(\mu^{C}_t).\]
Since $(d^*,\bm{\mu^{C}})$ is feasible, the above inequality must be the equality. Thus, $\phi(\mu^{*C}_t) = \phi(\mu^C_t)$ for every $t \in \mathcal{T},$ and the DM is indifferent at every time $t \in \mathcal{T}$ under $(d^*,\bm{\mu^{C}})$. This means $\bm{\mu^{C}}$ is a maximal belief path corresponding to $\tau^*$, implying both increasing and extremal paths. These together conclude $d^* \in \mathcal{D}^{IIM},$ and $d^* \succeq_{CX} d,$ as desired.  

Finally, we show that $d \succeq_{CX} d^{DET}$. Let $d,d^{DET} \in \mathcal{D}_s$. We need to show, for every concave function $f \in \mathcal{T} \to \mathbb{R},$ $\Ex_{\tau \sim d}[f(\tau)] \leq \Ex_{\tau \sim d^{DET}}[f(\tau)].$ For each $f,$ define an augmented function $f_\mathbb{R}$ as a linear interpolation of $f$ on $\mathbb{R}.$ Note that $f_\mathbb{R}$ is still concave if $f$ is concave. By Jensen's inequality, we must have
\begin{align*}
    \Ex_{\tau \sim d}[f(\tau)] = \Ex_{\tau \sim d}[f_\mathbb{R}(\tau)] \leq f_\mathbb{R}(s) = \Ex_{\tau^{DET} \sim d}[f(\tau)],
\end{align*}
where the last inequality follows from that $f_\mathbb{R}$ is linear on the domain $[\lfloor s\rfloor, \lfloor s\rfloor+1].$ Thus, $d \succeq_{CX} d^{DET}$, as desired.
\end{proof}

\subsection{Proof of Theorem \ref{thrm:convex_ext} (ii)}
We drop the boundary constraint and relax the obedience constraint as follows:
\begin{definition}
    A stopping time $\tau$ is \textit{pre-feasible} if $\tau$ satisfies the relaxed obedience constraint: $\Ex[c\tau \mid \tau > t] -ct \leq \phi^*$ for every $t \in \mathcal{T}$. For any $s \geq 0$, let
    \begin{align*}
    \mathcal{D}_{relax}(s) = \{d \in \Delta(\mathcal{T}): \Ex_{\tau \sim d}[\tau] = s, \text{$\tau \sim d$ is pre-feasible} \}
    \end{align*}
    be set of distributions of pre-feasible stopping times whose expected value is $s$. 
\end{definition}
It is clear that every feasible distribution of stopping times is pre-feasible: $\mathcal{D}(\mathcal{I}(s)) \subset \mathcal{D}_{relax}(s).$ We sketch the outline of omitted proofs in Section \ref{sec:stopping_time} is as follows. To prove Theorem \ref{thrm:convex_ext} (ii), we show that 1) extreme points of $\mathcal{D}_{relax}(s)$ have the property of block structures, and 2) extreme points of $\mathcal{D}_{relax}(s)$ are in $\mathcal{D}(\mathcal{I}(s))$ using a supporting single-jump belief path.

The following Lemmas \ref{lemma:IC-binding} and \ref{lem: relax=original} imply Theorem \ref{thrm:convex_ext} (ii).
\begin{lemma}\label{lemma:IC-binding}
For every $s>0$, $\text{Ext}(\mathcal{D}_{relax}(s)) = \mathcal{D}(\mathcal{I}^{\text{BLOCK}}(s)) \cap \mathcal{D}_{relax}(s)$. Equivalently,
for every $d(\tau) \in \text{Ext} (\mathcal{D}_{relax} (s))$, there does not exist times $r_1$ and $r_1$ with $r_1<r_2$ such that all of the followings hold:
    \begin{enumerate}[nosep]
        \item[(i)] The stopping time is $r_1$ and $r_2$ with positive probability, but not between these times: $\Pr(\tau =r_1), \Pr(\tau =r_2) > 0$, and $\Pr(\tau \in (r_1,r_2)) = 0$
        \item[(ii)] $\tau$ is after $r_2$ with positive probability: $\Pr(\tau > r_2)>0$.
        \item[(iii)] The DM is not indifferent at $r_1$ under $\tau$: $\Ex[c\tau\mid \tau>r_1]-cr_1 < \phi^*.$
    \end{enumerate}
   Moreover, $\text{Ext}(\mathcal{D}_{relax}(s)) = \text{Ext} \big( \text{Ext}(\mathcal{D}_{relax}(s))\big).$
\end{lemma}
\begin{proof}[Proof of Lemma \ref{lemma:IC-binding}]
      Suppose towards a contradiction that there exist $r_1<r_2 \in \mathcal{T}$ satisfying (i), (ii), and (iii). Define $\alpha \in (0,1)$ as the solution of $\alpha r_1 + (1-\alpha) \Ex[\tau \mid \tau > r_2] = r_2$. This is well-defined because $\Pr(\tau >r_2) = 0$. For $\epsilon$ in a small neighborhood of $0$, we define a new stopping time $\tau^{\epsilon}$ as follows:
    \begin{align*}
        \Pr(\tau^\epsilon = t)= \begin{cases}
            \Pr(\tau = t) &t<r_1 \\
            \Pr(\tau = r_1)-\alpha\epsilon &t=r_1 \\
            0 &r_1<t<r_2 \\
            \Pr(\tau = r_2)+\epsilon &t=r_2 \\
            \Big(1-\frac{(1-\alpha)\epsilon}{\Pr(\tau > r_2)} \Big)\Pr(\tau = t) & t>r_2
        \end{cases}
    \end{align*}

    It is easy to check that $\tau^{\epsilon}$ corresponds to a valid probability distribution. An important observation is that $\tau^\epsilon | \tau^\epsilon > r_2 \overset{d}{=} \tau | \tau > r_2$. We now verify that the relaxed obedience constraints remain fulfilled.

   \noindent  \underline{Case 1:} the relaxed obedience constraint at $t<r_1$. Note that $\Pr(\tau^\epsilon > t) = \Pr(\tau > t)$. Consider that
    \begin{align*}
        \Ex[\tau^\epsilon 1\{\tau^\epsilon > t\}] - \Ex[\tau 1\{\tau > t\}]  = r_2-(\alpha r_1 + (1-\alpha) \Ex[\tau \mid \tau >r_2]) = 0
    \end{align*}
    by the definition of $\alpha$. Thus, $\Ex[\tau^\epsilon | \tau^\epsilon > t] = \Ex[\tau | \tau > t]$. Since $\tau$ satisfies the relaxed obedience constraint at $t$, so does $\tau^\epsilon$.
    
   \noindent  \underline{Case 2:} the relaxed obedience constraint at $t \in [r_1,r_2)$. Since $\Ex[c\tau\mid \tau>r_1]-cr_1 > \phi^*$ (condition (iii)), $\tau^\epsilon$ must satisfy the relaxed obedience constraint at $r_1$ for small $\epsilon$. The relaxed obedience constraint at $t \in (r_1,r_2)$ is then implied by that at $r_1$.

   \noindent  \underline{Case 3:} the relaxed obedience constraint at $t \geq r_2$. This clearly holds because $\tau^\epsilon | \tau^\epsilon > r_2 \overset{d}{=} \tau | \tau > r_2$ and  $\tau$ satisfies the relaxed obedience constraint at $t$.

     Therefore, $\tau^\epsilon$ satisfies the relaxed obedience constraints, implying $d(\tau^\epsilon) \in \mathcal{D}_{relax}(s)$. This is a contradiction because $d(\tau) = d(\tau^\epsilon)/2 + d(\tau^{-\epsilon}) / 2$ but $d(\tau) \in \text{Ext} (\mathcal{D}_{relax}(s))$. It is easy to see that $\text{Ext}(\mathcal{D}_{relax}(s)) = \text{Ext} \big( \text{Ext}(\mathcal{D}_{relax}(s))\big)$ because a convex combination of two different block structures cannot be another block structure (with the same expected time).
\end{proof}
\begin{lemma} \label{lem: relax=original}
Suppose $\mu_0 \in \Phi^*$ or we are in the binary environment. For every $s>0$, $\text{Ext}(\mathcal{D}_{relax}(s)) \subset \mathcal{D}(\mathcal{I}(s)) .$
\end{lemma}
\begin{proof}[Proof of Lemma \ref{lem: relax=original}]
    Suppose $d^{*} \in \text{Ext}(\mathcal{D}_{relax}(s))$. To show that $d^{*} \in \mathcal{D}(\mathcal{I}(s))$, we will find a supporting belief path $(\mu_t)_{t \in \mathcal{T}}$ such that $(d^{*},(\mu_t)_{t \in \mathcal{T}})$ is feasible.   Let $t_0 := \min \{t \in \mathcal{T}: d^{*}(t) > 0 \}$ be the first time that the DM has any chance to obtain full information. Define a jumping belief $\mu_{t_0} \coloneqq (1-\beta)\mu_0 \oplus \beta \mu^*$, where $\beta$ satisfies $(1-\beta) \phi(\mu_0) + \beta \phi^* = \Ex[c\tau \mid \tau>t_0]-ct_0$, and $\mu^* \in \Phi^*$.   We construct a single-jump belief path $(\mu_t)_t$ with a jump time $\bar{t} = t_0$ and a jump destination $\bar{\mu} = \mu_{t_0}$.
    
    To complete the proof, we show that $(d^{*}(\cdot),(\mu_t)_t)$ satisfies the obedience and boundary constraints.
To see this, when $t < t_0$, the obedience constraint at time $0$ is sufficient since the DM receives no information between times $0$ and $t_0$. The obedience constraint at time $t=t_0$ holds by the definition of $\beta$ and the concavity of $\phi$. Consider any time $t > t_0.$ The original and relaxed obedience constraints at time $t$ coincide if $\mu_{t_0} = \mu^*$, which is true if $d^{*}$ has at least one indifferent block. If $d^{*}$ has only a terminal block with a terminal time $\bar{t}$, the DM receives no information between times $t_0$ and $\bar{t}$, so the the obedience constraint at time $t$ is implied by that at time $t_0.$  

For boundary constraints, it is sufficient to check at time $t_0$ because it is the only time the belief path moves. Suppose $d$ yields a stopping time $\tau$. It is easy to verify that $\frac{\phi(\mu_t)}{\phi(\mu^*)} = \min_{\theta \in \Theta} \frac{\mu_0(\theta)}{\mu_t(\theta)}$ when $|A|=2$ and $|\Theta|=2$. The obedience constraint at time $0$ implies
\begin{align*}
\phi(\mu_0) &\geq \Pr(\tau > t_0) \Ex[c\tau \mid \tau>t_0] 
\geq \Pr(\tau > t_0) \big( (1-\beta) \phi(\mu_0) + \beta \phi^* \big).
\end{align*}
Moreover,
\begin{align*}
    \min_{\theta \in \Theta} \frac{\mu_0(\theta)}{\mu_t(\theta)} = \min_{\theta \in \Theta} \frac{\mu_0(\theta)}{(1-\beta)\mu_0(\theta) + \beta \mu^*(\theta)} = \frac{ \phi(\mu_0)}{(1-\beta) \phi(\mu_0) + \beta \phi^*}.
\end{align*}
 Thus,
\begin{align*}
        \frac{\Pr(\tau>t_0)}{\Pr(\tau>t_0-1)} = \Pr(\tau>t_0) \leq \frac{ \phi(\mu_0)}{(1-\beta) \phi(\mu_0) + \beta \phi^*}  =\min_{\theta \in \Theta} \frac{\mu_0(\theta)}{\mu_t(\theta)},
    \end{align*}
as desired. 
\end{proof}

\subsection{Proof of Proposition \ref{prop:step_and_S-shaped}}
We will prove Proposition \ref{prop:step_and_S-shaped} (i) and deter the proof of a generalization of Proposition \ref{prop:step_and_S-shaped} (ii) to Online Appendix \ref{onlineappendix: S-curve}.
\begin{proof}[Proof of Proposition \ref{prop:step_and_S-shaped} (i)]
    
First we show that, for any designer-optimal stopping time $\tau,$ $\text{supp } \tau \cap [T,\infty) \ne \emptyset$. Suppose a contradiction that $\text{supp } \tau \subset (0,T)$. Then $\Ex[f(\tau)] = 0$ which is suboptimal for the designer because, if $\tau'$ is a Poisson stopping time with a high arrival rate so that the obedience constraints hold, then $\Ex[f(\tau')] > 0.$

Next, we show $\text{supp } \tau \cap [T-\phi^*/c,T] \ne \emptyset$. Suppose that $t_1 = \min \text{supp } \tau \cap [T,\infty)$. Let $t_0 = \max \text{supp } \tau \cap [0,t_1).$ This implies $(t_0,t_1) \cap \text{supp } \tau = \emptyset$. The obedience condition at time $t_0$ implies
\begin{align*}
       \phi^* \geq \phi(\mu_{t_0}) \geq \Ex[c(\tau-t_0) \mid \tau>t_0] \geq c(t_1-t_0),
\end{align*}
which implies $t_0 \geq t_1 - \phi^*/c \geq T-\phi^*/c$. Since $t_1 = \min \text{supp } \tau \cap [T,\infty) > t_0$, $t_0<T$. Thus, $t_0 \in \text{supp } \tau \cap [T-\phi^*/c,T],$ as desired.

Next, we show that the DM is indifferent at every time $t<T-\phi^*/c.$ Pick $t_0 \in \text{supp } \tau \cap [T-\phi^*/c,T]$. Because $f(t_0)-f(t_0-1) = 0$, we can directly use Pivot Lemma to show that DM must be indifferent at $t_0-1$ and $t_0-1 \in \text{supp } \tau$. Applying Pivot Lemma (backward) inductively, we obtain that the DM must be indifferent at every time $t<t_0$, as desired.

Finally, we show that that The DM's belief reaches the basin before time $t< T-\phi^*/c$. We introduce the following lemma. 

\begin{lemma} \label{prop: converge_>2states}
Let $|\Theta| =n $. There exists a constant $C>0$ such that the following statement is true: fixing $\tau$, suppose that $T>0$ satisfies
\begin{align*}
    \sum_{t=0}^T (\log \Pr(\tau >t) - \log \Pr(\tau > t+1))^{n-1} > C.
\end{align*}
Moreover, assume that $\{\mu \in \Delta(\Theta) : \phi(\mu) \geq \phi(\mu_0) \} \subset \text{int } \Delta(\Theta)$. Then, $\mu_t \in \Phi^*$ for every $t>T$.
\end{lemma}
\begin{proof}
    See Online Appendix \ref{onlineappendix: convergence_to_basin}
\end{proof}
Because the DM is indifferent at every time $t< T -\phi^*/c$, we obtain the following
\begin{align*}
    \Pr(\tau> t+1 \mid \tau>t) \phi(\mu_{t+1}) + c  = \phi(\mu_t),
\end{align*}
implying
\begin{align*}
    \Pr(\tau> t+1 \mid \tau>t) = \frac{\phi(\mu_t)-c}{\phi(\mu_{t+1})} \leq \frac{\phi(\mu_{t+1})-c}{\phi(\mu_{t+1})} \leq  \frac{\phi^*-c}{\phi^*} \eqqcolon p^* < 1.
\end{align*}
Set $T_0 = \frac{2C}{(-\log p^*)^{n-1}}$ and consider sufficiently large enough $T$ such that $T> T_0+\phi^*/c$. Thus,
\[\sum_{t=0}^{T_0} (\log \Pr(\tau >t) - \log \Pr(\tau > t+1))^{n-1} \geq \frac{2C}{(-\log p^*)^{n-1}} \cdot (-\log p^*)^{n-1} = 2C > C.\]
Lemma \ref{prop: converge_>2states} implies that $\mu_t \in \Phi^*$ for every $t>T_0$, as desired.
\end{proof}

\section{
    Sequential optimality \& attention capture with suspense} 
    \label{appendix: time-consistent-modification}

\subsection{Proof of Theorem \ref{thrm:time_con}} We first define an information structure where DM is indifferent between continuing and taking action at every non-stopping history:
\begin{definition}\label{def:IN}
    A non-stopping history $H_t$ is indifferent or no-learning (IN) if the DM is indifferent between continuing at $H_t$ or stopping and taking her best option under belief $\mu|H_t$ i.e., 
        \[
            \sup_{\tau,a_{\tau}} \Ex^I[v(a_{\tau},\theta,\tau)|H_t] = \max_{a \in A} \Ex_{\mu|H_t}[v(a,\theta,\tau = t)];
        \] 
    $I$ is IN if every non-stopping history $H_t$ which realizes with positive probability is IN. 
\end{definition}

\noindent \textbf{Outline of proof.} The following sequence of results imply  Theorem \ref{thrm:time_con}:
    \begin{enumerate}[nosep]
        \item For all designer problems $f$ such that Assumption \ref{assumption: metric_exist} on the compactness of the set of feasible joint distribution is fulfilled, there exists a designer-optimal structure which is regular (to be made precise later) and has deterministic continuation beliefs (Lemma \ref{lemma: regular_sufficient}). 
        \item Regular information structures which have deterministic continuation beliefs can be modified to be IN while preserving the joint distribution of stopping actions and times (Lemmas \ref{lemma: indiff_modification} and \ref{lemma: modify_time_consistent})
        \item If an information structure $I$ is IN and designer-optimal,  $I$ is sequentially optimal. The converse is also true if the DM's payoff function can be written as $v(a,\theta,t) = u(a,\theta) - ct.$  (Lemma \ref{lemma: time_consistent_vs_sequential_optimal}).
        
    \end{enumerate}
 \begin{proof}[Proof of Theorem \ref{thrm:time_con}]We proceed via the steps outlined above. 

\noindent \underline{{\textbf{Step 1: Regular optimal structures with deterministic continuation beliefs exist.}}}

    \begin{definition} \label{definition: regular_info}
    An information structure $I$ is \textit{regular} if either 
    \begin{itemize}[nosep] 
        \item[(i)] $I$ has a terminal time $T$ i.e. $\Pr^I(\tau(I)>T) = 0$; or
        \item[(ii)] the set $\mathcal{T}_{IN} (i) \coloneqq \{t \in \mathcal{T} \mid \text{ $I$ is IN at every non-stopping history at time $t$}\}$
        is infinite.
    \end{itemize}
    \end{definition}

\begin{lemma} \label{lemma: regular_sufficient}
    Under Assumption \ref{assumption: metric_exist} with the designer's function $f$, there exists a regular information structure with deterministic continuation beliefs $I^*$ such that $I^*$ is a solution to $\max_{I \in \mathcal{I}} \Ex[f(a_\tau,\theta,\tau)]$
\end{lemma}

The proof of Lemma \ref{lemma: regular_sufficient} is involved and deferred to Online Appendix \ref{onlineappendix: time-consistent}. 

\noindent \underline{{\textbf{Step 2: Regular structures with deterministic continuation beliefs can be made IN.}}}

The next lemma provides a method to recursively modify continuation histories while preserving the joint distribution over outcomes. The proof proceeds by performing `surgery on the tree' and is in Online Appendix \ref{onlineappendix: time-consistent}. 
\begin{lemma} \label{lemma: indiff_modification}
    Fix a dynamic information structure $I$ and time $t \geq 1$.  Suppose $I$ is IN at every non-stopping history at time $t+1$. There exists a dynamic information structure $I'$ such that
    \begin{enumerate}[nosep]
        \item[(i)] $I$ and $I'$ induces the same joint distribution over actions, states, and stopping times;
        \item[(ii)] $I'$ is IN at every non-stopping history at time $t$; and
        \item[(iii)] for every $s \geq t+1$, if $I$ is IN at every non-stopping history at time $s$, so is $I'$.
    \end{enumerate}
\end{lemma}
Note that Lemma \ref{lemma: indiff_modification} implies the more intuitive inductive step that if $I$ is IN for all non-stopping histories after $t+1$, then there exists $I'$ such that $I'$ is IN for all non-stopping histories after time $t$. However, this form will be useful to prove the result for infinite structures. We now use Lemma \ref{lemma: indiff_modification} to modify regular structures with deterministic continuation beliefs such that they are IN. 
\begin{lemma} \label{lemma: modify_time_consistent}
    Suppose $\bar{I}$ is regular with deterministic continuation beliefs.
     Then there exists an information structure $I^*$ IN such that $\bar{I}$ and $I^*$ induce the same joint distribution over actions, states and stopping times. 
\end{lemma}
\begin{proof}[Proof of Lemma \ref{lemma: modify_time_consistent}]
    We start from a finite uniquely obedient information structure $\bar{I}$ with terminal time $T$. We use Lemma \ref{lemma: indiff_modification} to modify an information structure at the terminal time. As a result, we have the information structure that is IN at every history at time $T-1$. We apply the previous lemma again and obtain the information structure that is IN at every history at time $T-1$. Keep doing this until time $1$ and finally obtain an IN information structure.

    If $\mathcal{T}_{IN} \coloneqq \{t_1,t_2,\dots\}$ is infinite, assume $t_1<t_2<\dots$. We define $(I^n)_n$ iteratively: let $I_0 = \bar{I}$. For each $n \geq 1$,   we use Lemma \ref{lemma: indiff_modification} to modify an information structure $I^{n-1}$ at time $t_n$ and then $t_{n}-1\dots,t_{n-1}+1$ to finally get $I_n$. It is easy to show that $I^n$ is IN at every history at every time $t$ such that $1 \leq t \leq t_n$ and $d(I) = d(I^n)$ for every $n$. Moreover, if $m<n$, $I^m$ and $I^n$ coincide for every $t \leq t_m$. Thus, we can define $I^*$ such that $I^*$ and $I^m$ coincide for every $t<t_m$ and $m>1$. This implies $I^*$ is an IN information structure and $d(I^*) = d(\bar{I})$ as desired.
\end{proof}


\noindent \underline{{\textbf{Step 3: IN structures are sequentially optimal.}}} \\ 
In Steps 1 and 2 we showed optimal information structures can be made IN. We now show that optimal IN structures are sequentially optimal, which implies Theorem \ref{thrm:time_con}.

\begin{lemma} \label{lemma: time_consistent_vs_sequential_optimal}
    If a dynamic information structure $I$ is IN and optimal for the designer, then it is also sequentially optimal. The converse is also true if the DM's payoff function can be written as $v(a,\theta,t) = u(a,\theta) - ct.$
\end{lemma}
\begin{proof}[{Proof of Lemma \ref{lemma: time_consistent_vs_sequential_optimal}}]
Suppose towards a contradiction that $I$ is IN and optimal but not sequentially optimal for the designer. This means there exists a positive probability history $H_t$ such that, given that the DM pays attention until $H'_t$, the designer could do strictly better by offering an alternative information structure $I'_t$ instead of $I_t$ (which is $I$ after history $H_t$). Then modify $I$ by simply replacing the information structure after $H'_t$ with $I'_t$. This results in a new information structure $I'$.

We will show that the DM's value conditional on paying attention until $H_t$ under $I'$ (denoted by $U^{I'}(H_t)$) is higher than that under $I$ (denoted by $U^{I}(H_t)$). Since $I$ is IN, $U^I(H_t) = v^*_{t}(\mu_t \mid H_t)$. The DM's value conditional on paying attention until $H_t$ under $I'$ can be written as a solution to the following optimal stopping problem:
\begin{align*}
    U^{I'}(H_t) = \max_{\tau \geq t} \Ex^{I'}[v^*(\mu_\tau, \tau)  \mid H_t] &\geq  v^*(\mu_t \mid H_t,t) = U^I(H_t), 
\end{align*}
where the inequality is obtained by choosing $\tau = t^*$. 

This directly implies that the DM continues paying attention every time before history $H'_t$ because the continuation value at $H_t$ under $I'$ is greater than that under $I$. Moreover, at history $H_s$ where $s <t$ and $H_s \nsubseteq H_t$, payoffs remain unchanged. We have verified that the agent's continuation incentives at each history remain unchanged. But since the designer receives strictly more payoff at history $H_t$ under $I'$ and $H_t$ is reached with positive probability, $I'$ delivers strictly more expected payoff for the designer than $I$ does, which contradicts the optimality of $I$.

For the converse, if $I$ is not IN at some history $H_t$, then it is not sequentially optimal at $H_t$ by zero DM's surplus result from Proposition \ref{prop:no_surplus}, as desired.
\end{proof}
Steps 1, 2, and 3 imply Theorem \ref{thrm:time_con}.
\end{proof}

\subsection{Proof of Theorem \ref{thrm:suspenseoptimal}}
\begin{proof} [Proof of Theorem \ref{thrm:suspenseoptimal}]
    It suffiecs to consider information structures with degenerate stopping beliefs because this slackens every obedience constraint at which the DM prefers to continue paying attention.  Define a stochastic process $(\psi_t)_t$ where
    \[\psi_t \coloneqq \Ex[\phi_s(\mu_{t}) - \phi_s(\mu_{t+1}) \mid H_t] \geq 0.\] 
    Fixing $t$, observe $\psi_t = 0$ under $\{\tau < t\}$ since DM receives no further information after stopping. We can rewrite the obedience condition under suspense as follows:
    \begin{align*}
        \Ex\Big[\sum_{s=t}^\infty g(\psi_s) \mid H_t\Big ]\geq \Ex\Big[ c(\tau-t) \mid H_t\Big] \Longrightarrow \Ex\Big[\sum_{s=t}^\infty g(\psi_s) \mid \tau > t\Big ] \geq \Ex\Big [c(\tau-t) \mid \tau > t\Big].
    \end{align*}
    Thus, 
    \begin{align*}
        \Ex[c(\tau-t) \mid \tau>t]  &\leq \sum_{s=t}^\infty \Ex[g(\psi_s) \mid \tau>t] \\
        &= \sum_{s=t}^\infty \Ex[g(\psi_s) \mid \tau>s] \Pr(\tau>s \mid \tau>t) \\
        &\leq \Big(\sum_{s=t}^\infty \Pr(\tau>s \mid \tau>t)\Big)g\bigg(\frac{\Ex[\sum_{s=t}^\infty \psi_s \mid \tau>t]}{\sum_{s=t}^\infty \Pr(\tau>s \mid \tau>t)} \bigg). \tag{Jensen}
    \end{align*}
    Note $\sum_{s=t}^\infty \Pr(\tau>s \mid \tau>t) = \Ex[\tau-t \mid \tau>t]$ so obedience is implied by 
    \begin{align*}
        c \leq g\bigg( \frac{\Ex[\sum_{s=t}^\infty \psi_s \mid \tau>t]}{\Ex[\tau-t \mid \tau>t]}   \bigg) \Longrightarrow \frac{\Ex[\sum_{s=t}^\infty \psi_s \mid \tau>t]}{\Ex[\tau-t \mid \tau>t]} \geq g^{-1}(c)
    \end{align*}
    since $g$ is strictly increasing.
    Further observe from the law of iterated expectations, 
    \begin{align*}
        \Ex\Big[\sum_{s=t}^\infty \psi_s \mid H_t\Big]  &= \phi_S(\mu_t \mid H_t) 
        \Longrightarrow \Ex\Big[\sum_{s=t}^\infty \psi_s \mid \tau > t\Big] = \Ex\Big[\phi_S(\mu_t ) \mid \tau>t\Big] \leq \phi_S(\mu_t \mid \tau>t),
    \end{align*}
    which implies
    \begin{align*}
    g^{-1}(c) \cdot \Ex\Big[\tau-t \mid \tau>t\Big] \leq \Ex\Big[\sum_{s=t}^\infty \psi_s \mid \tau>t \Big] \leq \phi_S(\mu_t \mid \tau>t),     
    \end{align*}
    for every $t \in \mathcal{T}$. Call this condition  the \textit{relaxed obedience constraint.} Define $\mu^C_t = \mu_0 \mid \tau>t$. By the martingale condition
    \begin{align*}
        \Ex[\mu^C_t(\theta)] &= \Pr(\tau > t+1 \mid \tau >t) \mu^C_{t+1}(\theta) + \Pr(\tau=t+1 \mid \tau > t) (\mu(\theta) \mid \tau=t+1)\\
        &\geq  \Pr(\tau > t+1 \mid \tau >t) \mu^C_{t+1}(\theta).
    \end{align*}
    We call this condition a relaxed boundary constraint. Thus, the designer's optimal value is bounded above by the following relaxed program:
    \begin{align*}
        \bar{f}:=&\max_{\substack{\big(d_{\mathcal{T}}(\tau),(\mu_t^C)_{t}\big) 
            \\ 
            \in \Delta(\mathcal{T}) \times (\Delta(\Theta))^\mathcal{T}}} \Ex^I[f(\tau)] 
            \\
            & \text{s.t.} \quad \phi_S(\mu_t^C) \geq \Ex[g^{-1}(c) \tau \mid \tau>t]-g^{-1}(c)t \quad \forall t \in \mathcal{T} \tag{Relaxed obedience} \\
            & \quad \quad \Pr(\tau>t+1)\mu_{t+1}^C \leq \Pr(\tau > t)\mu_t^C \tag{Relaxed boundary}
        \end{align*}
    If $I^{**}$ as a sequentially optimal information structure of the above program. Consider any non-stopping history $H_t$ under $I^{**}$. With a sequential optimal modification, the relaxed obedience binds for every history $H_t$ under $I^{**}$, implying
    \begin{align*}
        \phi_S(\mu_t \mid H_t) = \Ex^I[\phi_S(\mu_{t+1}) \mid H_t] + g^{-1}(c) \Longrightarrow \psi_t \mid H_t = g^{-1}(c),
    \end{align*}
    which does not depend on $\mu_t$, for every non-stopping history $H_t$. Thus,
    \begin{align*}
        \Ex\Big[\sum_{s=t}^\infty g(\psi_s) \mid H_t\Big] = \Ex\Big[\sum_{s=t}^\infty c 1\{\tau> s\} \mid H_t \Big] = \Ex\Big[c(\tau-t) \mid \tau>t\Big],
    \end{align*}
    which means continuation incentives under $I^{**}$ are preserved. Thus, $I^{**}$ is a sender-optimal information structure under suspense, and the sender's optimal value under suspense must coincide with $\bar{f}.$  

    If $I^{**}$ is a sender-optimal information structure under suspense, then $I^{**}$ must be an optimal structure of the above program. Proposition \ref{prop:no_surplus} implies the relaxed obedience must bind at $t=0$. This means all earlier inequalities must bind when $t=0$, implying $  \Ex[\phi_S(\mu_t) -\phi_S(\mu_{t+1}) \mid H_t] = g^{-1}(c)$ for every non-stopping history $H_t.$ Since $\phi_S(\mu^S) = 0 $ for every stopping belief $\mu^S,$ we must have
    \[\phi_S(\mu_t \mid H_t) = \Ex\Big[\sum_{s=t}^{\tau-1}(\phi_S(\mu_s) - \phi_S(\mu_{s+1})) \mid H_t\Big] = g^{-1}(c) \Big(\Ex[ \tau \mid H_t]-t \Big),\]
    which implies the DM is indifferent at every non-stopping history $H_t.$ Thus, $I^{**}$ is IN and optimal for the designer under the case of instrumental utility. By Lemma \ref{lemma: time_consistent_vs_sequential_optimal}, $I^{**}$ is also sequentially optimal.

\end{proof}
\section{Proof of Optimal Attention Capture and Persuasion}\label{appendix:bangbang}
\subsection{Preliminaries for proof of Theorem \ref{thrm:attention_persuasion_separable} %
}
We first develop several useful results. Let $\bar \mu$ be the belief at which the DM is indifferent between actions $0$ and $1$. Recall $\mathcal{P}$ is the set of joint distributions over beliefs and stopping times. 
Define $\mathcal{P}^*$ as the set of designer-optimal feasible distributions, noting that it is without loss to consider $\mathcal{P}^* \subseteq \Delta(\{0,\bar \mu, 1\}\times \mathcal{T}) \cap \mathcal{P}$ since we can extremize stopping beliefs while preserving joint distributions over outcomes; Online Appendix \ref{onlineappendix:topology} formalizes this.
Note that each feasible distribution $p \in \mathcal{P}$ pins down the unique continuation belief path: if the DM has paid attention up to time $t$, her beliefs at $t$ are, by Bayes' rule, 
\[
\mu_t^C({p}) := \mathbb{P}_t^{{p}} (\mu_\tau \mid \tau > t) = \frac{\Ex_{(\mu^S,\tau) \sim p}[\mu^S 1\{\tau>t\}]}{\Pr_{(\mu^S,\tau) \sim p}( \tau>t)},
\tag{C-Belief}
\label{eqn:C_BELIEF}
\]
where we sometimes drop the dependence on $p$ when there is no ambiguity about the dynamic information structure. 

\begin{definition}[Switching distribution] Let $p \in \Delta(\{0,\bar{\mu},1\} \times \mathcal{T})$. For every $(\mu_1,t_1),(\mu_2,t_2) \in \{0,\bar{\mu},1\} \times \mathcal{T}$ such that $p_{\mu_1,t_1}$, $p_{\mu_2,t_2}>0$, define a new probability distribution $p^{\epsilon,(\mu_1,t_1;\mu_2,t_2)}$ for $\epsilon \in (0, \min \{p_{\mu_1,t_1}, p_{\mu_2,t_2}\}$) as follows: 
\begin{align*}
    p^{\epsilon,(\mu_1,t_1;\mu_2,t_2)}_{\mu_1,t_1} &= p_{\mu_1,t_1} - \epsilon \quad \quad 
    p^{\epsilon,(\mu_1,t_1;\mu_2,t_2)}_{\mu_2,t_2} = p_{\mu_2,t_2} - \epsilon \\
    p^{\epsilon,(\mu_1,t_1;\mu_2,t_2)}_{\mu_1,t_2} &= p_{\mu_1,t_2} + \epsilon \quad \quad 
    p^{\epsilon,(\mu_1,t_1;\mu_2,t_2)}_{\mu_2,t_1} = p_{\mu_2,t_1} + \epsilon.
\end{align*}
It is easy to see that $p$ and $p' = p^{\epsilon,(\mu_1,t_1;\mu_2,t_2)}$ share the same average belief and the same marginal distributions over actions and times. If both $p$ and $p'$ are feasible, then both of them yield the same designer's addivitely separable payoff.
\end{definition}
\begin{lemma}[Switching Lemma] \label{lemma: switching_lemma} Let $p$ be a feasible joint distribution and fix $T \in \mathcal {T}$. Then for sufficiently small $\epsilon > 0$,
  \begin{enumerate}
   \setlength\itemsep{0em}
  \item[(i)] if $\mu^C_T(p) > \bar{\mu}$, define $T_1 = \min \{t>T:p_{1,t} >0\}$ as the first time DM stops at belief $1$. Then $p^{\epsilon,(\mu_T,T;1,T_1)}$ is also feasible for every $\mu_T \in \{0,\bar{\mu}\}$;
  \item[(ii)] if $\mu^C_T(p) < \bar{\mu}$, define $T_0 = \min \{t>T:p_{0,t} >0\}$ as the first time DM stops at belief $0$. Then $p^{\epsilon,(\mu_T,T;0,T_0)}$ is also feasible for every $\mu_T \in \{1,\bar{\mu}\}$; and  
  \item[(iii)] $p^{\epsilon,(\mu_T,T;\bar{\mu},t)}$ is feasible for every $\mu_T \in \{0,1\}$ and $t>T.$  
  \end{enumerate}
\end{lemma}
The Switching Lemma gives a sufficient condition under which switching distribution does not hurt continuation incentives. We use this to appropriately push stoping beliefs $1$ and $\bar{\mu}$ to back in time. This implies a weaker version of Theorem \ref{thrm:attention_persuasion_separable}:

\begin{lemma}[Additively separable preferences] \label{prop:supermodular} Suppose $\hat f$ is additively separable. 
Define 
\begin{align*}
    \mathcal{P}^*_{full} &\coloneqq  \{p \in \mathcal{P}^* : \forall t\in\mathcal{T}, p_{\bar{\mu},t} = 0  \} \\
    \mathcal{P}^*_{bad} &\coloneqq  \Bigg \{p \in \mathcal{P^*}: 
    \substack{
    \text{\small{ (i) A terminal time $T$ exists and $p_{\bar \mu, T} > 0$}} \\  
    \text{\small{(ii) $p_{1,s} = p_{\bar{\mu},s} = 0$ for all $s < T$, and }} \\
    \text{\small{(iii) For every $t<T$ such that $p_{0,t}>0$, DM is indifferent}} \\ 
    \text{\small{between stopping and continuing at time $t$ upon receipt of $H^C_t$}}}
    \Bigg \}.
\end{align*}
Then $\mathcal{P}^*_{full} \cup \mathcal{P}^*_{bad} \neq \emptyset$. 
\end{lemma}
$ \mathcal{P}^*_{full}$ (``full information'') corresponds to distributions under which (i) the DM only stops when she obtains full information. $\mathcal{P}^*_{bad}$ (``bad news'') corresponds to distributions under which there is a fixed terminal time $T$ when the DM could stop with belief $\bar{\mu}$; at this terminal time, (ii) the DM only receives bad news for all times $t < T$ i.e., conditional on stopping before time $t$, the DM is certain that $\theta = 0$ and furthermore, (iii) on times $t < T$ where the DM {could} receive bad news (i.e., $p_{0,t} > 0$), the DM is indifferent between stopping and continuing if she does not receive bad news. 

The proof of Lemma \ref{prop:supermodular} is deferred to Online Appendix \ref{onlineappendix: attention_persuasion_separable}. We conclude the preliminaries with the following Pasting Lemma explained in the main text.

\begin{lemma}[Pasting Lemma] \label{lemma: pasting}
   Suppose $p$ be a feasible joint distribution with prior belief $\mu_0 > \bar{\mu}$, leaving DM surplus $\phi$. There exists a feasible joint distribution $p'$ with prior belief $\mu'_0 = \bar{\mu}$, which leaves DM's surplus $\phi$, and $I$ and $I'$ induce the same joint distribution over action and stopping time.
\end{lemma}

\subsection{Proof of Theorem \ref{thrm:attention_persuasion_separable}}

\begin{proof}[Proof of Theorem \ref{thrm:attention_persuasion_separable}]
    If $\mathcal{P}^*_{full} \ne \emptyset$, then a designer's optimal information structure can be obtained by implementing degenerate stopping beliefs (full information). Hence the DM's expected action coincides with her prior beliefs. By additive separability, it is sufficient for the designer to just consider the stopping time, which falls into the pure attention case. Now suppose $\mathcal{P}^*_{full} = \emptyset$. By Lemma \ref{prop:supermodular},  $\mathcal{P}^*_{bad} \ne \emptyset$ and consider any $p^* \in \mathcal{P}^*_{bad}$. Let $t_0$ be the first time the information structure provides a stopping message with positive probability before time $T$. Since the DM is indifferent between continuing and stopping at time $t_0$ and $\mu^C_{t_0} \geq \bar{\mu}$ by the definition of $\mathcal{P}^*_{bad}$, the designer could stop providing further information at $\mu^C_{t_0}$ and let the DM take action $1$ immediately while her dynamic incentives before time $t_0$ are still preserved.  Since $p^*$ is optimal for the designer, stop providing further information at time $t_0$ must be weakly suboptimal for the designer. We will show that the designer's utility does not change if the information structure ends at time $t_0$ instead. This concludes the proof because truncating the information structure at $t_0$ results one-shot persuasion.

 Suppose a contradiction that stopping providing further information at time $t_0$ is strictly suboptimal for the designer. We modify the information structure as follows: let $p' = (p^*)^{\epsilon,(0,t_0;\bar{\mu},T)}$ for small $\epsilon >0$. By Lemma \ref{lemma: switching_lemma}, $p'$ is a feasible distribution. Now observe that the designer's value function is additively separable hence both $p'$ and $p^*$ yield the same expected value. Thus, $p' \in \mathcal{P}^*$.

    By pasting lemma, we construct an information structure $I'\mid t> t_0$ with the prior belief $\bar{\mu}$ from $I\mid t>t_0$ with the prior belief $\mu^C_{t_0} > \bar{\mu}$ such that $I'\mid t> t_0$ and $I\mid t>t_0$ induce the same distribution of actions and stopping times. We further modify the information structure $p'$ by pasting $I'\mid t> t_0$ at the stopping belief $\bar{\mu}$ at time $t_0$ (this happens with positive probability because $p'_{\bar{\mu},t_0} >0$). This modified information structure must be strictly better for the designer than $p'$ because 1) the designer's utilities under $I'\mid t> t_0$ and $I\mid t>t_0$ are the same 2) the designer's utilities under stopping providing further information at $\bar{\mu}$ and $\mu^C_{t_0}$ at time $t_0$ are the same $f(1,t_0)$, and 3) we assumed earlier that the designer finds it suboptimal to stopping providing further information at $\mu^C_{t_0}$, which is a contradiction. 
\end{proof}

\newpage
\setstretch{1.29}
\setcounter{page}{1} 
\renewcommand{\thefootnote}{\fnsymbol{footnote}}

\begin{center}
    \Large{\textbf{Online Appendix to `Attention Capture'}} \\
    \large{\textbf{Andrew Koh and Sivakorn Sanguanmoo\footnote[3]{MIT Department of Economics; emails: \url{ajkoh@mit.edu} and \url{sanguanm@mit.edu}}}\\
    FOR ONLINE PUBLICATION ONLY}
\end{center}

\appendix 
\titleformat{\section}
		{\normalsize\bfseries\center\scshape}     
         {Online Appendix \thesection:}
        {0.5em}
        {}
        []
\renewcommand{\thesection}{\Roman{section}}

Online Appendix \ref{onlineappendix:topology} collects omitted technical details. Online Appendix \ref{onlineappendix: increasing_path+no_surplus} collects the remaining proofs of auxiliary results used in the main Appendix. Online Appendix \ref{appendix:nonlinearcost} discusses optimal attention capture under nonlinear time preferences. 

\section{Technical details}
\label{onlineappendix:topology} 

\subsection{Generalized martingale and obedience constraints} \label{onlineappendix: generalized_constraint}
From Step A of the proof Theorem \ref{thrm:reduction}, it is without loss to consider information structures with a unique continuation belief if the joint distribution over of stopping times and stopping beliefs are objects of interest. If the DM prefers to continue at these unique continuation beliefs, call such information structures uniquely obedient. Each uniquely obedient dynamic information structure $I$ induces a probability distribution $p$ over $\Delta(\Theta) \times \mathcal{T}$, where, for every Borel set $B \subset \Delta(\Theta) \times \mathcal{T},$ $p(B) = \Pr^I(\mu_{\tau(I)},\tau(I)) \in B)$ i.e., $p$ is a probability distribution of stopping times and stopping beliefs of $I$. Note that $p$ also pins down the unique continuation belief path: if the DM has paid attention up to time $t$, her beliefs at $t$ are, by Bayes' rule, 
\[
\mu_t^C({p}) := \mathbb{E}_t^{{p}} (\mu_\tau \mid \tau > t) = \frac{\Ex_{(\mu^S,\tau) \sim p}[\mu^S 1\{\tau>t\}]}{\Pr_{(\mu^S,\tau) \sim p}( \tau>t)},
\]
where we sometimes drop the dependence on $p$ when there is no ambiguity about the dynamic information structure. We say that a distribution $p$ over $\Delta(\Theta) \times \mathcal{T}$ is feasible, if some uniquely obedient information structure induces $p$. The next lemma gives conditions under which $p$ is feasible.
\begin{lemma} \label{lemma: martingale_obedience}
    A distribution ${p}$ over $ \Delta(\Theta) \times \mathcal{T}$ is feasible if and only if the following constraints hold:
    \begin{enumerate}
    \setlength\itemsep{0em}
        \item[(i)] (Martingale constraint) $\mu_0 = \Ex_{(\mu^S,\tau) \sim p} [\mu^S]$.
        \item[(ii)] (Obedience constraint)  For every $t \in \mathcal{T}$,
\[\underbrace{\Pr_{(\mu^S,\tau) \sim p}( \tau>t) \cdot v^*(\mu^C_t,t)}_{\eqqcolon STOP^{{p}}_t}\leq \underbrace{\Ex_{(\mu^S,\tau) \sim p}[v^*(\mu^S,\tau) 1\{\tau>t\}]}_{\eqqcolon CONT^{{p}}_t}.\]
        
    \end{enumerate}  
    
\end{lemma}

\subsection{Extremized information structures} \label{onlineappendix: extremized_info}
  We further reduce the dimension of feasible distributions by considering only extremal beliefs. Recall DM's action space $A$ and state space $\Theta$ are finite. 
 For each $A' \subseteq A$, define the set of beliefs under which every action in $A'$ is one of the DM's best responses: \[X_{t,A'} \coloneqq \{\mu \in \Delta(\Theta) : A' \subseteq a^*(\mu,t)  \}. \]
Since $A$ is finite, $X_{t,A'}$ is a polytope. Each extreme point of $X_{t,A'}$ is called an \textbf{extremal belief}. We denote the set of all extremal beliefs as follows:
\[X^{EXT}_t \coloneqq \bigcup_{A' \subseteq A} \text{Ext} (X_{t,A'}).\]

An \textbf{extremized information structure} of $\pi_t$ is $\pi^*_t \in \Delta(X_t^{EXT})$ such that (i) $\pi^*_t$ preserves the mean of $\pi_t$; (ii) DM's expected utility under $\pi_t$ and $\pi^*_t$ are the same. 
\begin{lemma} \label{lem: sufficient_signal_structure} 
    For any $\pi_t \in \Delta(\Delta(\Theta))$, there exists an extremized information structure of $\pi_t$ denoted $\pi^*_t$. Moreover, for any designer's utility function $f_t: A \to \mathbb{R}$ inducing the indirect utility function $h_t: \Delta(\Theta) \to \mathbb{R}$, $\Ex_{\mu \sim \pi_t} h_t(\mu) \leq \Ex_{\mu \sim \pi^*_t} h_t(\mu)$ . 
\end{lemma}

The proof of Lemma \ref{lem: sufficient_signal_structure} is omitted and  straightforward, and variants of it has been used in (static) information design \citep*{bergemann2015limits}. 

Lemma \ref{lem: sufficient_signal_structure} implies it is sufficient for the designer to consider feasible distribution $p$ over $\Delta(\Theta) \times \mathcal{T}$ to the set of pairs of time and corresponding extremal beliefs $X^{EXT}(\mathcal{T}) = \{(\mu_t,t): t \in \mathcal{T}, \mu_t \in X^{EXT}_t\}$. This is because an extremized information structure 1) preserves the DM's continuation utility (hence, incentive to continue) and hence 2) weakly increases the designer's value. Thus, it is without loss for the designer to choose a feasible distribution $p$ from the set of feasible distributions $\mathcal{P}^{EXT} \coloneqq \mathcal{P} \cap \Delta(X^{EXT}(\mathcal{T}))$. For each $p \in \mathcal{P}^{EXT}$, we denote \[p_{\mu^*,t} = \Pr_{(\mu,\tau) \sim p}(\mu = \mu^*,\tau = t )\] for every $t \in \mathcal{T}$ and $\mu^* \in X^{EXT}_t$.

\subsection{Topology over the set of feasible distributions $\mathcal{P}^{EXT}$}
\label{onlineappendix: topology_space}
Assume as in the main text that costs are additively separable $v(a,\theta,t) = u(a,\theta) -ct$. It is easy to see that the extremal beliefs do not change over time. Define $X^{EXT} = X^{EXT}_t$. 
We will create a topology of feasible stopping times and beliefs over $X^{EXT} \times \mathcal{T}$. We start with the following lemma.

\begin{lemma} \label{lemma:small_inf_cost} Let $\delta = 1-\frac{c}{\phi^*}.$
 There exists a constant $K$ such that $\Pr(\tau = t) < K \delta^t$ for every $t \in \mathcal{T}$ and every feasible stopping time $\tau$.
\end{lemma}
\begin{proof}[Proof of Lemma \ref{lemma:small_inf_cost}]
Let $C_t \coloneqq \Ex[c\tau \mid \tau >t ]- ct$. We know from the obedience constraint that $C_t$ is bounded above by $\phi^*$. Consider that 
\begin{align*}
    C_t+ct 
    &=\Ex[c\tau \mid \tau > t+1] \Pr(\tau > t+1 \mid \tau>t) + c(t+1) (1 - \Pr(\tau > t+1 \mid \tau>t)) \\ 
    &= c(t+1) + C_{t+1}\Pr(\tau>t+1 \mid \tau>t),
\end{align*}
which implies that $\Pr(\tau>t+1 \mid \tau >t) = \frac{C_t-c}{C_{t+1}},$
for every $t \in \mathcal{T}$. Note also that $C_{t} \geq c $. Thus,
\begin{align*}
    \Pr(\tau > t-1) &= \prod_{s=0}^{t-2} \Pr(\tau > s+1 \mid \tau > s) \\
    &= \prod_{s=0}^{t-2} \frac{C_s-c}{C_{s+1}} 
    = \frac{C_0-c}{C_{t-1}} \prod_{s=1}^{t-2} \frac{C_s-c}{C_s} 
    \leq \frac{\phi^*-c}{c} \bigg(1- \frac{c}{\phi^*} \bigg)^{t-2},
\end{align*}
as desired.
\end{proof}
\begin{corollary} \label{corollary: dominated_conv}
    Suppose $(p^i)_{i \in \N} \subset \mathcal{P}^{EXT}$ pointwise converges to $p^* \in [0,1]^\mathbb{N}$. For every function $f: \Delta(\Theta) \times \mathcal{T} \to \mathbb{R}$ such that $\lim_{t \to \infty} \sup_{\mu \in \Delta(\Theta)}|f(\mu,t)| \delta_f^t$ = 0 for some $\delta_f \in (\delta,1]$ , we must have  
        \begin{align*}
            \lim_{i \to \infty} \sum_{t \in \mathcal{T}} \sum_{\mu \in X^{EXT}}
            |p^i(\mu,t)-p^*(\mu,t)| |f(\mu,t)| = 0.
        \end{align*}
\end{corollary}
\begin{proof} [Proof of Corollary \ref{corollary: dominated_conv}]
    From Lemma \ref{lemma:small_inf_cost}, we observe that $ |p^i(\mu,t)-p^*(\mu,t)| |f(\mu,t)| \leq 2M\delta^t|f(\mu,t)| \eqqcolon h(\mu,t)$. Consider that
    \begin{align*}
        \sum_{t \in \mathcal{T}} \sum_{\mu \in X^{EXT}} h(\mu,t) = 2M\sum_{\mu \in X^{EXT}} \sum_{t \in \mathcal{T} }\delta^t|f(\mu,t)| < \infty,
    \end{align*}
    which follows by our condition that $\lim_{t \to \infty} |f(\mu,t)| \delta_f^t = 0$ for some $\delta_f \in (\delta,1]$.  Because $\big\lvert p^i_{\mu,t}-p^*_{\mu,t} \big\rvert f(\mu,t)$ converges pointwise to $0$ and $f$ is integrable in $l^1$, the lemma is directly followed by dominated convergence theorem.
\end{proof}
Pick any $\bar{\delta} \in (\delta,1)$. We introduce a metric $\Delta$ of $\mathcal{P}^{EXT}$ as follows: for any probability measures $p^1,p^2 \in \mathcal{P}^{EXT}$, 
\[\Delta(p^1,p^2) \coloneqq \sum_{t=1}^\infty \sum_{\mu \in X^{EXT}}
            |p^1_{\mu,t}-p^2_{\mu,t}| \bar{\delta}^{-t}. \]
This metric is well-defined because the obedience constraint at $t=0$ implies 
\begin{align*}
    \sum_{t=1}^\infty \sum_{\mu \in X^{EXT}} \big\lvert p^1_{\mu,t} - p^2_{\mu,t} \big \rvert \bar{\delta}^{-t}  &\leq \sum_{t=1}^\infty \sum_{\mu \in X^{EXT}} p^1_{\mu,t} \bar{\delta}^{-t} + \sum_{t=1}^\infty \sum_{\mu \in X^{EXT}}  p^2_{\mu,t}  \bar{\delta}^{-t} 
    \leq 2 \sum_{t=1}^\infty M (\delta / \bar{\delta})^t,
\end{align*}
which must be finite because $\bar{\delta} > \delta$.
It is easy to verify that $\Delta$ is a metric. We use the metric $\Delta$ to construct a topological space of $\mathcal{P}^{EXT}$ and obtain the following lemma.

\begin{lemma} \label{prop: D-cpct}
$\mathcal{P}^{EXT}$ is compact under the metric $\Delta$.
\end{lemma}

\begin{proof}[Proof of Lemma \ref{prop: D-cpct}]
Consider any sequence $(p^i)_{i \in \mathbb{N}} \subset \mathcal{P} \subset [0,1]^\mathbb{N}$. Since $[0,1]^\mathbb{N}$ is compact under the product topology, there exists a subsequence $(p^{i_n})_{n \in \N}$ that converges pointwise to some $p^* \in [0,1]^\mathbb{N}$. We will show that 1) $p^* \in \mathcal{P}$ and 2) $(p^{i_n})_{n \in \N}$ converges to $p^*$ under the metric $\Delta$.

1) To show that $p^* \in \mathcal{P}^{EXT}$, we show the following:
\begin{itemize}[nosep]
    \item $p^*$ satisfies the martingale constraint, which also implies $p^*$ is a probability distribution. For every $\theta \in \Theta$, using Corollary \ref{corollary: dominated_conv} with $f(\mu,t) = \mu(\theta)$ and $\delta_f = 1-\epsilon$ gives
    \[\sum_{t \in \mathcal{T}}\sum_{\mu \in X^{EXT}} p^*_{\mu,t}\mu(\theta) = \lim_{n \to \infty }\sum_{t \in \mathcal{T}}\sum_{\mu \in X^{EXT}} p^{i_n}_{\mu,t} \mu(\theta) = \mu_0(\theta),\]
    as desired.
     \item $p^*$ satisfies the obedience constraints at every time $t \in \mathcal{T}$. Fix action $a \in A$, $f(\mu,s) = \sum_{\theta \in \Theta} \mu(\theta) u(a,\theta,s)1\{s>t\}$ and $\delta_f = 1-\epsilon$ satisfies the condition in Corollary 2 because $u(a,\theta,s)$ is linear in $s$. Thus,
    \begin{align*}
    \sum_{s=t+1}^\infty \sum_{\mu^S \in X^{EXT}_s(\Theta)} \sum_{\theta \in \Theta} u(a,\theta,t)\mu^S(\theta) p^*_{\mu^S,s} 
    =
    \lim_{n \to \infty}\sum_{s=t+1}^\infty \sum_{\mu^S \in X^{EXT}} \sum_{\theta \in \Theta} u(a,\theta,t)\mu^S(\theta) p^{i_n}_{\mu^S,s}.
    \end{align*}
    Similarly, using Corollary \ref{corollary: dominated_conv} with $f(\mu,t) = u^*(\mu,s)$ gives
    \[\sum_{s=t+1}^\infty \sum_{\mu^S \in X^{EXT}} p^*_{\mu^S,s}u^*(\mu^S,s) = \lim_{n \to \infty} \sum_{s=t+1}^\infty \sum_{\mu^S \in X^{EXT}} p^{i_n}_{\mu^S,s}u^*(\mu^S,s).\]
    With the obedience constraint at $t$ for every $p^{i_n}$, these together imply the obedience constraint at $t$ for $p^*$, as desired.
\end{itemize}

2) Applying Corollary \ref{corollary: dominated_conv} with $f(\mu,t) = \bar{\delta}^{-t}$ and $\delta_f = \bar{\delta} - \epsilon $, we obtain $\lim_{n \to \infty} \Delta(p^{i_n},p^*) = 0$. Thus, $(p^{i_n})_{n \in \N}$ converges to $p^*$ under metric $\Delta$, as desired.

From 1) and 2), any sequence $(p^i)_{i \in \N} \subset \mathcal{P}^{EXT}$ has a convergent subsequence in $\mathcal{P}^{EXT}$ under the metric $\Delta$. This implies $\mathcal{P}^{EXT}$ is compact under the metric $\Delta$.
\end{proof}
Note that we can apply a similar proof to show that $\mathcal{D}_s$ is compact under the metric $\Delta_{\mathcal{T}},$ defined as 
\[\Delta_{\mathcal{T}}(d^1,d^2) \coloneqq \sum_{t\in \mathcal{T}}
            |d^1(t)-d^2(t)| \bar{\delta}^{-t}, \]
 for every $d^1,d^1 \in \mathcal{D}_s.$
\begin{proposition} \label{prop: additive_topology}
If costs are additively separable as in the main text: $v(a,\theta,t) = u(a,\theta) -ct$ with the designer's function $f: A \times \mathcal{T} \to \mathbb{R}^+$ such that there exists $\delta_f \in (1-\frac{c}{\phi^*},1)$ such that  $\text{lim sup}_{t \to \infty} f(a,t)\delta_f^t < \infty$ for  every $a \in A$. Then, Assumption \ref{assumption: metric_exist} holds.  
\end{proposition}
\begin{proof} [Proof of Proposition \ref{prop: additive_topology}]
    Let $\delta = 1-\frac{c}{\phi^*}$. We choose $\bar{\delta} \in (\delta,\delta_f)$ and create a metric $\Delta$ using $\bar{\delta}$.  We showed earlier that $\mathcal{P}^{EXT}$ is compact under metric $\Delta$. Because $\text{lim sup}_{t \to \infty} f(a,t)\delta_f^t < \infty$ for every $a \in A$ and $A$ is finite, there exists a constant $M$ such that $f(a,t)\bar{\delta}^t<M$ for every $a \in A$ and $t \in \mathcal{T}$.    
    Recall that $h(\mu,t) \coloneqq  \max_{a \in a^*(\mu)} f(a^*(\mu),t)$ for every $\mu \in X^{EXT}$, which implies $h(\mu,t)\bar{\delta}^t \leq M$. 
    For any $\epsilon >0$ and $p^1,p^2 \in \mathcal{P}^{EXT}$, if $\Delta(p^1,p^2) < \epsilon/M$, then
    \begin{align*}
        \bigg\lvert  \sum_{t\in\mathcal{T}} \sum_{\mu \in X^{EXT}}p^1_{\mu,t}h(\mu,t) &- \sum_{t\in\mathcal{T}} \sum_{\mu \in X^{EXT}}p^2_{\mu,t}h(\mu,t) \bigg\rvert \\
        &\leq \sum_{t \in \mathcal{T}}\sum_{\mu \in X^{EXT}} |p^1_{\mu,t}-p^2_{\mu,t}|h(\mu,t) \leq M \Delta(p^1,p^2) < \epsilon. 
    \end{align*}
    This implies the map $p \mapsto \sum_{t\in\mathcal{T}} \sum_{\mu \in X^{EXT}}p_{\mu,t}h(\mu,t)$ is continuous under metric $\Delta$. Thus, Assumption \ref{assumption: metric_exist} holds under metric $\Delta,$ as desired.
\end{proof}

This proposition implies the existence of a solution to the optimization problem for the designer under the regularity assumption, and the set of the designer's optimal and extremized information structure, says $\mathcal{P}^* \subset \mathcal{P}^{EXT}$, is compact.  Finally, noting that the function $p \mapsto p_{\bar{\mu},t}2^{-t}$ is well-defined and continuous under the metric $\Delta$ we have the following corollary. 

\begin{corollary} \label{corollary: min_exists}
When $|A| = 2$ and $|\Theta| = 2$ with the set of extremal beliefs $X^{EXT} = \{0,\bar{\mu},1\}$, the optimization problem $\min_{p \in \mathcal{P}^*} \sum_{t \in \mathcal{T}} p_{\bar{\mu},t}2^{-t}$ has a solution.
\end{corollary}

\subsection{Proof of the existence of an undominated belief path}
\label{onlineappendix: existence_undominated_belief}

\begin{proof}[Proof of Lemma \ref{prop: existence_maximal}]
Our first step is, fixing a distribution of stopping time $d(\tau)$, to study the set of belief paths that satisfy Proposition \ref{prop: IC+boundary-condition} in order to verify that $\tau$ is a feasible stopping time. We endow $\mathcal{W}$ with the product topology of the weak topology on $\Delta(\Theta)$. Because $\Delta(\Theta)$ endowed with the weak topology is compact and metrizable, so is $\mathcal{W}$ by Tychonoff's theorem. We show that $\mathcal{W}(\tau)$ is sequentially compact; hence, compact. Consider $((\mu^i_t)_{t \in \mathcal{T}})_{i=0}^\infty \subset \mathcal{W}(\tau)$. Because $\mathcal{W}$ is compact, there exists a convergent subsequence $((\mu^{i_n}_t)_{t \in \mathcal{T}})_{n=0}^\infty$ as $n \to \infty$. Suppose that $(\mu^{i_n}_t)_{t \in \mathcal{T}} \to (\mu^{*}_t)_{t \in \mathcal{T}} \in \mathcal{W}$ as $n \to \infty$. This implies $\mu_t^{i_n} \to \mu^*_t$ as $n \to \infty$ under the weak topology on $\Delta(\Theta)$ for every $t \in \mathcal{T}$. It suffices to show that $(\mu^{*}_t)_{t \in \mathcal{T}} \in \mathcal{W}(\tau).$ Because $\phi$ is continuous under the weak topology on $\Delta(\Theta)$, for every $t \in \mathcal{T}$, we have
\begin{align*}
    \phi(\mu^*_t) = \lim_{n \to \infty} \phi(\mu^{i_n}_t) \geq  \Ex[c\tau \mid \tau>t]-ct,
\end{align*}
where the inequality follows by $(\mu^{i_n}_t)_{t \in \mathcal{T}} \in \mathcal{W}(\tau)$. This implies the obedience constraint holds for the belief path $(\mu^{*}_t)_{t \in \mathcal{T}}$. Moreover, we have $\Pr(\tau>t+1)\mu^{i_n}_{t+1}(\theta) \leq \Pr(\tau>t)\mu^{i_n}_{t}(\theta)$ for every $t \in \mathcal{T}$. Therefore,
\begin{align*}
    \Pr(\tau>t+1)\mu^{*}_{t+1}(\theta) = \lim_{n \to \infty} \Pr(\tau>t+1)\mu^{i_n}_{t+1}(\theta) \leq \Pr(\tau>t)\mu^{i_n}_{t}(\theta) = \Pr(\tau>t)\mu^{*}_{t}(\theta),
\end{align*}
which implies the boundary constraint. This implies $(\mu^*_t)_{t \in \mathcal{T}} \in \mathcal{W}(\tau)$, as desired.

Define $F: \mathcal{W} \to \mathbb{R}$ such that $F((\mu_t)_{t \in \mathcal{T}}) = \sum_{t=0}^\infty \frac{1}{2^t}\phi(\mu_t)$. $F$ is well-defined because $|F\big((\mu_t)_{t \in \mathcal{T}}\big)| \leq \sum_{t=0}^\infty \frac{1}{2^t}\phi^* \leq 2\phi^*$. Moreover, for every $\epsilon>0$ and every  $T > 1+ \log (\phi^*/\epsilon)/\log 2 $, we have
$\sum_{t=T}^\infty \frac{1}{2^t}\phi(\mu_t) \leq \frac{1}{2^{T-1}}\phi^* < \epsilon$. This implies $\sum_{t=0}^\infty \frac{1}{2^t}\phi(\mu_t)$ is uniformly convergent. Therefore, $F$ is continuous under $\mathcal{W}$. Since $\mathcal{W}(\tau)$ is compact and nonempty, we can find
\begin{align*}
 (\mu^*_t)_{t \in \mathcal{T}}  \in \text{argmax}_{(\mu_t)_{t \in \mathcal{T}} \in \mathcal{W}(\tau)} F((\mu_t)_{t \in \mathcal{T}}).
\end{align*}
Suppose that there exists $(\mu_t)_{t \in \mathcal{T}} \in \mathcal{W}(\tau)$ such that $\phi(\mu_t) \geq \phi(\mu^*_t)$ for every $t \in \mathcal{T}$ and the inequality is strict for some $t \in \mathcal{T}$. This directly implies $F((\mu_t)_{t \in \mathcal{T}}) > F((\mu^*_t)_{t \in \mathcal{T}})$, which contradicts the optimality of $(\mu^*_t)_{t \in \mathcal{T}}$. Therefore, $(\mu^*_t)_{t \in \mathcal{T}}$ is a maximal belief path under $\tau$. 
\end{proof}

\setcounter{footnote}{0}

\section{Remaining proofs of technical results} \label{onlineappendix: increasing_path+no_surplus}

\subsection{Proof of auxillary results used to show Theorem \ref{thrm:convex_ext}} \label{onlineappendix: convex_frontier}
We will prove Lemma \ref{lem: convex_frontier_exist} and \ref{lem: convex_indiff}

\begin{proof}[Proof of Lemma \ref{lem: convex_frontier_exist}]

Fix any $d \in \mathcal{D}_s.$ We first show that the upper contour set $UP(d)= \{d' \in \mathcal{D}_s: d' \succeq_{CX} d\}$ is compact under the metric $\Delta_{\mathcal{T}}$. 

Since $\mathcal{D}_s$ is compact under $\Delta_{\mathcal{T}}$, it suffices to show that $UP(d)$ is closed. Suppose a sequence of $(d_i)_{i \in \mathbb{N}} \subset \mathcal{D}_s$ converges to $d^* \in \Delta(\mathcal{T})$ under the metric $\Delta_{\mathcal{T}}.$ Since $\mathcal{D}_s$ is closed, $d^* \in \mathcal{D}_s.$ For any $s \in \mathcal{T}$, define $f_s:\mathcal{T} \to \mathbb{R}$ such that $f_s(t) = (t-s)\cdot 1\{t \geq s\}$. By Corollary \ref{corollary: dominated_conv}, we must have
\begin{align*}
\lim_{i \to \infty} \sum_{t=0}^\infty f_s(t)d_i(t) = \sum_{t=0}^\infty f_s(t)d^*(t).
\end{align*}
This implies 
\[\Ex_{\tau^* \sim d^*} [f_s(\tau^*)] = \Ex_{\tau_i \sim d_i} [f_s(\tau_i)] \geq \Ex_{\tau \sim d} [f_s(\tau)], \]
for every $s \in \mathcal{T},$ where the inequality follows from the convexity of $f_s$ and $d_i \succeq_{cx} d$. We must have $d^* \succeq_{cx} d$, implying $d^* \in UP(d)$. Thus, $UP(d)$ is closed, as desired.

Define a function $h: \mathcal{T} \to \mathbb{R}$ such that $h(t) = \delta^{-t},$ where $\delta = 1-\frac{c}{2\phi^*}.$ The proof of Proposition \ref{prop: additive_topology} shows that the map $d \in \mathcal{D}_s \mapsto \sum_{t \in \mathcal{T}} d(t)h(t)$ is continuous under metric $\Delta_{\mathcal{T}}$. Thus, an answer to the following maximization problem exists:
    \begin{align*}
        \max_{d' \in UP(d)} \Ex_{\tau \sim d'}[h(\tau)].
    \end{align*}

    Note that there is no $d' \in \mathcal{D}_s$ such that $d' \succ_{CX} d^*$. To see this, if $d' \succ_{CX} d^*$, then $d' \succ_{CX} d$, which implies $d' \in UP(d)$. Since $h$ is a strictly convex function, $\Ex_{\tau \sim d'}[h(\tau)] > \Ex_{\tau \sim d^*}[h(\tau)]$, which contradicts the optimality of $d^*.$
\end{proof} 

\begin{proof} [Proof of Lemma \ref{lem: convex_indiff}]
    Suppose $I$ is an information structure under which the DM strictly prefers to continue paying attention at time $t_0$ and $t_0+1 \in \text{supp } \tau,$ where $\tau$ is the induced stopping time.  This means there exists $\theta_0 \in \Theta$ such that $\Pr^I(\mu_{t_0+1} = \delta_{\theta_0
    }) >0.$  Define a new information structure as follows: 
    \begin{enumerate}
        \item For every $t<t_0$, let $I_{t} = I'_t$ and define a continuation belief $\mu'^C_{t} = \mu^C_t.$
        \item At time $t_0$, modify $I_{t_0}$ as follows:
        \begin{align*}
            I'_{t_0}((\delta_{\theta_0},S) \mid (\mu'^C_t,C)_{t \leq t_0-1}) &= I_{t_0}((\delta_{\theta_0},S) \mid (\mu^C_t,C)_{t \leq t_0-1}) + \epsilon  \\
            I'_{t_0}((\delta_{\theta},S) \mid (\mu'^C_t,C)_{t \leq t_0-1}) &= I_{t_0}((\delta_{\theta},S) \mid (\mu^C_t,C)_{t \leq t_0-1}) \quad \forall \theta \ne \theta_0  \\
            I'_{t_0}((\mu'^C_{t_0},C) \mid (\mu'^C_t,C)_{t \leq t_0-1}) &= I_{t_0}((\mu^C_{t_0},C) \mid (\mu^C_t,C)_{t \leq t_0-1})-\epsilon, 
        \end{align*}
        where $\mu'^C_{t_0}$ is an appropriate belief so that the martingale condition holds. 
        \item To modify $I_{t_0+1}$, we introduce a new augmented message $C_{delay}$ and define a new information structure $I'_{t_0+1}$ as follows: let $P_{t_0} = I_{t_0}((\mu^C_{t_0},C) \mid (\mu^C_t,C)_{t \leq t_0-1})-\epsilon$, define
        \begin{align*}
            I'_{t_0+1}((\delta_{\theta_0},S) \mid (\mu'^C_t,C)_{t \leq t_0}) &= \frac{1}{P_{t_0}} \cdot(P_{t_0}+\epsilon-\delta)I_{t_0+1}((\delta_{\theta_0},S) \mid (\mu^C_t,C)_{t \leq t_0})  - \frac{\epsilon}{P_{t_0}} \\
            I'_{t_0+1}((\delta_\theta,S) \mid (\mu'^C_t,C)_{t \leq t_0}) &= \frac{1}{P_{t_0}} \cdot(P_{t_0}+\epsilon -\delta)I_{t_0+1}((\delta_\theta,S) \mid (\mu^C_t,C)_{t \leq t_0}) \quad \forall \theta \ne \theta_0  \\
            I'_{t_0+1}((\mu'^C_{t_0+1},C) \mid (\mu'^C_t,C)_{t \leq t_0}) &= \frac{1}{P_{t_0}} \cdot(P_{t_0}+\epsilon -\delta)I_{t_0+1}((\mu^C_{t_0+1},C) \mid (\mu^C_t,C)_{t \leq t_0})  \\
            I'_{t_0+1}((\mu'^{C}_{t_0},C_{delay}) \mid (\mu'^C_t,C)_{t \leq t_0}) &= \frac{\delta}{P_{t_0}}
        \end{align*}
        where we set $\mu'^C_{t_0+1} = \mu^C_{t_0+1}$ and $\delta = \frac{\epsilon}{\Pr(\tau \in[t_0+1,t_1) \mid \tau \geq t_0)}$.
        \item For any $t \geq t_0+1$ with history $(\mu'^C_s,C)_{s \leq t}$, set 
        \[I'_{t+1}(\cdot \mid (\mu'^C_s,C)_{s \leq t}) = I_{t+1}(\cdot \mid (\mu^C_s,C)_{s \leq t}).\] 
        \item For any $t \in [t_0+1,t_1-1)$ with history $H_t \in \{(\mu_s,m_s)_{s \leq t}: \exists s' \text{ s.t. }m_{s'} = C_{delay}\}$, set \[I'_{t+1}(\cdot \mid H_t) = I_{t}(\cdot \mid (\mu^C_s,C)_{s \leq t-1}) .\]
        \item For history $H_{t_1-1} \in \{(\mu_s,m_s)_{s \leq t_1-1}: \exists \text{ s.t. }m_{s'} = C_{delay}\}$ set 
        \begin{align*}
            I'_{t_1}((\mu,S) \mid H_{t_1-1}) &= I_{t_1-1}((\mu,S) \mid (\mu^C_s,C)_{s \leq t_1-2}) + P_{t_1}I_{t_1}((\mu,S) \mid (\mu^C_s,C)_{s \leq t_1-1}) \\
            I'_{t_1}((\mu,C) \mid H_{t_1-1}) &= P_{t_1}I_{t_1}((\mu,C) \mid (\mu^C_s,C)_{s \leq t_1-1}),
        \end{align*}
        where $P_{t_1} = I_{t_1-1}((\mu^C_t,C) \mid (\mu^C_s,C)_{s \leq t_1-2})$.
        \item For any $t  \geq t_1$ with history $H_t \in \{(\mu_s,m_s)_{s \leq t}: \exists s' \text{ s.t. }m_{s'} = C_{delay}\}$, set \[I'_{t+1}(\cdot \mid H_t) = I_{t+1}(\cdot \mid (\mu^C_s,C)_{s \leq t}) .\]
    \end{enumerate}
    
    It is easy to see that the martingale condition holds for every $t \in \mathcal{T}$. We show the DM's optimal stopping time under $I'$ to continue paying attention when she sees message $C$ or $C_{delay}.$ Suppose this induces stopping time $\tau'$. From the construction of $I'$, we must have
    \begin{align*}
        \Pr(\tau' = t) = \begin{cases}
            \Pr(\tau = t) &\text{if } t<t_0 \\
            \Pr(\tau = t) + \epsilon A  &\text{if } t=t_0 \\
            (1-\delta A)\Pr(\tau = t) - \epsilon A  &\text{if } t=t_0 + 1 \\
            (1-\delta A)\Pr(\tau = t) + \delta A\Pr(\tau = t-1)  &\text{if } t \in ( t_0 + 1,t_1), \\
            \Pr(\tau = t_1) + \delta A\Pr(\tau = t_1-1)  &\text{if } t = t_1, \\
            \Pr(\tau = t)  &\text{if } t > t_1,
        \end{cases}
    \end{align*}
    where $A = \Pr^I(\tau \geq t_0).$ 
    The obedience constraints clearly hold at every history $(\mu_s)_{s \leq t}$ when $t>t_1$ by obedience constraints under $I$. We will show that the obedience constraints hold at every history $((\mu_s,m_s))_{s \leq t}$ in the support of $\tau'$ when $t\in [t_0+1,t_1)$ and $m_{t_{0}+1} = C_{delay}$. It is easy to see that $\mu'_t = \mu^C_{t-1}$. By the obedience constraint at $t-1$ under $\tau$,
    \begin{align*}
        \phi(\mu'_t) = \phi(\mu^C_{t-1}) \geq \Ex[c(\tau-(t-1)) \mid \tau> t-1].
    \end{align*}
    Consider that
    \begin{align*}
        &\Ex[c(\tau'-t) \mid (\mu_s,m_s)_{s \leq t}] \\
        &= \Ex[c(\tau-(t-1)) 1\{\tau \in (t-1,t_1]\} \mid \tau > t-1] + \Ex[c(\tau-t) 1\{\tau>t_1\}\mid \tau>t-1] \\
        &\leq \Ex[c(\tau-(t-1)) \mid \tau >t-1] \\
        &\leq \phi(\mu'_t),
    \end{align*}
    which implies the obedience constraint at the history $((\mu_s,m_s))_{s \leq t}$ in the support of $\tau'$ when $t\in [t_0+1,t_1)$ and $m_{t_{0}+1} = C_{delay}$. The DM still prefers to continue paying attention at history $(\mu'^C_t)_{t \leq t_0}$ with small perturbation $\epsilon>0$ because the obedience constraint at time $t_0$ under $I$ slacks. It suffices to check the DM's incentive at time $t<t_0$. Since $\mu^C_t = \mu'^C_t$ for every $t<t_0$, the stopping utilities under both $I$ and $I'$ before time $t_0$ are the same. The difference of the continuation utilities under $I$ and $I'$ at $t$ is
    \begin{align*}
        \Ex[c\tau' \cdot 1\{\tau' >t\}] - \Ex[c\tau \cdot 1\{\tau >t\}] = -\epsilon A + \delta A \Pr(\tau \in  [t_0+1,t_1)) = 0,
    \end{align*}
    implying the continuation utilities at $t$ under $I$ and $I'$ are the same. Thus, the DM continues paying attention whenever she sees message $C$ or $C_{delay}.$ These together implies $d(\tau') \in \mathcal{D}_s$, where $s = \Ex[\tau].$ 
     For (i), the designer's utility under $I'$ is more than that under $I$ because
    \begin{align*}
    \Ex[f(\tau')] - [f(\tau)] &= -A\epsilon (f(t_0+1) - f(t_0) ) + A\epsilon \Ex^I[f(\tau+1) -f(\tau) \mid \tau \in [t_0+1,t_1) ]   \\
    &> 0,
    \end{align*}
     where the last line is from the assumption of $f$, which proves (i). 
    For (ii), consider every strictly convex function $h$. We must have
    \begin{align*}
    \Ex[h(\tau')] - \Ex[h(\tau)] &= -A\epsilon (h(t_0+1) - h(t_0) ) + A\epsilon \Ex^I[h(\tau+1) -h(\tau) \mid \tau \in [t_0+1,t_1) ]   \\
    &> 0.
    \end{align*}
    This implies $d(\tau') \succ_{CX} d(\tau),$ which proves (ii).
\end{proof}

\subsection{Proof of auxillary results used to show Proposition \ref{prop:step_and_S-shaped} (i)}
\label{onlineappendix: convergence_to_basin}
Before proving Lemma \ref{prop: converge_>2states}, we introduce a definition of a feasible set of $\mu_{t+1}$ given $\mu_t$ and stopping time $\tau$ derived by the boundary constraint. 

\begin{definition}
For every nonnegative number $r \leq 1$ and belief $\mu \in \Delta(\Theta)$, define $F(\mu,r) \subset \Delta(\Theta)$ such that
\begin{align*}
    F(\mu,r) = \{\mu' \in \Delta(\Theta) \mid r\mu'(\theta) \leq \mu(\theta) \quad \forall \theta \in \Theta\}.
\end{align*}
\end{definition}
With this definition, the boundary constraint is equivalent to $\mu_{t+1} \in F\Big(\mu_{t}, \frac{\Pr(\tau > t+1)}{\Pr(\tau > t)} \Big)$ for every $t \in \mathcal{T}$. We now begin the proof of Lemma \ref{prop: converge_>2states}  
\begin{proof} [Proof of Lemma \ref{prop: converge_>2states}]
Consider any feasible stopping time $\tau$. Suppose that $\mu_T \notin \Phi^*.$

Choose an undominated belief path and a distribution of stopping time $(\mu_t)_{t \in \mathcal{T}}$. From Lemma \ref{lem:increasing_extremal}, we must have $\phi(\mu_t)$ is increasing in $t \in \mathcal{T}$ and the boundary constraint binds for every $t<T$. Consider any $t_0 \in \mathcal{T}$. We will show that, for every $t<T$ and $\lambda \in [0,1)$ such that $t > t_0+1$, we have $\lambda \mu_{t_0+1} + (1-\lambda)\mu_{t} \notin \text{ int } F\Big(\mu_{t_0}, \frac{\Pr(\tau > t_0+1)}{\Pr(\tau > t_0)} \Big).$

Assumes a contradiction that there are $t_1 \in \mathcal{T}$ and $\lambda \in [0,1)$ such that $t_1>t_0+1$ and $\lambda \mu_{t_0+1} + (1-\lambda)\mu_{t_1} \in \text{ int } F\Big(\mu_{t_0}, \frac{\Pr(\tau > t_0+1)}{\Pr(\tau > t_0)} \Big)$. This means $\frac{\Pr(\tau > t_0+1)}{\Pr(\tau > t_0)}  < 1$, so the property of an extremal path implies $\mu_{t_0+1} \in \text{Bd } F\Big(\mu_{t_0}, \frac{\Pr(\tau > t_0+1)}{\Pr(\tau > t_0)} \Big)$, so $\mu_{t_0+1} \ne \mu_{t_0}$.

We define a new belief path $(\mu'_t)_{t \in \mathcal{T}}$ as follows:
\begin{align*}
    \mu'_t = \begin{cases}
    \lambda \mu_{t} + (1-\lambda) \mu_{t_1}, &\text{if $t_0 < t \leq t_1$} \\
    \mu_t, &\text{otherwise.}
    \end{cases}
\end{align*}
We will show that a pair of a belief path and a distribution of stopping time $((\mu'_t)_{t \in \mathcal{T}},d(\tau))$ is feasible. The obedience constraint is still the same for $t \notin \{t_0+1,\dots,t_1\}$. If $t \in \{t_0+1,\dots,t_1\}$, we have
\begin{align*}
    \phi(\mu'_t) \geq \lambda \phi(\mu_t) + (1-\lambda)\phi(\mu_{t_1}) \geq \phi(\mu_t),
\end{align*}
where the inequality follows by the fact that $\phi(\mu_t)$ is increasing in $t \in \mathcal{T}$. This directly implies the obedience constraint for $t \in \{t_0+1,\dots,t_1\}$. The boundary constraint is still the same for $t \in \{0,\dots,t_0-1\} \cup \{t_1,\dots\}$. The boundary constraint holds when $t = t_0$ by the construction of $\lambda$. For any $t \in \{t_0+1,\dots,t_1-1\}$, we have
\begin{align*}
    \min_{\theta \in \Theta} \frac{\mu'_t(\theta)}{\mu'_{t+1}(\theta)} = \min_{\theta \in \Theta} \frac{\lambda\mu_t(\theta)+(1-\lambda)\mu_{t_1}(\theta)}{\lambda\mu_{t+1}(\theta)+(1-\lambda)\mu_{t_1}(\theta)} \geq  \min_{\theta \in \Theta} \frac{\mu_t(\theta)}{\mu_{t+1}(\theta) }\geq \frac{\Pr(\tau>t+1)}{\Pr(\tau>t)},
\end{align*}

This concludes that the belief path $(\mu'_t)_{t \in \mathcal{T}}$ is also an undominated belief path corresponding to the stopping time $\tau$ because $\phi(\mu'_t) \geq \phi(\mu_t)$ for every $t \in \mathcal{T}$. However, the above inequality us strict for $t = t_0$ because $\mu_{t_0+1} \ne \mu_{t_0}$, which implies that  $(\mu'_t)_{t \in \mathcal{T}}$ does not satisfy the property of a maximal path. This contradicts with the fact that $(\mu'_t)_{t \in \mathcal{T}}$ is an undominated belief path. Therefore, for every $t_0,t_1 \in \mathcal{T}$ such that $t_1>t_0+1$, we must have $\lambda\mu_{t_0+1}+(1-\lambda)\mu_{t_1} \notin \text{int }F\Big(\mu_{t_0}, \frac{\Pr(\tau>t+1)}{\Pr(\tau>t)}\Big)$.

For any $\Theta' \subset \Theta$, define
\begin{align*}
    \mathcal{T}_{\Theta'} = \bigg\{t \leq T \bigg \rvert \Theta' = \Big\{ \theta \in \Theta \Big \rvert \mu_{t+1}(\theta) = \frac{\Pr(\tau > t)}{\Pr(\tau>t+1)} \mu_t(\theta) \Big\} \bigg\}
\end{align*}
Note that $\mathcal{T}_{\emptyset} = \emptyset$ because the boundary constraint must bind. Moreover, for every $t \in \mathcal{T}_\Theta$, we must have $\Pr(\tau>t) = \Pr(\tau>t+1)$ and $\mu_t = \mu_{t+1}$. Therefore, 
\begin{align*}
    &\sum_{\Theta' \subsetneq \Theta} \sum_{t \in \mathcal{T}_{\Theta'}} (\log \Pr(\tau >t) - \log \Pr(\tau > t+1))^{n-1}\\
    &=\sum_{t=0}^T (\log \Pr(\tau >t) - \log \Pr(\tau > t+1))^{n-1}
    > C.
\end{align*}
Thus, there is a nonempty set $\Theta' \subset \Theta$ such that $\sum_{t \in \mathcal{T}_{\Theta'}} (\log \Pr(\tau >t) - \log \Pr(\tau > t+1))^{n-1} > \frac{C}{2^n}$. Consider any $t_0<t_1 \in \mathcal{T}_{\Theta'}$, we will show that there exists $\theta \in \Theta'$ such that $\mu_{t_1}(\theta) \geq \frac{\Pr(\tau > t_0)}{\Pr(\tau>t_0+1)} \mu_{t_0}(\theta)$. Suppose a contradiction that $\mu_{t_1}(\theta) < \frac{\Pr(\tau > t_0)}{\Pr(\tau>t_0+1)} \mu_{t_0}(\theta)$ for every $\theta \in \Theta'$. Because $\mu_{t_0+1}(\theta) < \frac{\Pr(\tau > t_0)}{\Pr(\tau>t_0+1)} \mu_{t_0}(\theta)$ for every $\theta \notin \Theta'$ we can find a sufficiently small $1-\lambda>0$ such that $\lambda\mu_{t_0+1}(\theta) + (1-\lambda)\mu_{t_1}(\theta) < \frac{\Pr(\tau > t_0)}{\Pr(\tau>t_0+1)} \mu_{t_0}(\theta)$ for every $\theta \notin \Theta'$. For $\theta \in \Theta$, we have 
\begin{align*}
    \frac{\Pr(\tau > t_0)}{\Pr(\tau>t_0+1)} \mu_{t_0}(\theta) > \lambda\mu_{t_0+1}(\theta) + (1-\lambda)\mu_{t_1}(\theta).
\end{align*}
This implies $\lambda\mu_{t_0+1}+(1-\lambda)\mu_{t_1} \in \text{int } F\Big(\mu_{t_0}, \frac{\Pr(\tau>t+1)}{\Pr(\tau>t)}\Big)$, which is a contradiction. Thus, there exists $\theta \in \Theta'$ such that $1 \geq \mu_{t_1}(\theta) \geq \frac{\Pr(\tau > t_0)}{\Pr(\tau>t_0+1)} \mu_{t_0}(\theta)$. This implies
\begin{align*}
    \varprod_{\theta \in \Theta'}\bigg(\log \mu_{t_0}(\theta), \log \bigg(\frac{ \Pr(\tau > t_0)}{\Pr(\tau>t_0+1)} \mu_{t_0}(\theta)\bigg) \bigg) \cap \varprod_{\theta \in \Theta'}\bigg(\log\mu_{t_1}(\theta), \log \bigg(\frac{\Pr(\tau > t_1)}{\Pr(\tau>t_1+1)} \mu_{t_1}(\theta)\bigg) \bigg)
\end{align*}
is the empty set for every $t_0<t_1 \in \mathcal{T}_{\Theta'}$.

Because $\big\{ \mu \in \Delta(\Theta) \mid \phi(\mu) \geq \phi(\mu_0) \big\} \subset \text{int } \Delta(\Theta)$ and $\phi(\mu_t)$ is increasing in $t$, we have $\mu_t \in  \big\{ \mu \in \Delta(\Theta) \mid \phi(\mu) \geq \phi(\mu_0) \big\} \subset \text{int } \Delta(\Theta).$ Therefore, for each $\theta \in \Theta$, there is $\underbar{$\mu$}_\theta > 0$ such that $\mu_t(\theta) \geq \underbar{$\mu$}_\theta$ for every $t \in \mathcal{T}$. This implies that for every $t \in \mathcal{T}_{\Theta'}$ we have
\begin{align*}
    \varprod_{\theta \in \Theta'}\bigg(\log \mu_{t}(\theta), \log \bigg( \frac{\Pr(\tau > t)}{\Pr(\tau>t+1)} \mu_{t}(\theta) \bigg) \bigg) \subset \varprod_{\theta \in \Theta'}[\log \underbar{$\mu$}_\theta, 0],
\end{align*}
This implies
\begin{align*}
    \bigcup_{t \in \mathcal{T}_{\Theta'}}\varprod_{\theta \in \Theta'}\bigg(\log \mu_{t}(\theta), \log \bigg( \frac{\Pr(\tau > t)}{\Pr(\tau>t+1)} \mu_{t}(\theta) \bigg) \bigg) \subset \varprod_{\theta \in \Theta'}[\log \underbar{$\mu$}_\theta, 0].
\end{align*}
We showed that two different boxes inside the union have disjoint interiors. This implies
\begin{align*}
     \prod_{\theta \in \Theta'} |\log \underbar{$\mu$}_\theta|&= Vol \bigg( \varprod_{\theta \in \Theta'}[\log \underbar{$\mu$}_\theta, 0]\bigg)  \\
    &\geq \sum_{t \in \mathcal{T}_{\Theta'}} Vol\bigg(\varprod_{\theta \in \Theta'}\bigg(\log \mu_{t}(\theta), \log \bigg( \frac{\Pr(\tau > t)}{\Pr(\tau>t+1)} \mu_{t}(\theta) \bigg) \bigg) \bigg) \\
    &= \sum_{t \in \mathcal{T}_{\Theta'}} (\log\Pr(\tau>t) - \log\Pr(\tau>t+1))^{|\Theta'|} \\
    &\geq \bigg(\sum_{t \in \mathcal{T}_{\Theta'}} (\log\Pr(\tau>t) - \log\Pr(\tau>t+1))^{n-1}\bigg)^{|\Theta'|/(n-1)} \\
    &> \bigg( \frac{C}{2^n} \bigg)^{|\Theta'|/(n-1)}
\end{align*}
where the second last equality follows from the fact that $|\Theta'| \leq n-1$. We can simply choose
\[C = 2^n \max_{\Theta' \subset \Theta} \Big(\prod_{\theta \in \Theta'} |\log \underbar{$\mu$}_\theta|\Big)^{(n-1)/|\Theta'|}, \]
which makes the above inequality false. Therefore, with the chosen $C$, a maximal belief path must stay at $\Phi^*$ for every $t>T$, as desired. 
\end{proof}

\subsection{Attention capture for S-shaped functions in general cases} \label{onlineappendix: S-curve}
We will first provide results of attention capture for S-shaped functions that generalize Proposition \ref{prop:step_and_S-shaped} (ii) and then prove Proposition \ref{prop:step_and_S-shaped} (ii). We first define conditional concavification.

\begin{definition}
(Conditional concavification) Suppose $f$ is S-shaped. For every time $t \in \mathcal{T}$, if $s^*(t) \coloneqq \min \big\{s>t : f(s+1)-f(s) < \frac{f(s)-f(t)}{s-t}\big\}$ exists, define the conditional concavification from $t$ of $f$, $\text{cconv}_t(f): \{s\in\mathcal{T}: s \geq t\} \rightarrow \mathbb{R}_+$ such that
\begin{align*}
    \text{cconv}_t(f)(s) = \begin{cases}
    \frac{s^*(t)-s}{s^*(t)-t}f(t) + \frac{s-t}{s^*(t)-t}f(s^*(t)), &\text{for } s \leq s^*(t)\\
    f(s), &\text{otherwise.}
    \end{cases}
\end{align*}
\end{definition}

We now state the following result of attention capture for S-shaped functions under a general prior.

\begin{proposition} \label{prop: general_S_curve}
    If $f$ is a S-shaped function, every optimal dynamic information structure $(d,\bm{\mu}^C)$ must be such that there exist $t_1<t_2 \leq t_3 \in \mathcal{T}$ such that
    \begin{enumerate}
        \item $\text{supp }(d) = \{1,\dots,t_1\} \cup \{t_2,t_3\}$,
        \item  Obedience constraints bind for all $t \leq t_1$,
        \item $t_3 \in [s^*(t_1+1),s^*(t_1)]$ and either
        \begin{itemize}
            \item $t_2 = t_1+1$, or
            \item $t_3-t_2 \leq 1$ and $s^*(t_1+1) \leq t_2$. 
        \end{itemize}
    \end{enumerate}
\end{proposition}

Before proving Proposition \ref{prop: general_S_curve}, we develop several lemmas establishing properties of $cconv_t(h)$ and $s^*$.
\begin{lemma} \label{lemma: concavity}
Let $T$ be the inflection point of $f$, meaning $f(t) - f(t-1) \lessgtr f(t+1) - f(t)$ if and only if $t \lessgtr T$. Then, for every $t\in\mathcal{T}$, $s^*(t)>T$, $\text{cconv}_t(f)$ is concave, and $\text{cconv}_t(f) \geq f$ over the domain of $\text{cconv}_t(f)$.
\end{lemma}

\begin{proof}[Proof of Lemma \ref{lemma: concavity}]
We will show that $\text{cconv}_t(f)(s)$ has decreasing differences. Within this proof, we abuse notation $s^*(t)$ by $s^*$. By the definition of $s^*$, we must have $f(s^*)-f(s^*-1) \geq \frac{f(s^*-1)-f(t)}{s^*-1-t} > f(s^*+1) - f(s^*)$, which implies $s^*>T$. This means $f(s+1)-f(s)<f(s)-f(s-1)$ for every $s>s^*$. Note also that $\text{cconv}_t(f)$ is linear over the domain $\{t,\dots,s^*\}$. Consider that
\begin{align*}
    \text{cconv}_t(f)(s^*)-\text{cconv}_t(f)(s^*-1) &= \frac{f(s^*)-f(t)}{s^*-t} \\
    &> f(s^*+1)-f(s^*)  \\
    &= \text{cconv}_t(f)(s^*+1)-\text{cconv}_t(f)(s^*),
\end{align*}
which implies the concavity of $\text{cconv}_t(f)$. To see that $\text{cconv}_t(f) \geq f$ over the domain of $\text{cconv}_t(f)$, it suffices to consider the domain of $\{t,\dots,s^*\}$. We will show by induction on $\Delta \in \{0,\dots,s^*-t\}$ that $\text{cconv}_t(f)(s^*-\Delta) \geq f(s^*-\Delta)$. It is clear that the inequality is true when $\Delta = 0$. Assume that the inequality is true for some $\Delta$. By the definition of $s^*$, we must have $f(s^*-\Delta) - f(s^*-\Delta-1) \geq \frac{f(s^*-\Delta) -f(t)}{s^*-\Delta -t}$, which implies
\begin{align*}
    f(s^*-\Delta-1) &\leq \Big( 1 - \frac{1}{s^*-\Delta-t}\Big)f(s^*-\Delta) + \frac{1}{s^*-\Delta-t}f(t)\\
    &\leq \Big( 1 - \frac{1}{s^*-\Delta-t}\Big)\text{cconv}_t(f)(s^*-\Delta) + \frac{1}{s^*-\Delta-t}\text{cconv}_t(f)(t) \\
    &= \text{cconv}_t(f)(s^*-\Delta-1),
\end{align*}
where the inequality follows the linearity of $\text{cconv}_t(f)$ over the domain $\{t,\dots,s^*\}$, as desired.
\end{proof}

\begin{lemma} [Properties of $s^*$] \label{lemma: property_s*}
The following are properties of $s^*$:
\begin{enumerate}
\setlength\itemsep{0em}
    \item[(i)] For every $t<t' \in \mathcal{T}$, we have $s^*(t') - t' \leq s^*(t)-t$. This becomes a strict inequality if $t < T.$
    \item[(ii)] For every $t,t_1,t_2 \in \mathcal{T}$ such that $t\leq t_1 < t_2 \leq s^*(t)$, we have $f(t_1) < \frac{t_2-t_1}{t_2-t} f(t) + \frac{t_1-t}{t_2-t}f(t_2).$
    
\end{enumerate}
\begin{proof}[Proof of Lemma \ref{lemma: property_s*}]

(i) It is sufficient to show the statement in the case of $t'=t+1$. If $s^*(t)-t>1$, then $t < t+1 < s^*(t)$. From (i), we have $s^*(t+1) \leq s^*(t)$, which implies that $s^*(t+1)-(t+1) < s^*(t)-t$. On the other hand, if $s^*(t)-t = 1$, then $f(t+2)-f(t+1) < f(t+1)-f(t)$, which means $t \geq T$. This implies that $f(t+3)-f(t+2) < f(t+2)-f(t+1)$, so $s^*(t+1) = t+2$. Therefore, $s^*(t+1)-(t+1)=1=s^*(t)-t$, as desired. From the proof, we get the strict inequality when $t<T$.

(ii) Consider any $t' \in \{t+1,\dots, s^{*}(t)\}$. 
The definition of $s^*(\cdot)$ implies that 
$f(t') - f(t'-1) > \frac{f(t'-1)-f(t+1)}{t'-t-2}$, which implies the inequality in (iii) when $t_2-t_1=1$. If $t'<s^{*}(t)$, we also have $(t'-t)f(t') -(t'-t-1)f(t'+1)   < f(t+1)$. Thus,
\begin{align*}
    (t'-t)f(t'-1) - (t'-t-2)f(t'+1) < 2f(t+1), 
\end{align*}
which also implies the inequality in (ii) when $t_2-t_1=2$. With a simple induction, we obtain the inequality in (ii) for an arbitrary value of $t_2-t_1$.
\end{proof}
\end{lemma}

We are ready to prove Proposition \ref{prop: general_S_curve}.

\begin{proof} [Proof of Proposition \ref{prop: general_S_curve}]
 Suppose $t_* = \max (\text{supp } \tau \cap [0,T]).$ We proceed with the following steps.

\textbf{Step 1: The DM must be indifferent at every time $t < t_*$.} Since $t_* \in \text{supp }\tau$, set $t_0 = t_*-1$ and $t_1 = t_*+1$. Consider that
\begin{align*}
    \Ex[f(\tau+1)-f(\tau) \mid \tau \in [t_0+1,t_1)] &= f(t_0+2)-f(t_0+1) \\
    &= f(t_*+1)-f(t_*) \\
    &> f(t_*)-f(t_*-1).
\end{align*}
Pivot Lemma implies that the DM must be indifferent at time $t_*-1$. We can do this (backward) inductively to show that the DM must be indifferent at every time $t<t_*$.

\textbf{Step 2: Show $\text{supp } \tau \cap (T,\infty) \ne \emptyset$.} Suppose a contradiction that $\text{supp } \tau \cap (T,\infty) = \emptyset$. Thus, $\max(\text{supp } \tau) = t_*$, which is impossible because the DM is indifferent at time $t_*-1$ but $\Ex[c(\tau-(t_*-1))\mid \tau>t_*-1] = c < \phi^*.$ Therefore, we must have $\text{supp } \tau \cap (T,\infty) \ne \emptyset$.

\textbf{Step 3: Aggregating concave part: $|\text{supp } \tau \cap (T,\infty)| \leq 2$, and $\max \text{supp } \tau - \min (\text{supp } \tau \cap (T,\infty)) \leq 1 $.} This is directly implied by Theorem \ref{thrm:convex_ext} (i) on the domain $(T,\infty)$ and the assumption that $f$ is concave on the domain $(T,\infty).$

Suppose $\text{supp } \tau \cap (T,\infty) = \{t_{pter},t_{ter}\}$, where $t_{ter}\in \{t_{p_{ter}},t_{pter}+1\}.$

\textbf{Step 4: Show $t_{pter} \geq s^*(t_*+1)$.} Suppose a contradiction that  $s^*(t_*+1) \geq  t_{pter}+1$. For small $\epsilon > 0$, we define a new stopping time $\tau'$ and belief path $(\mu'_t)_t$ as following
\begin{align*}
    \Pr(\tau' = t) = \begin{cases}
        \Pr(\tau = t), &\text{if $t\leq t_*$}\\
        \Pr(\tau = t) + \epsilon, &\text{if $t=t_*+1$}\\
        \Pr(\tau = t) - (t_{pter}-t_*)\epsilon, &\text{if $t=t_{pter}$}\\
        \Pr(\tau = t) + (t_{pter}-t_*-1)\epsilon, &\text{if $t=t_{pter}+1$} \\
        0, &\text{otherwise,}
    \end{cases}
    \quad
    \mu'_t = \begin{cases}
        \mu_t, &\text{if $t<t_*+1$}\\
        \mu_{t_*}, &\text{if $t \geq t_*+1$}.
    \end{cases}
\end{align*}
    Boundary constraints clearly hold at every time $t \in \mathcal{T}$. The obedience constraint holds at $t_*+1$ because, since $\max \text{supp } (\tau') = t_{pter}+1$,
    \begin{align*}
        \Ex[c(\tau'-(t_*+1)) \mid \tau'>t_*+1] \leq c(t_{pter}-t_*) \leq \Ex[c(\tau-t_*) \mid \tau>t_*] \leq \phi(\mu_{t_*}).
    \end{align*}
    This also implies the obedience constraint holds at every time $t \geq t_*+1$. It suffices to show that obedience constraints at time $t \leq t_*$ hold. This is clear because the difference of the (unconditional) continuation utilities under $\tau$ and $\tau'$ at $t \leq t^*$ is
    \begin{align*}
        \Ex[c\tau\cdot 1\{\tau>t\}] - \Ex[c\tau'\cdot 1\{\tau'>t\}] = 0,
    \end{align*}
    so obedience constraints still hold for $t \leq t^*$. Thus, $\tau'$ is feasible. However, the designer's utility under $\tau'$ is more than that under $\tau$ because
    \begin{align*}
        \frac{\Ex[f(\tau')] - \Ex[f(\tau)]}{\epsilon} = f(t_*+1) + (t_{pter}-t_*-1)f(t_{pter}+1) - (t_{pter}-t_*)f(t_{pter}) > 0,
    \end{align*} 
    where the inequality is implied by Lemma \ref{lemma: property_s*} (ii), and $s^*(t_*+1) \geq t_{pter}+1$, which is a contradiction

\textbf{Step 5: Show $t_{ter} \leq s^*(t_*)$.} Suppose a contradiction that $t_{ter} > s^*(t_*).$ For small $\epsilon>0$, we define a new stopping time $\tau'$ and belief path $(\mu'_t)_t$ as following
\begin{align*}
    \Pr(\tau' = t) &= \begin{cases}
        \Pr(\tau = t), &\text{if $t< t_*$}\\
        \Pr(\tau = t) - \epsilon, &\text{if $t=t_*$}\\
        \Pr(\tau = t) + (t_{ter}-t_*)\epsilon, &\text{if $t=t_{ter}-1$}\\
        \Pr(\tau = t) - (t_{ter}-t_*-1)\epsilon, &\text{if $t=t_{ter}$} \\
        0, &\text{otherwise,}
    \end{cases} \\
    \mu'_t& = \begin{cases}
        \mu_t, &\text{if $t<t_*$}\\
        p\mu_{t_*} + (1-p)\mu_{t_*-1}, &\text{if $t \geq t_*+1$},
    \end{cases}
\end{align*}
where $p\in[0,1]$ will be specified later. We will choose $p$ so that $(\tau',(\mu'_t)_t)$ is feasible. Boundary constraints clearly hold at every time $t\in\mathcal{T}$. With the same argument as before, the obedience constraints still hold for $t<t^*$ because the continuation utility at $t<t^*$ does not change. Thus, it suffices to check the obedience constraint and the boundary constraint at time $t_*$. 
Let $q = \Pr^I(\tau>t_*|\tau>t_*-1)$, so $q+\epsilon = \Pr^{I'}(\tau>t_*|\tau>t_*-1)$. We set $p = \frac{q}{q+\epsilon} \cdot \frac{1-(q+\epsilon)}{1-q}\in (0,1)$. First, we show the boundary constraint at time $t_*$ under $\tau'$, which is
\begin{align*}
    (p\mu_{t_*}(\theta)+ (1-p)\mu_{t_*-1}(\theta))(q+\epsilon) &\leq \mu_{t_*-1}(\theta) \quad \forall \theta \in \Theta.
\end{align*}
The inequality is true if and only if 
\begin{align*}
    p(q+\epsilon)\mu_{t_*}(\theta) \leq (1-(1-p)(q+\epsilon))\mu_{t_*-1}(\theta).
\end{align*}
The boundary constraint at time $t_*$ under $I$ implies $q\mu_{t_*}(\theta) \leq \mu_{t_*-1}(\theta)$ for every $\theta \in \Theta$. Thus, 
\begin{align*}
    p(q+\epsilon)\mu_{t_*}(\theta) \leq pq(q+\epsilon)\mu_{t_*-1}(\theta) = (1-(1-p)(q+\epsilon))\mu_{t_*-1}(\theta),
\end{align*}
 which implies the boundary constraint. Next, we show the obedience constraint at time $t_*$ under $\tau'$, which is
 \begin{align*}
     \phi(p\mu_{t_*}+ (1-p)\mu_{t_*-1}) &\geq \Ex[c(\tau' -t^*) \mid \tau'>t^*].
 \end{align*}
Since $q\mu_{t^*}(\theta) \leq \mu_{t_*-1} (\theta),$ Jensen's inequality implies
\begin{align*}
    \phi(\mu_{t_*-1}) \geq q\phi(\mu_{t_*}) + (1-q)\phi\Big(\frac{\mu_{t_*-1} - q\mu_{t^*}}{1-q}\Big) \geq q\phi(\mu_{t_*}).
\end{align*}
By Jensen's inequality, we have
\begin{align*}
     \phi(p\mu_{t_*}+ (1-p)\mu_{t_*-1}) &\geq p\phi(\mu_{t_*})+ (1-p)\phi(\mu_{t_*-1}) \\
     &\geq (p+(1-p)q)\phi(\mu_{t_*}) \\
     &= \frac{q}{q+\epsilon} \phi(\mu_{t_*}).
 \end{align*}
The obedience constraint at $t_*$ under $\tau$ implies
\begin{align*}
    \phi(\mu_{t_*}) \geq \Ex[c(\tau-t_*) \mid \tau>t_*] = \frac{\Ex[c(\tau-t_*) 1\{\tau>t_*\}]}{q\Pr(\tau > t_*-1)} = \frac{\Ex[c(\tau'-t_*) 1\{\tau'>t_*\}]}{q\Pr(\tau' > t_*-1)}.
\end{align*}
Thus,
\begin{align*}
    \Ex[c(\tau' -t^*) \mid \tau'>t^*] = \frac{\Ex[c(\tau'-t_*) 1\{\tau'>t_*\}]}{(q+\epsilon)\Pr(\tau' > t_*-1)} \leq \frac{q}{q+\epsilon}\phi(\mu_{t_*}) \leq \phi(p\mu_{t_*}+ (1-p)\mu_{t_*-1}),
\end{align*}
which implies the obedience constraint at time $t_*$ under $\tau'$. Thus, $\tau'$ is a feasible stopping time. However, the designer's utility under $\tau'$ is more that that under $\tau$ because
\begin{align*}
        \frac{\Ex[f(\tau')] - \Ex[f(\tau)]}{\epsilon} =  (t_{ter}-t_*)f(t_{ter}-1) - (f(t_*) + (t_{ter}-t_*-1)f(t_{ter})) > 0,
    \end{align*} 
where the inequality is implied by $t_{ter} >s^*(t_*)$ and $cconv_{t_*}(f)$ is concave, which is a contradiction.

\textbf{Step 6:} Show if the DM is not indifferent at $t_*$, then $t_{ter} = t_{pter}.$ Suppose a contradiction that $t_{pter} = t_{ter}-1.$ We can use the same stopping time $\tau'$ and a belief path $(\mu'_t)_t$ from Step 5 but $\epsilon<0$ and $p=1$ instead. The boundary constraint at time $t_*$ holds because $(q+\epsilon)\mu_{t_*}(\theta) < q \mu_{t_*}(\theta) \leq \mu_{t_*-1}(\theta)$. The obedience constraint at time $t_*$ holds with $\epsilon$ close to $0$ because of the slackness of the obedience constraint at time $t_*$ under $\tau.$ Thus, $\tau'$ is a feasible stopping time. However, the designer's utility under $\tau'$ is more that that under $\tau$ because
\begin{align*}
        \frac{\Ex[f(\tau')] - \Ex[f(\tau)]}{\epsilon} =  (t_{ter}-t_*)f(t_{ter}-1) - (f(t_*) + (t_{ter}-t_*-1)f(t_{ter})) < 0,
    \end{align*} 
where the inequality is implied by  $t_{ter} \leq s^*(t_*)$ and Lemma \ref{lemma: property_s*} (ii).
\end{proof}

We now prove Proposition \ref{prop:step_and_S-shaped} (ii), which is a special case of Proposition \ref{prop: general_S_curve}.

\begin{proof}[Proof of Proposition \ref{prop:step_and_S-shaped} (ii)] Define $\tilde{t} = \min \{t \in \mathcal{T} : s^*(t)-t \leq \phi^*/c\}$.
Choose a feasible stopping time $\tau$ that maximizes $\Ex[f(\tau)]$. Using Theorem \ref{thrm:convex_ext} (ii), we know from the Proposition \ref{prop: general_S_curve} that $\tau$ must be generated by an increasing sequence $1,\dots,t_1$ with a pair of terminal time $(t_2,t_3)$, where $t_1<t_2 \leq t_3.$ If $t_3-t_2 = \phi^*/c$, then the DM is indifferent between continuing paying attention and stopping at time $t_2$ and $t_2=t_1+1$. Thus, an increasing sequence $1,\dots,t_2$ with a pair of terminal times $(t_3,t_3)$ also generates $\tau$. It is sufficient to consider the case that $t_3-t_2< \phi^*/c$. We will show that $t_2=t_3 = t_1+\phi^*/c$ 

Suppose a contradiction that $t_2 < t_3$. Define stopping times $\tau_1,\tau_2$ generated by the same increasing sequence $1,\dots,t_1$ but with different terminal stopping times $(t_2-1,t_3)$ and $(t_2+1,t_3)$. the stopping time $\tau_1$ is well-defined because $t_2-1 \geq t_1$ and $t_3-(t_2-1) \leq \phi^*/c$. The stopping time $\tau_2$ is also well-defined because $t_3>t_{2}$. We obtain the following equations:
\begin{align*}
    \Ex[f(\tau) \mid \tau>t_1] &= \frac{t_3-\phi^*/c - t_1}{t_3-t_2}f(t_2)+ \frac{\phi^*/c + t_1-t_2}{t_3-t_2} f(t_3)  \\
    \Ex[f(\tau_1) \mid \tau_1>t_1] &= \frac{t_3-\phi^*/c - t_1}{t_3-t_2+1}f(t_2-1)+ \frac{\phi^*/c + t_1-t_2+1}{t_3-t_2+1} f(t_3),\\
    \Ex[f(\tau_2) \mid \tau_2>t_1] &= \frac{t_3-\phi^*/c - t_1}{t_3-t_2-1}f(t_2+1)+ \frac{\phi^*/c + t_1-t_2-1}{t_3-t_2-1} f(t_3),.
\end{align*}
Note that, if $t_2 = t_1+1$, the second equation still holds. To see this, we have $\phi^*/c+t_2 > t_3 \geq \phi^*/c+t_1$, so $t_3 =\phi^*/c+t_1$, which implies that the coefficient of $f(t_2-1)$ is equal to $0$. Because $\tau,\tau_1,$ and $\tau_2$ are identical until time $t_1$, the optimality of $\tau$ implies
\begin{align*}
    \Ex[f(\tau_1) \mid \tau>t_1] \leq \Ex[f(\tau) \mid \tau>t_1] \
    &\Longrightarrow (t_3-t_2)f(t_2-1) + f(t_3) \leq (t_3-t_2+1)f(t_2) \\
     \Ex[f(\tau_2) \mid \tau>t_1] \leq \Ex[f(\tau) \mid \tau>t_1]  
    &\Longrightarrow (t_3-t_2)f(t_2+1) \leq (t_3-t_2-1)f(t_1)+f(t_3).
\end{align*}
Because $t_3-t_2 >0$, these two inequalities altogether imply $f(t_2+1)-f(t_2) < f(t_2)-f(t_2-1)$. By the definition of S-shaped, we obtain $t^* < t_2 < t_3$, so $f$ is concave in the interval $[t_2,t_3]$. Consider a stopping time $\tau'$ generated by the same increasing sequence $1,\dots,t_1$ but with a pair of terminal stopping times $(t_{1}+\phi^*/c, t_{1}+\phi^*/c)$. We obtain $\Ex[f(\tau') \mid \tau'>t_1] = f(t_1+\phi^*/c),$ which implies that
\begin{align*}
 &\Ex[f(\tau') \mid \tau'>t_1] - \Ex[f(\tau) \mid \tau'>t_1] 
 = f(t_1+\phi^*/c) - \big(\lambda f(t_2)+ (1-\lambda) f(t_3)\big) >0
\end{align*}
by the concavity of $f$ in the interval $[t_2,t_3],$ where $\lambda = \frac{t_3-\phi^*/c - t_1}{t_3-t_2}$. Because $\tau$ and $\tau'$ are identical up until time $t_1$, we have $\Ex[f(\tau')] > \Ex[f(\tau)]$, which contradicts the optimality of $\tau$. Therefore, we have $t_2=t_3$ which implies $t_2=t_3 = t_1+\phi^*/c$.

We want $t_1$ such that $t_1+\phi^*/c \in [s^*(t_1+1),s^*(t_1)]$. This is equivalent to $s^*(t_1)-t_1 \geq \phi^*/c$ but $s^*(t_1+1) - (t_1+1) < \phi^*/c$. From Lemma \ref{lemma: property_s*} (i), $s^*(t)-t$ is strictly decreasing in $t \in [0,T]$ and $s^*(T) - T = 0$. This means there exists a unique $t_1$ such that $t_1+\phi^*/c \in [s^*(t_1+1),s^*(t_1)]$, which is $t_1 = \tilde{t} \coloneqq \max\{t \in \mathcal{T}: s^*(t)-t \geq \phi^*/c\}$ (if the set is empty, set $\tilde{t} = 0$).
\end{proof}

\subsection{Proof of auxillary results used to show Theorem \ref{thrm:time_con}}\label{onlineappendix: time-consistent}
\begin{proof}[Proof of Lemma \ref{lemma: regular_sufficient}]
    Consider that $\mathcal{P}$ is both convex and compact. Moreover, the function \[p \mapsto \Ex_{(\mu^S,\tau) \sim p }[h(\mu^S,\tau)]\] is linear and continuous. By Bauer's maximum principle, there exists $p^* \in \text{Ext} (\mathcal{P})$ such that $p^*$ is a solution to the designer's optimization problem. Suppose that an extremal and deterministic information structure $I^*$ induces $p^*$. It is without loss to assume that $p^*$ has the support of a countable set because of countable linear constraints of $p$.  We will show that $I^*$ is regular. Assume a contradiction that $I^*$ does not have a terminal time and $\mathcal{T}_{IN}(\bar{I})$ is finite. Suppose that $\max \mathcal{T}_{IN}(\bar{I}) = T^*$. Because $I^*$ does not have a terminal time, there exist $(\mu^1,t_1),\dots,(\mu^{n+2},t_{n+2}) \in  \Delta(\Theta) \times \mathcal{T}$ such that $t_1,\dots,t_{n+2} > T^*$ and $p^*_{\mu^i,t_i} > 0$. We choose $k_1,\dots,k_{n+2}$ satisfying the following conditions:
    \begin{align*}
        \begin{cases}
            \sum_{i=1}^{n+2} k_i \mu^i = 0 \\
            \sum_{i=1}^{n+2} k_i u^*(\mu^i,t_i) = 0,
        \end{cases}
    \end{align*}
    and $(k_1,\dots,k_{n+2}) \ne (0,\dots,0)$.
    A non-zero solution exists because the system of linear equations has $n+2$ variables but only $n+1$ equations. This implies $\sum_{i=1}^{n+2} k_i = \sum_{i=1}^{n+2} \sum_{\theta \in \Theta} k_i \mu^i(\theta)= 0.$ For any $\epsilon \in \mathbb{R}$, we define a probability distribution $p^\epsilon$ as
    $p^\epsilon_{t_i,\mu^i} \coloneqq p^*_{t_i,\mu^i} + k_i\epsilon$ and $p^\epsilon_{t,\mu} \coloneqq p^*_{t,\mu}$ for every $(t,\mu) \ne (t_i,\mu^i)$. We will show that $p^\epsilon \in \mathcal{P}$ when $|\epsilon|$ is small enough. It is clear that  $p^\epsilon_{t,\mu} > 0$ when $|\epsilon|$ is small enough because $p^*_{t_i,\mu^i} > 0$. $p^\epsilon_{t,\mu}$ also satisfies the martingale condition by $\sum_{i=1}^{n+2} k_i \mu^i = 0$. It remains to show $p^\epsilon_{t,\mu}$ satisfies the obedience constraints. Key observations are
    \begin{enumerate}
        \item The continuation beliefs $\mu^C_t$ are still the same for every $t\leq T^*$ and $t>t_{n+2}$: 
        \[\mu^C_t(p^*) = \frac{\sum_{s=t+1}^\infty\sum_{\mu^S \in \Delta(\Theta)} \mu^S p^*_{\mu^S,s}}{\sum_{s=t+1}^\infty\sum_{\mu^S \in \Delta(\Theta)} p^*_{\mu^S,s}} = \frac{\sum_{s=t+1}^\infty\sum_{\mu^S \in \Delta(\Theta)} \mu^S p^\epsilon_{\mu^S,s}}{\sum_{s=t+1}^\infty\sum_{\mu^S \in \Delta(\Theta)} p^\epsilon_{\mu^S,s}} = \mu^C_t(p^\epsilon), \]
        where the second equality follows from $\sum_{i=1}^{n+2} k_i \mu^i = 0$ and $\sum_{i=1}^{n+2} k_i = 0$.
    \item The continuation utility is still the same for every $t \leq T^*$ and $t>t_{n+2}:$
    \[\sum_{s=t+1}^\infty \sum_{\mu^S \in \Delta(\Theta)} p^*_{\mu^S,s}u^*(\mu^S,s) = \sum_{s=t+1}^\infty \sum_{\mu^S \in \Delta(\Theta)} p^\epsilon_{\mu^S,s}u^*(\mu^S,s)\]
    because $\sum_{i=1}^{n+2} k_i u^*(\mu^i,t_i) = 0.$
    \end{enumerate}
    Two key observations imply that the obedience constraint still holds for every $t \leq T^*$ and $t > t_{n+2}$. We assumed in the beginning the DM is not indifferent between stopping and continuing paying attention at every time $t>T^*$. Thus, the obedience constraints at time $t \in (T^*,t_{n+2}]$ slack under $p^*$. Thus, with sufficiently small $|\epsilon|>0$, the obedience constraints at time $t \in (T^*,t_{n+2}]$ still hold under $p^\epsilon$. Therefore, $p^\epsilon \in \mathcal{P}$ for  sufficiently small $|\epsilon|>0$. However, $p^* = \frac{p^\epsilon + p^{-\epsilon}}{2}$ and $p^\epsilon, p^{-\epsilon} \in \mathcal{P}$. This is a contradiction because $p^*$ is an extreme point of $\mathcal{P}$, as desired.
    \end{proof}

\begin{proof}[Proof of Lemma \ref{lemma: indiff_modification}]

Suppose $I$ is IN at every non-stopping at time $t+1$. Fix a history $H^*_t \in \text{supp } (I)$ such that the DM prefers to continue paying attention at $H^*_t$. Define 
\[\text{supp } I_{t+1}(\cdot|H_{t}^*) =: \Big\{(\mu_{t+1} (i),m_{t+1} (i))\Big\}_{i \in M^*}
\]
which are the belief-message pairs at time $t+1$ which realize with positive probability from history $H_t$ labelled with $M^*$. For every $i \in M^*$, define $V_{t+1} (i)$ as the DM's optimal expected utility conditional on paying attention until the time-$t+1$ history $(H^*_t,(\mu_{t+1} (i),m_{t+1} (i)))$ under $I$.  Define
    \begin{align*}
        S \coloneqq \Big\{(p_i)_i &\in \mathbb{R}_{\geq 0}^{|M^*|} \bigg \lvert \sum_{i\in M^*} p_i = 1, v^*\Big(\sum_{i\in M^*} p_i \mu_{t+1} (i),t \Big) \leq \sum_{i \in M^*} p_iV_{t+1} (i) \Big\}.
    \end{align*}
    as the distributions over $M^*$ such that the DM's incentive to stop under the time-$t$ belief is less than her expected continuation utility. 
    
    We first argue $S$ is non-empty, convex, and closed. $S$ is non-empty because we can pick the original distribution $p^*$ under $I$ where $p^*_i = \mathbb{P}^I(\mu_{t+1} = \mu_{t+1} (i), m_{t+1} = m_{t+1} (i) \mid H^*_t)$ for all $i \in M^*$. Furthermore, since $v^*_t$ is convex and continuous, $S$ is convex and closed. 

    Another important observation is that $|\text{supp }(p)| > 1$ for every $p \in S$. To see this, since $I$ is IN at every non-stopping at time $t+1$, $V_{t+1} (i) = v^*(\mu_{t+1} (i),t+1) < v^*(\mu_{t+1} (i),t)$ because of the impatience assumption, which implies $p$ with a singleton support cannot be in $S$. We next argue that for any $p \coloneqq (p_i)_i \in \text{Ext} (S) \subseteq S$, 
    \begin{align*}
      v^*\Big(\sum_{i\in M^*} p_i \mu_{t+1} (i),t \Big) =
      \sum_{i \in M^*} p_iV_{t+1} (i)  \label{indiff_condition_1} \tag{$\diamondsuit$}
    \end{align*}
    i.e., the DM is indifferent. To see this, suppose, towards a contradiction, that
     the LHS of the above equation is strictly less than the RHS in (\ref{indiff_condition_1}). We showed earlier that $|\text{supp } p| > 1$. Let $i_1,i_2 \in M^*$ be such that $p_{i_1},p_{i_2} > 0$. We can perturb $p$ for $\epsilon > 0$ to $p^+$ and $p^-$ as follows:
    \begin{align*}
        &p^+_{i_1} = p_{i_1} + \epsilon, \quad p^+_{i_2} = p_{i_2} - \epsilon, \quad p^+_{i} = p_{i},   \\
        &p^-_{i_1} = p_{i_1} - \epsilon, \quad
        p^-_{i_2} = p_{i_2} + \epsilon, \quad p^-_{i} = p_{i},
    \end{align*}
    for every $i \ne i_1,i_2$. Because we assumed the LHS of (\ref{indiff_condition_1}) is strictly less than the RHS, for sufficiently small $\epsilon$, we must have $p^+,p^- \in S$. But then $p = \frac{p^+ +p^-}{2}$ which this contradicts the assumption that $p$ is an extreme point of $S$, as desired.

    Recall that we denoted the original distribution with $p^*$ and by construction $p^* \in S$. Since $S$ is convex and closed, there exists a finite set $\{p^n\}_n \subset \text{Ext}(S)$ such that $p^*$ can be written as a convex combination of $(p^n)_n$: 
    \[p^* = \sum_{n} \lambda_n p^n \quad \text{where} \quad \sum_{n} \lambda_n = 1, \lambda_n \in [0,1].\] 
    We now abuse notation slightly by associating histories with just their realized belief paths (and omitting accompanying messages). Let $ H^*_t = (\mu^*_1,\dots,\mu^*_t)$. For any $s<t$ and history $H_t^*$, we denote $H_s^* \subset H_t$ as the first $s$ realized beliefs of $H_t$.
    
    We modify the information structure $I$ to $I^*$ as follows:
    \begin{enumerate}[nosep]
        \item $I^*$ and $I$ are identical except for the information structure continuing from $H_{t-1}^*$. 
        \item {Modify time-$t$ beliefs and transition probabilities from $t-1$ to $t$:} \\For the unique history $H^*_{t-1} = (\mu^*_1,\dots,\mu^*_{t-1}) \subset H_t^*$  define the new posterior beliefs $(\mu^n_t)_n$ as follows: for each $n$, 
        \[
        \mu_t^n = \sum_{i \in M^*} p^n_{i} \mu_{t+1} (i) \quad \text{with conditional probability $\lambda_n \mathbb{P}^I(\mu^*_t|H_{t-1}^*)$.}
        \]
        where recall $p^i \in \text{Ext}(S)$ and $\sum_n \lambda_n p^n = p^*$. 
        \item {Modify transition probabilities from $t$ to $t+1$:} \\
        At time $t$ at history $H_t^n := (\mu^*_1,\dots,\mu^*_{t-1},\mu^n_t)$, for each $i \in M^*$,
        \[
        \mathbb{P}^{I^*} \Big( \mu_{t+1} = \mu_{t+1} (i) \Big| H_t^n\Big) = p^n_i
        \]
        
        \item {Preserve structure after $t+1$:} After history $H_{t+1} = (\mu^*_1,\dots,\mu^*_{t-1},\mu^n_t,\mu_{t+1} (i))$ for each $i$, $I^*$ is identical to $I$. 
    \end{enumerate}
    
\begin{figure}[h!]
\centering
\caption{Illustration of Steps 1-4}
    {\includegraphics[width=1\textwidth]{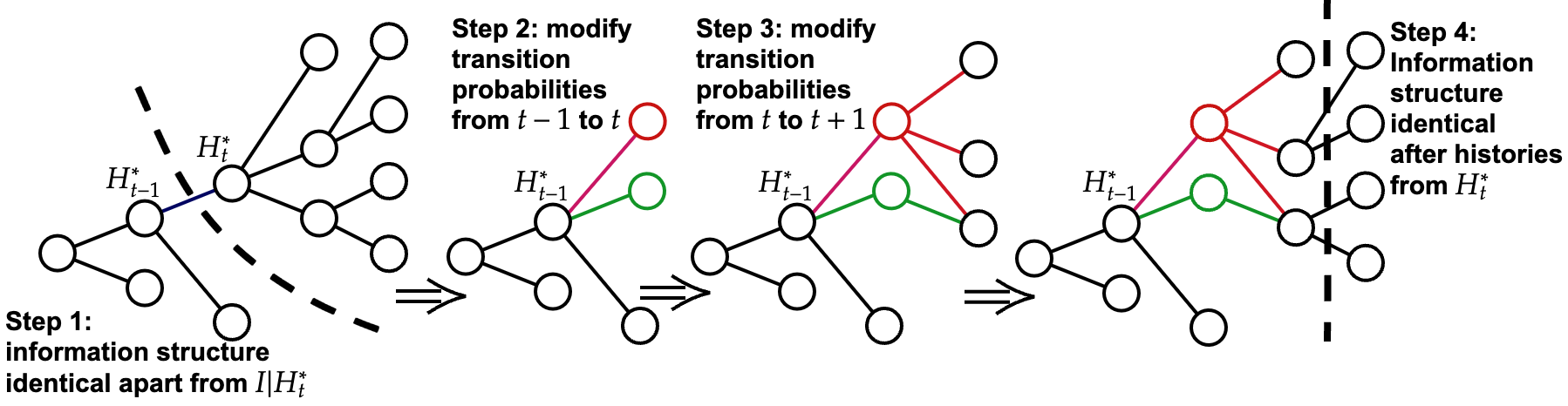}}
    \label{fig:IN_modificationsteps}
\end{figure}

    This modification is depicted in Figure \ref{fig:IN_modificationsteps}.
    Part 1 states that $I^*$ and $I$ are identical except for the information structure continuing from history $H_{t-1}^*$ onwards. Part 2 modifies the transition probabilities from $H^*_{t-1}$ to the new beliefs $(\mu^n_t)_n$. Part 3 modifies the transition probabilities from the new beliefs $(\mu^n_t)_n$ to the original time-$t+1$ histories $(H^*_t, (\mu_{t+1} (i), m_{t+1} (i)))_{i \in M^*}$. Part 4 states that after each time-$t+1$ history the continuation information structure under $I^*$ and $I$ are identical. 

    It remains to verify that this modification (i) is a valid information structure; (ii) preserves the joint distribution; and (iii) replaces $H^*_t$ with non-stopping histories $(H^n_t)_n$ which are IN. Note that the martingale condition at history $H^*_{t-1}$ still holds:
    \begin{align*}
          \sum_{n} \lambda_n \bigg( \sum_{i \in M_s} p^n_{i} \mu_{t+1} (i) \bigg) = \mu^*_t = \sum_{i \in M^*} p^*_{i} \mu_{t+1} (i).
    \end{align*}
    The martingale condition at histories $H_t^n = (\mu^*_1,\dots,\mu^*_{t-1},\mu^n_t)$ also clearly hold, implying $I'$ is a valid information structure. Because $p^n \in \text{Ext}(S)$, then either $|\text{supp } p^n| = 1$ or (\ref{indiff_condition_1}) holds such that the DM is indifferent between continuing and stopping to act. Therefore, $I^*$ is IN at history $H_t = (\mu^*_1,\dots,\mu^*_{t-1},\mu^n_t)$ for every $n$. Finally, observe that by construction, $I$ and $I^*$ yield the same joint distribution over actions states and stopping times as the DM's beliefs at stopping histories are unchanged. We have performed the modification for a given $H_t^*$. Now successively perform this operation until all non-stopping histories at time $t$ are modified to get $I'$ which implies the result. 
\end{proof}

\subsection{Proof of auxillary results used to show Theorem \ref{thrm:attention_persuasion_separable}} \label{onlineappendix: attention_persuasion_separable}

\begin{proof}[Proof of Lemma \ref{lemma: switching_lemma} (switching lemma)] 
    Proof of (i): Write $p' := p^{\epsilon,(\mu_T,T;1,T_1)}$. It is straightforward to verify that obedience constraints are unchanged for times $s<t$ and times $s \geq T_1$. Now consider any time $s$ where $s \in [T,T_1)$. The martingale condition implies that $\mu^C_T(p)$ is a positive linear combination of $\mu^C_s(p)$ and stopping beliefs between time $T$ and $s$ which are at most $\bar{\mu}$ by the definition of $T_1$. Since $\mu^C_T(p) > \bar{\mu}$, $\mu^C_s(p)$ must also be strictly greater than $\bar{\mu}$. Similarly, for small enough $\epsilon$ we also have $\mu^C_s(p') > \bar{\mu}$. This means action $1$ is the optimal action at time $s$ under both $p$ and $p'$. This implies
     \begin{align*}
     &\Pr_{(\mu,\tau) \sim p}(\tau>s) \cdot (\mu^C_s(p')-\mu^C_s(p)) = - (1-\mu_T) \epsilon \\
     \implies &STOP^{p'}_s - STOP^{p}_s = -\epsilon (1-\mu_T) (u(1,1)-u(1,0))
     \end{align*}
     which follows from equation \ref{eqn:C_BELIEF}. Furthermore,
     \begin{align*}
      CONT^{p'}_s - CONT^{p}_s = (u^*(\mu_T) - u(1,1))\epsilon   
     \end{align*}
      which then implies $CONT^{p'}_s - STOP^{p'}_s \geq CONT^p_s - STOP^{p}_s.$ Since $p$ is feasible, the obedient constraint at time $s$ holds under $p'$ so it is feasible too. Part (ii) follows from a symmetric argument. 
      
      Proof of (iii): Write $p' = p^{\epsilon,(\mu_T,T;\bar{\mu},t)}$. Without loss of generality, assume that $\mu_T=0$. With the same argument as before, it is sufficient to verify obedience constraints at time $s \in [T,t).$ Observe 
      \begin{align*}
          \Pr_{(\mu,\tau) \sim p}(\tau>s) \cdot(\mu^C_s(p')-\mu^C_s(p)) = -\bar{\mu}\epsilon 
          \implies &STOP^{p'}_s - STOP^{p}_s \leq  ( u(0,0)-u(1,0))\bar{\mu}\epsilon.
     \end{align*}
     and furthermore,
     \begin{align*}
      CONT^{p'}_s - CONT^{p}_s = ( u(0,0) - u^*(\bar{\mu}))\epsilon.   
     \end{align*}
     Putting these inequalities, together we then have $CONT^{p'}_s - STOP^{p'}_s \geq CONT^p_s - STOP^{p}_s.$ Since $p$ is feasible, the obedient constraint at time $s$ holds under $p'$ so it is feasible too.
\end{proof}

Before proving Lemma \ref{prop:supermodular}, we start with the following lemma.

\begin{lemma}\label{lemma:supermodular_pf_lemma1} Define $\mathcal{P}^*_{ter} \coloneqq  \{p \in \mathcal{P^*}: \text{A terminal time $T(p)$ exists and $p_{\bar \mu, T(p)} > 0$}\}$. Then $\mathcal{P}^*_{full} \cup \mathcal{P}^*_{ter} \ne \emptyset$
\end{lemma}

 \begin{proof}[Proof of Lemma \ref{lemma:supermodular_pf_lemma1}]
     By Corollary \ref{corollary: min_exists} in Online Appendix \ref{onlineappendix:topology}, the optimization problem \[\min_{p \in \mathcal{P}^*} \sum_{t=1}^\infty 2^{-t}p_{\bar{\mu},t}\] has a solution. Suppose that $p^* \in \mathcal{P}^*$ solves such an optimization problem. We will show that $p^*$ must satisfy either condition in Proposition 1. If $p^* \notin \mathcal{P}^*_{full}$, there exists $t \in \mathcal{T}$ such that $p_{\bar{\mu},t} >0$. 
     
     Define $T := \min \{t \in \mathcal{T}: p_{\bar{\mu},t}>0 \}$ as the first time that under the information structure $p^*$ the DM stops with belief $\bar \mu$, and $T_{\mu} := \min \{t \geq T: p_{\mu,t} > 0\}$ for $\mu \in \{0,1\}$ as the times after $T_{\bar \mu}$ under which the information structure leaves the DM with stopping belief either $0$ or $1$. Note that if the information structure does not end at time $T_{\bar \mu}$, this implies $\{t>T : p_{1,t} > 0\}$ and $\{t>T : p_{0,t} > 0\}$ are both non-empty. To see this, observe that otherwise the DM takes the same action for every stopping time $t>T_{\bar \mu}$, which cannot incentivize her to pay attention at time $T_{\bar \mu}$. As such, it will suffice to show that we cannot have $T_0 > T$ or $T_1 > T$. 
     We proceed by considering cases.
     
     \noindent \underline{{Case 1: ${\mu^{p^*}_T < \bar{\mu}}$ or $\big({\mu^{p^*}_T = \bar{\mu}}$ but $CONT^{p^*}_T > STOP^{p^*}_T \big)$.}} Suppose towards a contradiction that $T_0>T$.
     Define $p' = (p^*)^{\epsilon, (\bar{\mu},T;0,T_0)}$. By Switching Lemma (Lemma \ref{lemma: switching_lemma}) $p'$ is also a feasible distribution.\footnote{This is also true for the latter case with a small perturbation $\epsilon$ because the obedience constraint at $T$ under $p^*$ slacks.} Now observe $p' \in \mathcal{P}^*$ because $p$ and $p'$ induce the same marginal distributions of times and actions and $p \in \mathcal{P}^*$.
     However,
     $\sum_{t=1}^\infty 2^{-t}p'_{\bar{\mu},t} - \sum_{t=1}^\infty 2^{-t}p^*_{\bar{\mu},t} = (2^{-T_0} - 2^{-T}) \epsilon < 0,$
     which contradicts the fact that $p^*$ solves the optimization problem $\min_{p \in \mathcal{P}^*} \sum_{t=1}^\infty 2^{-t}p_{\bar{\mu},t}$.
         
     \noindent  \underline{{Case 2: $\mu^{p^*}_T > \bar{\mu}$.}} Suppose towards a contradiction that $T_1 > T$. Define $p' = (p^*)^{\epsilon, (\bar{\mu},T;1,T_1)}$. We can apply the argument from Case 1 to show a contradiction.
     
     \noindent  \underline{{Case 3: $\mu^{p^*}_T = \bar{\mu}$ and $STOP^{p^*}_T = CONT^{p^*}_T$.}} We modify the information structure as follows: let $p'$ be a distribution over $\Delta(\{0,\bar{\mu},1\} \times \mathcal{T})$ such that, with $\epsilon$ in a small neighborhood around $0$, 
     \begin{align*}
         p'_{\bar{\mu},T} &= p^*_{\bar{\mu},T} - \sum_{\mu',t'>T} p^*_{\mu',t'}\epsilon, \\
         p'_{\mu,t} &= p^*_{\mu,t} + p^*_{\mu,t} \epsilon \quad \text{for all } \mu \in \{0,\bar{\mu},1\}, t>T,
     \end{align*}
     and $p'_{\mu,t} = p^*_{\mu,t}$ for every other pair of $(\mu,t)$. Intuitively speaking, we decrease the probability of stopping at belief $\bar{\mu}$ at time $T$ but increase the continuation probability at time $T$ instead. It is easy to see that $p'$ satisfies a martingale condition. Obedience constraints also clearly hold because the information structure after time $T$ has not changed, and the continuation value has not changed: for $t<T$,
     \begin{align*}
     CONT^{p^*}_t - CONT^{p'}_t =   \bigg(\sum_{s>T} \sum_{\mu \in \{0,\bar{\mu},1\}}(v^*(\mu)-cs)p^*_{\mu,s} - \sum_{\mu',t'>T} p^*_{\mu',t'} (v^*(\bar{\mu})-cT)\bigg) \epsilon = 0,
     \end{align*}
     where the equality follows $STOP^{p^*}_T = CONT^{p^*}_T$. Therefore, $p'$ is a feasible distribution. Note that $\Ex^{p'}[f(a_\tau,\tau)]-\Ex^{p^*}[f(a_\tau,\tau)]$ is linear in $\epsilon$. Since $p' \in \mathcal{P}^*$, we have $\Ex^{p'}[f(a_\tau,\tau)] = \Ex^{p^*}[f(a_\tau,\tau)]$ for every $\epsilon $ in a small neighborhood around $0$. But taking a small enough $\epsilon >0$,
     \begin{align*}
         \sum_{t=1}^\infty 2^{-t}p'_{\bar{\mu},t} - \sum_{t=1}^\infty 2^{-t}p^*_{\bar{\mu},t} = \sum_{m,t>T} (2^{-t}-2^{-T}) p^*_{\bar{\mu},t}\epsilon < 0,
     \end{align*}
     which contradicts the fact that $p^* \in \text{argmin}_{p \in \mathcal{P}^*} \sum_{t=1}^\infty 2^{-t}p_{\bar{\mu},t}$.
 \end{proof}

We finally prove Lemma \ref{prop:supermodular}.
\begin{proof} [Proof of Lemma \ref{prop:supermodular}]
     If $\mathcal{P}^*_{full} \ne \emptyset$, there is nothing to show. Next suppose that $\mathcal{P}^*_{full} = \emptyset$. Lemma \ref{lemma:supermodular_pf_lemma1} implies $\mathcal{P}^*_{per} \ne \emptyset$. Pick $p^* \in \mathcal{P}^*_{per}$ and let $T$ be the terminal time of $p^*$. 
     
     Since $p^*$ is a designer's optimal information structure, it must solve the following optimization problem:
     
    \begin{align*}
    &\max_{(p_{0,t},p_{1,t})_{t=1}^{T},p_{\bar{\mu},T}} \Ex^p[\hat{f}(a,\tau)] = \sum_{t=1}^T p_{0,t}f(0,t) + \sum_{t=1}^T p_{1,t}f(1,t) + p_{\bar{\mu},t}f(1,T) \\
    &\text{s.t.} \quad \sum_{t=1}^T p_{0,t} + \sum_{t=1}^T p_{1,t} + p_{\bar{\mu},T} = 1 \tag{Total probability} \\
    &\quad
    \sum_{t=1}^T p_{1,t} + \bar{\mu}p_{\bar{\mu},t} = \mu_0 \tag{Martingale} \\
    &\quad v^*(\mu_0) \leq \sum_{t=1}^T p_{0,t}(u(0,0)-ct) + \sum_{t=1}^T p_{1,t}(u(1,1)-ct) + p_{\bar{\mu},T}(\bar{u}-cT)  \tag{Obedience-0} \\
    &\quad\text{Obedience-}(a,t) \text{ $\forall$ }  a \in \{0,1\} \text{ $\forall$ } t\in\{1,\dots,T-1\} \tag{Obedience-$(a,t)$}\\
    &\quad p_{0,t}\geq 0, p_{1,t} \geq 0, p_{\bar{\mu},T} \geq 0 \text{ for all } t\in\{1,\dots,T\},  \tag{Non-negativity}
    \end{align*}
    
    where $\bar{u} = \frac{u(0,0)u(1,1)}{u(0,0)+u(1,1)}$. Note that all constraints of this optimization problem are linear, and the optimization problem is linear. Define $\mathcal{R}$ and $\mathcal{R}^*$ as the polytope of $((p_{0,t},p_{1,t})_{t=1}^{T},p_{\bar{\mu},T})$ which satisfies the above constraints and solves the above optimization problem, respectively. By standard arguments, we have $\text{Ext } \mathcal{R}^* \subset \text{Ext } \mathcal{R},$ and $\text{Ext } \mathcal{R}^* \subset \mathcal{P}^*$. Pick $p^{**} \in \mathcal{R}^*$ that solves the optimization problem $\min_{p \in \mathcal{R}^*}\sum_{s=1}^T 2^{-s}p_{1,s}$ and $p \in \text{Ext } \mathcal{R}^*$. This exists because extreme points attain the optimum. This implies $p \in \text{Ext } \mathcal{R}^* \subset \text{Ext } \mathcal{R} \cap \mathcal{P}^*.$

\noindent     \textbf{Step 1: $p^{**}_{0,T}, p^{**}_{1,T}, p^{**}_{\bar{\mu},T} > 0$.}
    
    It is clear that $p^{**}_{\bar{\mu},T} > 0$ because $\mathcal{P}^*_{full} = \emptyset$. If $p^{**}_{0,T} = 0$, then the DM takes the same action $1$ at time $T$, so there is no value of the information at time $T$ which violates the obedience constraint at time $T-1$. Therefore, $p^{**}_{0,T}>0$. Similarly, $p^{**}_{1,T}>0.$ 

\noindent     \textbf{Step 2: For each $1 \leq t < T$, at most two of the following four constraints bind: Obedience-$(0,t)$, Obedience-$(1,t)$, $p^{**}_{0,t} \geq 0$, $p^{**}_{1,t} \geq 0$} 
    
    Suppose that both Obedience-$(0,t)$ and  Obedience-$(1,t)$ bind. This means the utilities of taking action $0$ and $1$ at time $t$ are the same, which implies $\mu_t = \bar{\mu}$, and the DM must be indifferent between continuing and stopping at time $t$. If $p^{**}_{0,t} = 0$, then the DM's utility after time $T$ is equivalent to that when she always takes action $1$. Thus, there is no value of information after time $t$, which contradicts the obedience constraint at time $t-1$. Therefore, $p^{**}_{0,t} > 0,$ and, similarly,  $p^{**}_{1,t} > 0,$. This means only two constraints bind, as desired. Suppose that both $p^{**}_{0,t}=p^{**}_{1,t}=0$. This means the designer does not send any message at time $t.$ Then any obedience constraint at time $t$ cannot bind; otherwise, the obedience constraint at time $t-1$ would be violated, as desired. 

\noindent     \textbf{Step 3: For each $1 \leq t < T$ Obedience-0 binds and exactly two of the following four constraints bind: (i) Obedience-$(0,t)$; (ii) Obedience-$(1,t)$; (iii) $p_{0,t} \geq 0$; and (iv) $p_{1,t} \geq 0$} 
    
    Noting that the optimization problem had $2T+1$ variables, then since $p^{**} \in \text{Ext }\mathcal{R}$, $2T+1$ constraints must bind \citep[Proposition 15.2]{simon2011convexity}. From Step 1 we know that none of the non-negativity constraints at time $T$ bind. From Step 2, we know that for each $1 \leq t < T$ at most two of the following four constraints: Obedience-$(0,t)$, Obedience-$(1,t)$, $p_{0,t} \geq 0$, $p_{1,t} \geq 0$. If exactly two of them bind for all $1 \leq t < T$, then in addition to the Total probability, Martingale, and Obedience-0 constraints, we have $3 + 2(T-1) = 2T+1$ constraints as desired. 
    
 \noindent    \textbf{Step 4: $p^{**}$ satisfies (i) if $p^{**}_{1,t} + p^{**}_{0,t} > 0$ then DM is indifferent at $t$; and (ii) if $\mu^{C}_t(p^{**}) \ne \bar{\mu},$ then $p^{**}_{1,t} = 0$.}
    
    \begin{enumerate}
        \item[(i)] If $t<T$ such that $p^{**}_{1,t}+p^{**}_{0,t}>0$, then only one of two constraints $p_{1,t}\geq 0$ and $p_{0,t} \geq 0$ bind. From Step 3, either Obedience-$(0,t)$ or Obedience-$(1,t)$ must bind, which means the DM is indifferent at time $t$.
        \item[(ii)] If $t<T$ such that $\mu^C_t(p^{**}) \ne \bar{\mu}$, then at most one of two constraints Obedience-$(0,t)$ or Obedience-$(1,t)$ bind. From Step 3, one of $p_{1,t}$ and $p_{0,t}$ equals $0$. Suppose that $p_{1,t} \ne 0$. From the previous point, the DM must be indifferent at time $t$ and $p_{0,t} = 0$. This together implies $\mu^C_t(p^{**}) < \bar{\mu}$. Let $T_0 = \min\{s>t : p^{**}_{0,s} > 0\}$ and $p' = (p^{**})^{\epsilon,(1,t;0,T_0)}$. By Switching Lemma (Lemma \ref{lemma: switching_lemma}) $p'$ is also a feasible distribution and induces the same marginal distributions of actions and stopping times, which implies $p' \in \mathcal{P}^{**}.$ However,
        \begin{align*}
        \sum_{s=1}^T 2^{-s} p'_{1,s} - \sum_{s=1}^T 2^{-s} p^{**}_{1,s} = (2^{-T} - 2^{-t}) \epsilon < 0,
        \end{align*}
        which contradicts the fact that $p^{**} \in \text{arg min}_{p \in \mathcal{R}^*} \sum_{s=1}^T 2^{-s}p_{1,s}$.
    \end{enumerate}

    Note that part (i) or Step 4 is exactly equal to condition (iii) of $\mathcal{P}^*_{bad}$ so it remains to show that for all $t < T$, $p^{**}_{1,t} = 0$ so that we have a pure bad news dynamic information structure (condition (ii) of $\mathcal{P}^*_{bad}$).
    
 \noindent    \textbf{Step 5: truncate $p^{**}$.} 
    
    If $p^{**}_{1,t} = 0$ for every $t<T$, then $p^{**} \in \mathcal{P}^*_{bad}$, as desired. Suppose that $p{**}_{1,t} > 0$ for some $t<T$. Define $T_0:= \min\{t \leq T: p^{**}_{1,t} > 0\}$. Step 4 implies $\mu^C_t(p^{**}) = \bar{\mu}$. We modify the information structure as follows: $p' = p^{\epsilon,(1,T_0 ; \bar{\mu},T)}$.    
    Switching lemma (Lemma \ref{lemma: switching_lemma}) implies that $p'$ is also a feasible distribution. Note that $p' \in \mathcal{P}^*$ because both $p^{**}$ and $p'$ yield the same joint distribution of actions and times (since we break DM indifference in favour of action $1$). 
    
    Now observe that under $p^{**}$, when the DM is indifferent at time $T_0$ with belief $\bar{\mu}$ (from Step 4), the designer provides more information until time $T$. On the other hand, under $p'$, when the DM reaches the stopping belief $\bar{\mu}$ at time $T_0$, the designer prefers to stop providing further information. To see this, assume, towards a contradiction, that this is not the case so that the designer strictly prefers to provide further information at time $T_0$ when the continuation belief is $\bar \mu$. We have shown that under $p'$, the designer obtains the same utility and furthermore, $p'_{\bar \mu, T_0} > 0$. But since the designer strictly prefers to continue at time $T_0$, it cannot find it optimal to have the DM stop at time $T_0$ with belief $\bar \mu$ which contradicts the optimality of $p'$ and hence $p^{**}$.

    Truncate $p^{**}$ at time $T_0$ by constructing $p^{***}$ as follows:
    \begin{align*}
        p^{***}_{m,t} \coloneqq \begin{cases}
            p^{**}_{m,t} &t<t_0 \\
            p^{**}_{m,t} & t=t_0, m \in \{0,1\} \\
            \sum_{s>t_0} \sum_{m} p^{**}_{m,s} & t=t_0, m = \bar{\mu} \\
            0  & t>t_0,
        \end{cases}
    \end{align*}
    and by the previous argument, $p^{***}$ and $p^{**}$ yield the same designer's utility. Thus, $p^{***} \in \mathcal{P}^*$. From the definition of $t_0$, it is simple to show that $p^{***} \in \mathcal{P}^*_{bad}$, as desired.
\end{proof}

\begin{proof}[Proof of Lemma \ref{lemma: pasting} (pasting lemma)]
    Let $p \in \Delta(\{0,\bar{\mu},1 \} \times \mathcal{T})$ be a feasible distribution with prior $\mu_0.$ Define a probability distribution $p^T$ as follows:
    \begin{align*}
       p^T_{\bar{\mu},t} &= p_{\bar{\mu},t} + p_{1,t}, \quad\quad
       p^T_{1,t}= 0,
    \end{align*}
    for every $t<T$ and every other probabilities are the same. Define 
    \[t^* \coloneqq \sup \{t \in \mathcal{T} \mid \Ex_{(\mu,\tau) \sim p^t}[\mu] \geq \bar{\mu}\}\]
    We define a new probability distribution $p'$ with prior belief $\bar{\mu}$ as follows:
    \begin{align*}
        p'_{\mu,t} &= p^{t^*}_{\mu,t}, \quad\quad \forall t \ne t^*, \quad \forall \mu \\
        p'_{\bar{\mu},t} &= p_{\bar{\mu},t} + \delta \\
        p'_{1,t} &= p_{1,t} - \delta,
     \end{align*}
    where $\delta > 0$ is an appropriate constant such that $\Ex_{(\mu,\tau) \sim p'}[\mu] = \bar{\mu}.$ $p'$ is still a valid probability distribution by the definition of $t^*.$ Clearly, $p$ and $p'$ yield the same joint distribution of action and stopping time. It is sufficient to show that $p'$ is a feasible distribution. 

    An important observation is that $\mu^C_t(p) > \mu^C_t(p')$ for every $t < t^*$ because we increase and decrease the stopping probability at the beliefs $\bar{\mu}$ and $1$ by the same amount at every time.
    Moreover, $\mu^C_t(p') > \bar\mu$ for every $t < t^*$ since $\mu_0 > \bar{\mu}$ and stopping beliefs at time in  $[0,t)$ under $p'$ are only $0$ and $\bar{\mu}$. This means action $1$ is optimal at continuation beliefs $\mu^C_t(p)$ and $\mu^C_t(p')$ for every $t < t^*$. For every $t<t^*$, we obtain
    \begin{align*}
       STOP^{p'}_t - STOP^{p}_t = 
       \Pr_{(\mu,\tau) \sim p}(\tau > t) \cdot \big(\Ex_{\theta \sim \mu^C_t(p') }u(1,\theta) - \Ex_{\theta \sim \mu^C_t(p) }u(1,\theta)\big). 
    \end{align*}
    However,
    \begin{align*}
        CONT^{p'}_t - CONT^{p}_t &= A \big(u(1,1) - \Ex_{\theta \sim \bar{\mu} }u(1,\theta) \big), 
    \end{align*}
    where $A$ is the change of the unconditional probability of stopping at belief $1$ after time $t$. The martingale constraint time $t$ implies   
    \begin{align*}
    \Pr_{(\mu,\tau) \sim p}(\tau > t) \mu^C_t(p') = \Pr_{(\mu,\tau) \sim p}(\tau > t) \mu^C_t(p) - A \cdot 1 + A \cdot \bar{\mu}.        
    \end{align*}
    By linearity of $\mu \mapsto \Ex_{\theta \sim \mu} u(1,\theta)$, we obtain
    \begin{align*}
      STOP^p_t - STOP^{p'}_t = CONT^p_t - CONT^{p'}_t \Rightarrow CONT^{p'}_t- STOP^{p'}_t  = CONT^{p}_t- STOP^{p}_t \geq 0   
    \end{align*}
    which $p'$ is a feasible distribution and induces the same DM surplus. 
\end{proof}

\newpage
\section{More general time preferences}\label{appendix:nonlinearcost}
In the main text we assumed the DM pays a constant cost per-unit time. Observe that Theorems \ref{thrm:reduction} and \ref{thrm:time_con} required impatience and nothing else, so they continue to hold as stated for more general time preferences. We now observe that results on designer-optimal structures (e.g., Theorems \ref{thrm:convex_ext} and \ref{thrm:attention_persuasion_separable}) extend readily to nonlinear waiting costs $c(\tau)$ where $c: \mathcal{T} \to \mathbb{R}$ is a strictly increasing function.\footnote{
Beyond additively separable time preferences, an earlier version of this paper developed some of our results for exponential discounting} As in the main text, suppose that the designer's payoff is a strictly increasing function $f$. This model is isomorphic (up to an integer constraint) to a model in which the DM pays a unit cost per-unit time, and the designer's value function if $f \circ c^{-1}$. To see this, observe that $f(\tau) = f (c^{-1}(c(\tau)))$
hence we can perform a time change by defining units of time in terms of $c(\tau)$. Hence, if $I$ is optimal for attention capture under $f$ with cost $c$ if and only if it is also optimal for attention capture under value $f \circ c^{-1}$ with cost $1$ per-unit time and the time-change $c$. 

This immediately implies that a model with nonlinear value of attention $f$ and linear constant per-unit cost can be mapped to a model with nonlinear $f: \mathcal{T} \to \mathbb{R}$ and $c: \mathcal{T} \to \mathbb{R}$ by considering $f \circ c^{-1}$. For instance, 
\begin{itemize}[nosep]
    \item[(i)] If $f$ is linear and $c$ is concave then $f \circ c^{-1}$ is convex hence the designer-optimal structure is the same form as when $f$ is convex and $c$ is linear as in Corollary \ref{cor:convexconcave} (i).  
    \item[(ii)] If $f$ is linear and $c$ is convex then $f \circ c^{-1}$ is concave hence the designer-optimal structure is the same form as when $f$ is concave and $c$ is linear as in Corollary \ref{cor:convexconcave} (ii).  
\end{itemize}

\clearpage 

\end{document}